%% file: 01_paper.tex
\documentclass[11pt,a4paper]{article}
\pdfoutput=1
\usepackage{jheppub}
\usepackage[utf8]{inputenc}

\usepackage[colorlinks=true]{hyperref}

\makeatletter
\newcommand{\hlfootnote}[1]{%
  \stepcounter{footnote}%
  \protected@xdef\@thefnmark{\thefootnote}%
  \hyperlink{hlfn.\the\c@footnote}{\@makefnmark}%
  \@footnotetext{\hypertarget{hlfn.\the\c@footnote}{}#1}%
}
\makeatother

\usepackage{colortbl}
\usepackage{color}
\usepackage[dvipsnames]{xcolor}
\usepackage{soul}
\usepackage{amsmath}
\usepackage{amssymb}
\usepackage{pifont}
\usepackage{mathtools}
\usepackage{esint}
\usepackage{pifont}
\usepackage{fnpct}
\usepackage{bbold}
\usepackage{bm}
\usepackage{graphbox}
\usepackage{comment}
\usepackage{feynmf}
\usepackage{cleveref}
\usepackage{bbm}
\usepackage{verbatim}   
\usepackage{subfigure}
\usepackage{caption}
\usepackage{float}

\usepackage{acronym}
\usepackage[export]{adjustbox}

\usepackage{amsfonts}
\usepackage{mathrsfs}
\usepackage{graphicx}
\usepackage{multirow}
\usepackage{slashed}
\usepackage{caption}

\usepackage{upgreek}
\usepackage{parskip}
\usepackage{scalerel}
\usepackage{bigints}
\usepackage{multicol}
\usepackage{multirow}
\usepackage{tensor}
\usepackage{ragged2e}
\usepackage{mathtools}
\usepackage{bbm} 
\usepackage{setspace}
\usepackage{ytableau}
\usepackage{dirtytalk}
\usepackage{diagbox}
\usepackage{enumitem}
\usepackage{scalefnt}

\usepackage{needspace}

\let\oldbibitem\bibitem
\renewcommand{\bibitem}{\Needspace{3\baselineskip}\oldbibitem}

\newcommand{\extable}{0.25ex}

\usepackage{imakeidx}
\makeindex[columns=3, title=Alphabetical Index, intoc]

\newcommand\numberthis{\addtocounter{equation}{1}\tag{\theequation}}

\newcommand\spa{\enskip} 
\newcommand{\esp}{\phantom{.}}

\newcommand{\txn}[1]{\textnormal{#1}}

\newcommand\eu[1]{\scaleobj{1.2}{ e^{ \scaleobj{1}{#1} } } }

\newcommand\me{\scaleobj{1.15}{\textsl{g}}} 

\newcommand{\del}{\partial}

\newcommand{\de}{\textnormal{d}}

\newcommand{\upsymb}[2]{
\mathrel{\raisebox{#1}{#2}}
}

\newcommand{\eps}{\varepsilon}
\newcommand{\epss}{\scaleobj{1.1}{\mathcal{\varepsilon}}}

\newcommand{\Rep}{\mathcal{R}}

\newcommand{\cmark}{\ding{51}}
\newcommand{\xmark}{\ding{55}}

\definecolor{myorange}{RGB}{227, 115, 0}
\definecolor{myred}{RGB}{185,0,0}
\newcommand\tcolor[1]{\textcolor{Blue}{#1}}

\newcommand\red[1]{\textcolor{Red}{#1}}
\newcommand\green[1]{\textcolor{OliveGreen}{#1}}

\newcommand{\simtworep}{  \txn{\textbf{S}}_2  }

\newcommand{\asimtworep}{  \txn{\textbf{A}}_2  }
\newcommand{\ccasimtworep}{  \overline{\txn{\textbf{A}}}_2  }

\newcommand{\dualccasimtworep}{  \txn{\textbf{A}}_{N-2}  }

\newcommand{\fundrep}{  \txn{\textbf{F}}_1  }
\newcommand{\ccfundrep}{  \overline{\txn{\textbf{F}}}  }

\title{On asymptotically and anomaly-free SU(N) chiral gauge theories for arbitrarily large N}

\author[]{Kort Beck,}
\emailAdd{kgr6@illinois.edu}

\author[]{Patrick Draper}
\emailAdd{pdraper@illinois.edu}

\affiliation[]{Illinois Center for Advanced Studies of the Universe \textnormal{\&}}
\affiliation[]{Department of Physics, University of Illinois at Urbana-Champaign, Urbana, IL 61801, USA}

\abstract{A large catalog of asymptotically free and anomaly-free SU$(N)$ chiral gauge theories that admit a large-$N$ limit was constructed by Eichten, Kang, and Koh. Here we compute the global structure of their symmetry groups and tabulate the 't Hooft anomalies of the faithful symmetries, including those that are visible only in backgrounds carrying fractional flux. Most theories in the catalog contain fundamentally charged matter and support no genuine one-form electric center symmetry. However,  fractional fluxes still arise  from the faithful quotient and contribute to the anomaly in the same way as a two-form background for a one-form symmetry. The anomalies are computed using two methods, first with the descent procedure, and second by placing the theory on a four-torus, which supports twisted fluxes for background gauge fields. We discuss a selection of candidate IR behaviors and present a general discussion of how matching the anomalies associated with fractional flux can constrain infrared descriptions beyond the imposition of ordinary zero-form anomaly matching at the level of triangle diagrams. Based on large-$N$ reasoning, we formulate a new proposal for the IR behavior of a pair of models, consistent with the ordinary zero-form 't Hooft anomaly matching conditions. In this case the matching condition involving fractional flux is satisfied automatically as a consequence of ordinary matching.}

\begin{document}
\maketitle

\newpage

\setcounter{page}{1}
\section{Introduction}

The infrared behavior of chiral gauge theories is difficult to study. Possible phases include confinement with or without chiral symmetry breaking, condensation of colored composites that give mass to some of the gauge bosons, flow to an infrared CFT, and perhaps others. In general, the most rigorous tool available to test candidate scenarios is 't Hooft anomaly matching~\cite{tHooft:1979}.

't Hooft anomalies are diagnosed by coupling the global symmetries to background gauge fields. It is not possible to gauge one or more of the symmetries simultaneously while preserving gauge invariance, and the partition function acquires a non-trivial phase under some subset of the transformations. In this sense a 't Hooft anomaly is an obstruction to gauging a global symmetry. They are also renormalization group invariants, and they must be reproduced at every energy scale by the relevant light degrees of freedom at that scale. Thus, a theory that possesses a 't Hooft anomaly cannot be trivially gapped \cite{tHooft:1979}. The matching of 't Hooft anomalies is a powerful tool to constrain the possible infrared phases of quantum field theories. 

When the global symmetry group of a theory involves both ordinary and generalized symmetries, there can be mixed anomalies between symmetries of different types. Anomalies involving only zero-form symmetries, which are gauged via coupling to ordinary one-form gauge fields, are sourced by integer field strength fluxes, i.e. instantons. By contrast, anomalies where a one-form symmetry is gauged via two-form gauge fields are usually associated with fractional fluxes. In recent years, 't Hooft anomaly matching involving generalized symmetries has been used to further constrain the low energy behavior of strongly coupled gauge theories; see for instance \cite{Gaiotto:Kapustin:Komargodski:Seiberg:2017,Shimizu:Yonekura:2017,Tanizaki:Misumi:Sakai:2017,Tanizaki:Kikuchi:Misumi:Sakai:2018,Anber:Poppitz:2018,Anber:Poppitz:2019,Anber:Poppitz:2018:2,Yonekura:2019,Wan:Wang:2019,Poppitz:Ryttov:2019,Cherman:Jacobson:Tanizaki:Mithat:2020,Cordova:Ohmori:2019,Komargodski:Ohmori:Roump:Seif:2021,Anber:Lohitsiri:Sulejmanpasic:2023,Hayashi:Tanizaki:Unsal:2025}. For ``ordinary" relativistic $4d$ gauge theories, the relevant investigation mostly concerns the global structure of the  symmetry groups. It is of particular interest to compute this structure and investigate the anomaly matching implications for chiral gauge theories~\cite{Bolognesi:Konishi:Luzio:2020,Bolognesi:Konishi:Luzio:2021,Bolognesi:Konishi:Luzio:2022,Bolognesi:Konishi:2023,Bolognesi:Konishi:Luzio:2023:review,Bolognesi:Konishi:Luzio:Orso:2025,Smith:Karasik:Lohitsiri:Tong:2022,Anber:Hong:Son:2022,Anber:Chan:2023}.

Here we develop this program further by studying a large catalog of $4d$ SU($N$) chiral gauge theories enumerated by Eichten, Kang, and Koh (EKK) in \cite{Eichten:Kang:Koh:1982}. These theories satisfy gauge anomaly cancellation and (in most cases) remain asymptotically free for arbitrarily large $N$. The matter content is given by $N_1$ left-handed Weyl fermions in the symmetric representation, $N_2$ in the anti-symmetric and $N_3$ in the anti-fundamental. For some combinations of $N_1$, $N_2$, and $N_3$, one or two of the color representations need to be complex-conjugated. There are 30 possible combinations of $N_i$ that render an anomaly-free and asymptotically free theory. To the best of our knowledge, only some particular models within this class have been studied in the literature as isolated cases. 

Our main goal is to compute the global symmetry structure and mixed anomalies across the catalog, and to assess the implications of anomaly matching for the candidate IR phases. In general the group that acts faithfully on the fermions is a quotient $G=\widetilde{G}/\mathcal{C}$, where $\widetilde{G}$ is the product of the gauge, flavor, and axial factors, and $\mathcal{C}$ is a product of redundant factors. These theories generically possess neither genuine global one-form symmetries nor discrete zero-form symmetries. Anomalies obstruct the gauging of $\widetilde{G}$ and therefore prevent $\mathcal{C}$ from appearing as a one-form symmetry of a gauged theory. However, the ordinary anomalies do  not restrict couplings to background fields; rather, the allowed backgrounds are determined by the quotient. Some background gauge fields for $G$ are not allowed backgrounds for $\widetilde{G}$, and these assign fractional fluxes to the factors of $\widetilde{G}$. The anomalies computed below include 't Hooft anomalies of $G$ that are sensitive to its global structure. 

The anomalies are computed in two formalisms. In the first approach, we use the Stora-Zumino descent procedure, beginning from the $6d$ anomaly functional and integrating down to $4d$. In this case the fractional fluxes are supplied by  two-form background gauge fields which are introduced as though the corresponding center symmetries were gauged. In the second approach, we place  the theory on a four-torus and introduce fractional fluxes via twisted configurations of the background gauge fields. That the two approaches lead to the same results is a consistency check of the computations.

These methods have been used in \cite{Bolognesi:Konishi:Luzio:2020,Bolognesi:Konishi:Luzio:2021,Bolognesi:Konishi:Luzio:2022,Bolognesi:Konishi:2023,Bolognesi:Konishi:Luzio:2023:review,Bolognesi:Konishi:Luzio:Orso:2025,Smith:Karasik:Lohitsiri:Tong:2022,Anber:Hong:Son:2022,Anber:Chan:2023} to study some of the simplest chiral gauge theories. For example, 2-index chiral gauge theories, or models with no fundamentally charged matter, have been studied in \cite{Anber:Hong:Son:2022,Anber:Chan:2023}. In the EKK catalog $N_3$ is fixed by $N$, and for some particular values of $N$ it happens that $N_3 = 0$. Since the matter content consists of fermions in two-index symmetric and anti-symmetric tensor representations only, these theories do admit a genuine one-form symmetry and sometimes an extra discrete zero-form symmetry. The presence of the discrete symmetry, in particular, leads to differences with the three-species models that are our focus.

Another tool used to gain control over strongly coupled gauge theories is supersymmetry. We will not leverage supersymmetry in this work, but see \cite{Craig:Essig:Hook:Torroba:2011,Craig:Essig:Hook:Torroba:2011:2,Csaki:Murayama:Telem:2021,Csaki:Murayama:Telem:2022,Kondo:Murayama:Sylber:2022,Csaki:Meade:Terning:2004,Goh:Murayama:Singh:Suter:Wong:2025} and references therein, where SUSY techniques were employed to study different classes of chiral gauge theories. There has also been considerable recent work on the old problem~\cite{Nielsen:Ninomiya:1980,Nielsen:Ninomiya:1981,Nielsen:Ninomiya:1981:2,Ginsparg:Wilson:1982,Kaplan:1992,Shamir:1993,Neuberger:1997,Luscher:1998,Luscher:1999,Eichten:Preskill:1986,Grabowska:Kaplan:2015,Luscher:2000,Golterman:2000,Neuberger:2001,Poppitz:Shang:2010} of putting chiral fermions on the lattice and coupling them to gauge fields~\cite{Wang:Weng:2018,Berkowitz:Cherman:Jacobson:2023,Seifnashri:2026,Lu:Seifnashri:Shao:2026,Thorngren:Preskill:Fidkowski:2026,Zakharov:Ueda:Verstraete:Beenakker:2026,Baig:Chen:Cherman:Neuzil:2026,Dang:Karur:Sen:2026,Lamm:Roggero:Singh:Spagnoli:2026,Lu:Shao:2026,Onoda:2026}. Lattice formulations are outside our scope, but a better understanding of the symmetry and anomaly structure of continuum theories might one day be useful for this problem.

This work is organized as follows. In Section \ref{section_Intro_to_theories} we review the class of $4d$ SU($N$) chiral gauge theories introduced in \cite{Eichten:Kang:Koh:1982}, whose matter content consists of $N_1$, $N_2$ and $N_3$ left-handed Weyl fermions in the symmetric, anti-symmetric and anti-fundamental representations, and which satisfy gauge anomaly cancellation and asymptotic freedom for arbitrarily large $N$. In Section \ref{section_global_symm_group_3_flavors} we determine the global symmetry group that acts faithfully on the fermions, showing that the candidate discrete zero-form symmetries are redundant with subgroups of the continuous symmetries, and determining the discrete identifications encoded in a set of cocycle conditions. In Section \ref{section_descent_anomalies} we compute the mixed 't Hooft anomalies through the Stora–Zumino descent procedure. In Section \ref{section_twisted_gauge_field} we recompute these anomalies with the complementary formalism in which fractional fluxes arise from twisted gauge field configurations on a four-torus. In Section \ref{fluxes_section} we evaluate the background fluxes in both approaches and establish their equivalence. Section \ref{section_witten_anomalies} briefly addresses the presence of Witten anomalies for some of these theories. In Section \ref{section_large_N} we discuss expectations from a large $N$ analysis. We review previous results in the literature for a particular theory among those studied here (denoted $R_3$). Extending this analysis to two new theories, $R_5$ and $R_{17}$, we   propose a symmetry breaking pattern and a  set of massless composite fermions that   match the ordinary triangle diagram anomalies. In Section \ref{section_anomaly_matching} we show that these candidate IR descriptions for $R_5$ and $R_{17}$ also reproduce the anomalies associated with unbroken symmetries and fractional fluxes, as a consequence of ordinary matching. Section \ref{fate_IR_dyna} discusses the fate of the IR dynamics for the theories more broadly, examining possible massless composite fermionic and bosonic operators that can be built out of the UV fermions,  constraints that follow from the $a$-theorem, and the possibility of Banks–Zaks fixed points. We collect our conclusions and outlook in Section \ref{concluding_remarks}. Finally, several technical results and supporting material are relegated to the appendices: explicit evaluation of the Dirac indices that encode the various mixed anomalies (Appendix \ref{appendix_fluxes}), the conventions followed when evaluating the three-loop beta function (Appendix \ref{appendix_beta_func}), a brief overview of models of the same kind as those studied here but that remain asymptotically free only for a finite range of $N$ (Appendix \ref{appendix_bar_R_models}), the minimization of the tree-level fermion bilinear potential as suggested by the most attractive channel hypothesis (Appendix \ref{appendix_clases_MAC}), and a catalog of various classes of chiral gauge theories that have been considered in the existing literature (Appendix \ref{appendix_clases_CGT}).

\input{02_su_n_CGT}

\input{03_global_symm_group}

\input{04_anomalies}

\input{05_anomalies_2}

\input{06_fluxes}

\input{05_02_witten_anomalies}

\input{07_Large_N}

\input{08_anomaly_matching}

\input{09_IR_dynamics}

\input{10_concluding_remarks}

\section*{Acknowledgments}

This work was supported in part by the U.S. Department of Energy, Office of Science, Office of High Energy Physics under award number DE-SC0015655.

\begin{center}
\section*{Appendix}
\end{center}

\appendix

\input{Appendix_0}

\input{Appendix_A}

\input{Appendix_bar_R_models}

\input{Appendix_B}

\input{Appendix_C}

\bibliography{refs}

\newpage
\thispagestyle{empty}
\phantom{a}

\end{document}

%% file: 02_su_n_CGT.tex
\setlength{\jot}{13pt}

\section{Asymptotically and anomaly-free SU(N) chiral gauge theories}
\label{section_Intro_to_theories}

A general representation $R_i$ for chiral gauge theories with gauge group SU$(N)$ that satisfy anomaly cancellation and asymptotic freedom for arbitrarily\hlfootnote{The initial value for $N$ such that a theory is asymptotically free depends on the multiplicity of the representation, see Tables \tcolor{\ref{table_rep_multiplictly_1}} and \tcolor{\ref{table_rep_multiplictly_2}}.} large $N$ in four space-time dimensions can be written as \tcolor{\cite{Eichten:Kang:Koh:1982}}
\begin{align}
R_i = 
N_1
\raisebox{-3.5pt}{
\scaleobj{0.75}{
\begin{ytableau}
~ & 
\end{ytableau}}}
\oplus N_2
\raisebox{3.5pt}{
\scaleobj{0.75}{
\begin{ytableau}
~ \\
~
\end{ytableau}}}
\oplus N_3
\raisebox{-3.2pt}{
\scaleobj{0.75}{
\begin{ytableau}
\overline{\phantom{\Big|aa.}} 
\end{ytableau}}}
\label{general_N1_N2_N3_rep}
\spa\spa.
\end{align}
The matter content in these theories, given in terms of Weyl fermions, is
\setlength{\jot}{9pt}
\begin{equation}
\begin{gathered}
\psi^{n_1 , ij} \esp \in
\raisebox{-3.5pt}{
\scaleobj{0.75}{
\begin{ytableau}
~ & 
\end{ytableau}}}
\spa\spa \simtworep
\spa\spa,\spa\spa
\lambda^{n_2 , ij} \esp \in
\raisebox{3.5pt}{
\scaleobj{0.75}{
\begin{ytableau}
~ \\
~
\end{ytableau}}}
\spa\spa \asimtworep
\spa\spa,\spa\spa
\eta^{n_3}_i \esp \in
\raisebox{-3.2pt}{
\scaleobj{0.75}{
\begin{ytableau}
\overline{\phantom{\Big|aa.}} 
\end{ytableau}}}
\spa\spa \ccfundrep
\spa\spa,
\\
i,j = 1, \esp... \esp , \esp N
\spa\spa,\spa\spa
n_i = 1, \esp... \esp , \esp N_i
\spa\spa.
\numberthis \label{matter_content_general_irep}
\end{gathered}
\end{equation}
\setlength{\jot}{13pt}
There are $N_1$ flavors $\psi$  in the two-index symmetric representation, $N_2$ flavors $\lambda$  in the two-index anti-symmetric representation and $N_3$ flavors $\eta$ in the anti-fundamental representation. We refer to each irreducible representation as a different species and $N_i$ as the flavor number. The $N_i$ are not arbitrary; only specific combinations guarantee that the theory is gauge anomaly-free and satisfies asymptotic freedom for arbitrarily large $N$. These combinations are shown in Table \tcolor{\ref{table_N1_N2_N3_values}}.  In general 
\begin{align}
N_3 = N_1(N+4) + N_2 (N-4) 
\spa\spa, 
\label{N3_as_function_of_N1_N2_N}
\end{align}
for the 30 possible $N_1$ and $N_2$ combinations. For all cases except the first row in Table \tcolor{\ref{table_N1_N2_N3_values}} $N\geq 5$. In 7 cases there is a fixed $N$ such that $N_3 = 0$, i.e. no fermions in the anti-fundamental representation; see Appendix \ref{appendix_4d_two_species_models}. Note also that for $R_9$, $R_8$, $R_{10}$, $R_{11}$, $R_{15}$, $R_{16}$, $R_{21}$, $R_{22}$ and $R_{26}$, $N_3 < 0$ for some values of $N$. In these cases the $\eta^i$ transform in the fundamental representation. In Table \tcolor{\ref{dynkin_indices}} we list the group theoretic data for the fermion representations. 

We can also consider adding vector-like matter to these theories, for instance, injecting some number of Dirac fermion species transforming under the one- and two-index tensor irreps of SU($N$). These theories remain anomaly-free automatically, and asymptotic freedom may be maintained if the number of species is not too large. Also, such additions would bring new global symmetries and typically more 't Hooft anomalies to be matched. We do not investigate such extensions here.

To the best of our knowledge, only particular combinations of $N_i$ have been studied in an isolated manner in the literature. For instance, the $R_1$ and $R_2$ models have been studied in \cite{Bars:Yankielowicz:1981,Bolognesi:Konishi:Luzio:2020,Smith:Karasik:Lohitsiri:Tong:2022,Karasik:Onder:Tong:2022}, and $R_3$ in \cite{Eichten:Peccei:Preskill:Zeppenfeld:1986}. See also \cite{Goity:Peccei:Zeppenfeld:1985,Eichten:Peccei:Preskill:Zeppenfeld:1986,Armoni:Shifman:2012,Bolognesi:Konishi:Shifman:2017,Sheu:Shifman:2022,Bolognesi:Konishi:Luzio:2022} for a series of analyses of the $R_7$ model. The cases with no matter in the anti-fundamental representation for which $N$ is fixed have been studied in \tcolor{\cite{Anber:Chan:2023,Anber:Hong:Son:2022}}, where the authors compute the mixed 't Hooft anomalies associated with the genuine one-form and discrete zero-form symmetries of these theories.\\

\renewcommand{\extable}{0.6ex}
\begin{minipage}[c]{0.5\textwidth}
\begin{center}
\begin{tabular}{||c c c c c||} 
\hline\hline
\rule{0pt}{3ex}  
\phantom{--}$R_i$\phantom{--}  & 
\phantom{-}$N_1$\phantom{-}  & 
\phantom{-}$N_2$\phantom{-}  &
\phantom{--}$N_3$\phantom{--}  &
\phantom{-}$\me_\gamma$\phantom{-}  \\ [0.5ex]
\hline\hline
\phantom{-} &  &  &   & \\ [-2ex]
$R_1$ &  1  &  0  &  $N + 4$  &  A   \\ [\extable]
$R_2$ &  0  &  1  &  $N - 4$  & A    \\ [\extable]
$R_3$ &  1  &  1  &  $2N$  & B        \\ [\extable]
$R_4$ &  1  &  2  &  $3N - 4$  & A   \\ [\extable]
$R_5$ &  1  &  3  &  $4N - 8$  & B   \\ [\extable]
$R_6$ &  1  &  4  &  $5N - 12$  & A    \\ [\extable]
$R_7$ &  1  &  -1  &  $8$  & B          \\ [\extable]
$R_8$ &  1  &  -2  &  $-N+12$  & A     \\ [\extable]
$R_9$ &  1  &  -3  &  $-2(N-8)$  & B     \\ [\extable]
$R_{10}$ &  1  &  -4  &  $-3N+20$  & A    \\ [\extable]
$R_{11}$ &  1  &  -5  &  $-4(N-6)$  & B    \\ [\extable]
$R_{12}$ &  2  &  1  &  $3N+4$  & A       \\ [\extable]
$R_{13}$ &  2  &  3  &  $5N-4$  & A     \\ [\extable]
$R_{14}$ &  2  &  -1  &  $N+12$  & A     \\ [\extable]
$R_{15}$ &  2  &  -3  &  $-N+20$ & A      \\ [\extable]
\hline
\end{tabular}
\end{center}
\end{minipage}
\begin{minipage}[c]{0.5\textwidth}
\begin{center}
\begin{tabular}{||c c c c c||} 
\hline\hline
\rule{0pt}{3ex}  
\phantom{--}$R_i$\phantom{--}  & 
\phantom{-}$N_1$\phantom{-}  & 
\phantom{-}$N_2$\phantom{-}  &
\phantom{--}$N_3$\phantom{--}  &
\phantom{-}$\me_\gamma$\phantom{-}  \\ [0.5ex]
\hline\hline
\phantom{-} &  &  &   & \\ [-2ex]
$R_{16}$ &  2  &  -5  &  $-3N+28$  & A \\ [\extable]
$R_{17}$ &  3  &  1  &  $4N+8$  & B \\ [\extable]
$R_{18}$ &  3  &  2  &  $5N+4$  & A \\ [\extable]
$R_{19}$ &  3  &  -1  &  $2N+16$  & B \\ [\extable]
$R_{20}$ &  3  &  -2  &  $N+20$  & A \\ [\extable]
$R_{21}$ &  3  &  -4  &  $-N+28$  & A \\ [\extable]
$R_{22}$ &  3  &  -5  &  $-2N+32$  & B \\ [\extable]
$R_{23}$ &  4  &  1  &  $5N+12$  & A \\ [\extable]
$R_{24}$ &  4  &  -1  &  $3N+20$  & A \\ [\extable]
$R_{25}$ &  4  &  -3  &  $N+28$ & A  \\ [\extable]
$R_{26}$ &  4  &  -5  &  $-N+36$ &  A \\ [\extable]
$R_{27}$ &  5  &  -1  &  $4N+24$ &  B \\ [\extable]
$R_{28}$ &  5  &  -2  &  $3N+28$ &  A \\ [\extable]
$R_{29}$ &  5  &  -3  &  $2N+32$ &  B \\ [\extable]
$R_{30}$ &  5  &  -4  &  $N+36$  & A \\ [\extable]
\hline
\end{tabular}
\end{center}
\end{minipage}
\captionof{table}{ \label{table_N1_N2_N3_values}
Matter content that yields an anomaly-free and asymptotically free theory for arbitrarily large $N$ \tcolor{\cite{Eichten:Kang:Koh:1982}}. Here  $R_i$ labels each theory. Negative $N_i$ indicates the appearance of $|N_i|$ copies of the associated complex-conjugate representation. The $\textnormal{g}_\gamma$ column indicates the allowed $\mathbbm{Z}_{\hat{N}_3}$ center symmetry for each theory (see Table \tcolor{\ref{table_3flavors_Z_gamma}}).  
}

\renewcommand{\extable}{1.6ex}
\begin{table}[h!]
\begin{center}
\begin{tabular}{||c c  c c c||} 
\hline\hline
\rule{0pt}{3ex}  
Fermion & 
Irrep  & Dimension&
$T_i$ &
$A_i$ \\ [0.3ex]
\hline\hline
\phantom{-}  &  &  & &  \\ [0.25ex]
$\psi$ 
& 
$\txn{\textbf{S}}_2$ 
& $\tfrac{1}{2}N(N+1)$ &  $\tfrac{1}{2}(N+2)$  &  $N + 4$   \\ [\extable]
$\lambda$ 
&  
$\txn{\textbf{A}}_2$
& $\tfrac{1}{2}N(N-1)$ &  $\tfrac{1}{2}(N-2)$  &  $N - 4$   \\ [\extable]
$\eta$ 
& 
$\overline{\txn{\textbf{F}}}$
\phantom{}
& $N$ &  $1/2$  &  $-1$   \\ [\extable]
\hline
\end{tabular}
\end{center}
\caption{
Dynkin indices $T_i$ and anomaly coefficients $A_i$ for the fermions. 
}
\label{dynkin_indices}
\end{table}

\renewcommand{\extable}{0.25ex}
\begin{center}
\begin{tabular}{||c c||} 
\hline\hline
\rule{0pt}{3ex}  
\phantom{---}$R_i$\phantom{---}  & 
\phantom{-}$ (\ell,N) $\phantom{-}  \\ [0.5ex]
\hline\hline
\phantom{-}   &   \\ [-2ex]
$R_1$ & $    
(2 , 3)  \spa,\spa  
(3,  4 \leq N \leq 8)   \spa,\spa  
(4,  9 \leq N \leq 30)  \spa,\spa  
(5,  N \geq 31) 
$\\ [\extable]
$R_2$ & $   
(13,  5)   \spa,\spa  
(10 , 6)   \spa,\spa     
(9,7)   \spa,\spa 
(8,  8 \leq N \leq 9)   
$\\ [\extable]
$ $ & $ 
(7,  10 \leq N \leq 13)   \spa,\spa  
(6,  14 \leq N \leq 35)   \spa,\spa  
(5 , N \geq 36)
$\\ [\extable]
$R_3$ &   $ (2, N \geq 5)  $ \\ [\extable]
$R_4$ &$    (2, N \leq 11) \spa,\spa (1, N \geq 12)   $ \\ [\extable]
$R_5$ &$   (1, N \geq 5)      $ \\ [\extable]
$R_6$ &$   (1, N \geq 5)     $ \\ [\extable]
$R_7$ &$    
(3,  5 \leq N \leq 10)   \spa,\spa 
(4,  11 \leq N \leq 40)   \spa,\spa 
(5 , N \geq 41)
$ \\ [\extable]
$R_8$ &$   
(2,  5 \leq N \leq 6)   \spa,\spa
(3,  7 \leq N \leq 41)  \spa,\spa
(2,  N \geq 42)
$  \\ [\extable]
$R_9$ &$    
(2,7)   \spa,\spa
(3,8)   \spa,\spa
(2,  9 \leq N \leq 39)   \spa,\spa 
(1 , N \geq 40)
$   \\ [\extable]
$R_{10}$ &$   
(2,10)   \spa,\spa
(1 , N \geq 11)
$ \\ [\extable]
$R_{11}$ &$   
(2,7)   \spa,\spa
(1 , N \geq 8)
$ \\ [\extable]
$R_{12}$ &$   (1, N \geq 5)    $  \\ [\extable]
\hline
\end{tabular}
\end{center}
\captionof{table}{ \label{table_rep_multiplictly_1}
Maximum representation multiplicity $\ell$ and values for $N$ such that each theory is asymptotically free. Except for the first row, $N$ is greater than or equal to 5.\\
}

\renewcommand{\extable}{0.25ex}
\begin{center}
\begin{tabular}{||c c||} 
\hline\hline
\rule{0pt}{3ex}  
\phantom{---}$R_i$\phantom{---}  & 
\phantom{-----}$ (\ell,N) $\phantom{-----}  \\ [0.5ex]
\hline\hline
\phantom{-}   &   \\ [-2ex]
$R_{13}$ &$   (1, N \geq 5)$  \\  [\extable]
$R_{14}$ &$   (1, N \leq 9) \spa,\spa (2, N \geq 10)$  \\  [\extable]
$R_{15}$ &$ 
(1,  5 \leq N \leq 12)   \spa,\spa
(2,  13 \leq N \leq 43)  
$  \\  [\extable]
$ $ &$  
(1, N \geq 44) 
$  \\  [\extable]
$R_{16}$ &$  (1, N \geq 5)$  \\  [\extable]
$R_{17}$ &$   (1, N \geq 5)$  \\  [\extable]
$R_{18}$ &$  (1, N \geq 7)$  \\  [\extable]
$R_{19}$ &$   (1, N \geq 5)$  \\  [\extable]
$R_{20}$ &$  (1, N \geq 5)$  \\  [\extable]
$R_{21}$ &$  (1, N \geq 6)$  \\  [\extable]
\hline
\end{tabular}
\begin{tabular}{||c c||} 
\hline\hline
\rule{0pt}{3ex}  
\phantom{--}$R_i$\phantom{--}  & 
\phantom{-----}$ (\ell,N) $\phantom{-----}  \\ [0.5ex]
\hline\hline
\phantom{-}   &   \\ [-2ex]
$R_{22}$ &$   (1, N \geq 6)$  \\  [\extable]
$R_{23}$ &$  (1, N \geq 19)$  \\  [\extable]
$R_{24}$ &$   (1, N \geq 9)$  \\  [\extable]
$R_{25}$ &$  (1, N \geq 11)$  \\  [\extable]
$R_{26}$ &$   (1, N \geq 12)$  \\  [\extable]
$R_{27}$ &$   (1, N \geq 33)$  \\  [\extable]
$R_{28}$ &$   (1, N \geq 35)$  \\  [\extable]
$R_{29}$ &$   (1, N \geq 37)$  \\  [\extable]
$R_{30}$ &$   (1, N \geq 39)$  \\  [\extable]
$ $ &  \\  [\extable]
\hline
\end{tabular}
\end{center}
\captionof{table}{ \label{table_rep_multiplictly_2}
Maximum representation multiplicity $\ell$ and $N$ values such that each theory is asymptotically free.
}

The pairs $(\ell,N)$ in Tables \ref{table_rep_multiplictly_1} and \ref{table_rep_multiplictly_2} differ marginally from those presented in \cite{Eichten:Kang:Koh:1982}. Here we list the values for which $\beta_0 < 0$ and $\beta_1 > 0$ when the inequality for $N$ is saturated such that asymptotic freedom is guaranteed at one loop (see Appendix \ref{appendix_beta_func}).\\

In the remainder of this manuscript, we do not write explicit absolute values for $N_2$ and $N_3$, which are sometimes needed. The context makes it clear when this is the case. In general, the sign of these values is relevant only for equations related to color physics such as (\ref{N3_as_function_of_N1_N2_N}).

%% file: 03_global_symm_group.tex
\setlength{\jot}{13pt}

\section{The faithful global symmetry group} \label{section_global_symm_group_3_flavors}

We begin by describing the naive continuous symmetries of the chiral gauge theories introduced in the previous section. We then investigate zero-form discrete global symmetries and show that they are all redundant with discrete subgroups of the continuous symmetries for $N_3\neq 0$. Finally, we investigate the global structure of the continuous symmetry group and identify discrete redundancies. These redundancies are critical for determining the  structure of the 't Hooft anomalies.\\ 

The classical zero-form symmetry group for the representation $R_i$ is

\begin{align}
G_i = 
\txn{SU}(N) \times 
\txn{U}_\psi(N_1) \times 
\txn{U}_\lambda(N_2) \times
\txn{U}_\eta(N_3) 
\spa\spa.
\end{align}

It includes the gauge transformations and a product of chiral rotations acting on each individual fermion species. Ignoring possible redundant transformations for the moment, a non-anomalous\hlfootnote{The expression in (\tcolor{\ref{general_flavor_group_3_fermions}}) is imprecise if some $N_i$ is zero, which is the case for general $N$ only for the $R_1$ and $R_2$ theories. For those theories there is only one non-anomalous U$(1)$ symmetry.} subgroup is

\begin{align*}
G_i
&= 
\txn{SU}(N) \times 
\txn{SU}(N_1) \times \txn{SU}(N_2) \times \txn{SU}(N_3) \times
\txn{U}_\alpha(1) \times \txn{U}_\beta(1)
\spa\spa.
\label{general_flavor_group_3_fermions}
\numberthis
\end{align*}

Under the U$_\alpha(1)$ and U$_\beta(1)$ symmetries, the fermion charges $q_i^\alpha$ and $q_i^\beta$  are such that these symmetries remain non-anomalous with the gauged SU$(N)$.  If all $N_i$ are non-zero, one particular solution is 
\setlength{\jot}{5pt}
\begin{align*}
&q^\alpha_i = 
\Big(
2 N_3 T_3 / \me_\alpha 
\esp,\esp 0 \esp,\esp
- 2N_1 T_1 / \me_\alpha
\Big)
\spa\spa,\spa\spa
\me_\alpha = \txn{gcd}\big( 2N_1 T_1 , 2N_3 T_3 \big)
\label{charges_U1_3_fermions_alpha}
\numberthis
\spa\spa,
\\
&q^\beta_i = 
\Big(
0
\esp,\esp 2N_3 T_3 / \me_\beta  \esp,\esp
- 2N_2 T_2 / \me_\beta
\Big)
\spa\spa,\spa\spa
\me_\beta = \txn{gcd}\big( 2N_2 T_2 , 2N_3 T_3 \big)
\numberthis
\spa\spa.
\label{charges_U1_3_fermions_beta}
\end{align*}
\setlength{\jot}{13pt}

This solution is not unique; one may take any linear combination of U$_\alpha(1)$ and U$_\beta(1)$. In general, we use the charge assignment in (\ref{charges_U1_3_fermions_alpha})  and (\ref{charges_U1_3_fermions_beta}) throughout this manuscript. In some cases however we use a different basis when convenient.\\

There might be an additional disconnected discrete transformation which we must include in the global symmetry group. Under a general $\txn{U}_\psi(1) \times \txn{U}_\lambda(1) \times \txn{U}_\eta(1)$ rotation, the fermions transform as 
\begin{align}
\psi \spa\longrightarrow\spa \eu{2\pi i \omega_1 }  \psi 
\spa\spa\spa,\spa\spa\spa
\lambda \spa\longrightarrow\spa \eu{2\pi i \omega_2 } \lambda
\spa\spa\spa,\spa\spa\spa
\eta \spa\longrightarrow\spa \eu{2\pi i \omega_3 } \eta  
\spa\spa,
\label{u(1)_disconnected_fermion_transform}
\end{align}

which induces a shift in the SU$(N)$ $\theta$ parameter given by
\begin{align}
\delta \theta = 
2\pi \Big(
N_1 2T_1 \omega_1 +
N_2 2T_2 \omega_2 +
N_3 2T_3 \omega_3 
\Big)
\spa\spa.
\end{align}

The case where the charges $\omega_i$ are those in equations (\tcolor{\ref{charges_U1_3_fermions_alpha}}) or (\tcolor{\ref{charges_U1_3_fermions_beta}}) corresponds to the trivial shift of the theta term $\delta \theta = 0$. In addition to this, there might be a discrete transformation that shifts the theta term by $2\pi \mathbbm{Z}$. In this case there is an extra disconnected discrete symmetry $\mathbbm{Z}_k \not\subset \txn{U}_\alpha(1) \times \txn{U}_\beta(1)$ in the global symmetry group. To find this zero-form discrete symmetry, we start by looking at a $\txn{U}_\gamma(1)$ symmetry defined by a parameter $\gamma$ that cannot be written as a $\txn{U}_\alpha(1) \times \txn{U}_\beta(1)$ transformation. For convenience, we can define one that is orthogonal in the sense that the charges $q^\gamma_i$ are proportional to the cross product between $q^\alpha_i$ and $q^\beta_i$. In other words, $\txn{U}_\alpha(1)$ and $\txn{U}_\beta(1)$ define a torus in the classical symmetry group, and $\txn{U}_\gamma(1)$ is normal to the torus and is anomalous in general. We look for a minimal non-anomalous discrete subgroup $\mathbbm{Z}_{g_\gamma}$ of it. First, we look for the smallest $\gamma$ such that it intersects $\txn{U}_\alpha(1) \times \txn{U}_\beta(1)$. Next we look for discrete values of $\gamma$ so that $\delta\theta = 2\pi n$. The solution is 
\begin{align}
\gamma=2\pi k/\me_\gamma
\spa\spa , \spa\spa
\me_\gamma = \gcd\big( 2 N_1 T_1, 2 N_2 T_2, 2 N_3 T_3 \big)
\spa\spa,
\end{align}

and thus in principle we must include this $\mathbbm{Z}_{\me_\gamma}$ in the non-anomalous global zero-form symmetry. If we were to include it explicitly, the charges for the fermions under this symmetry are Bézout coefficients, and thus satisfy Bézout's identity 

\begin{align}
q^\gamma_1 (2N_1 T_1) + q^\gamma_2 (2N_2 T_2)  + q^\gamma_3 (2N_3 T_3) 
=
\gcd\big( 2 N_1 T_1, 2 N_2 T_2, 2 N_3 T_3 \big)
\spa\spa.
\end{align}

This is a sufficient but not necessary condition; given a Bézout number $q^\gamma_i$, one can  choose a bigger $q^\gamma_i$ such that the $\mathbbm{Z}_{\me_\gamma}$ transformation is preserved, so that the action on the fermions stays the same.\\

Up to this point, the symmetry group reads
\begin{align}
G_i
=
\txn{SU}(N) \times \txn{SU}(N_1) \times \txn{SU}(N_2) \times \txn{SU}(N_3) \times 
\txn{U}_\alpha(1) \times \txn{U}_\beta(1) \times \mathbbm{Z}_{\me_\gamma}
\spa\spa.
\end{align}

However,  there may be redundant elements in the symmetry group that act trivially on the fermions or coincide with a gauge transformation. To avoid these redundancies we must quotient by the center symmetries (or sub-groups of them) of the gauge group $\mathbbm{Z}_N$ and flavor rotations $\mathbbm{Z}_{N_i}$

\begin{align*}
&\mathbbm{Z}_{N} \spa:\spa 
\eu{2\pi i n^c_i \esp m / N}
\spa\spa,\spa\spa 
m = 0, ... , N - 1
\spa\spa,
\numberthis
\\
&\mathbbm{Z}_{N_i} \spa:\spa 
\eu{2\pi i \esp m_i / N_i}
\spa\spa,\spa\spa 
m_i = 0, ... , N_i - 1
\spa\spa,
\numberthis
\end{align*}

$n_i^c = 2,\pm2,\mp1$ are the \textit{n}-alities or charges for the fermions under the color center symmetry. The redundancies are determined by a set of solutions $\{m,m_1,m_2,m_3,\alpha,\beta\}$ that satisfy the following consistency conditions:

\begin{align*}
\psi  \spa &:\spa
\eu{2\pi i n_1^c  \frac{m}{N}   }
\eu{2\pi i \frac{m_1}{N_1}}
\eu{ i \alpha q_1^\alpha } 
\eu{ i \beta q_1^\beta }
= 1 
\\
\lambda  \spa &:\spa
\eu{  2\pi i n_2^c \frac{m}{N}   }
\eu{2\pi i \frac{m_2}{N_2}}
\eu{ i \alpha q_2^\alpha } 
\eu{ i \beta q_2^\beta }
= 1 
\\
\eta  \spa &:\spa
\eu{  2\pi i n_3^c \frac{ m}{N}  }
\eu{2\pi i \frac{m_3}{N_3}}
\eu{ i \alpha q_3^\alpha } 
\eu{ i \beta q_3^\beta }
= 1 
\spa\spa.
\numberthis
\label{consistency_conditions_global_symm_section}
\end{align*}

The redundancies are as follows: \\


$\textbf{I. No extra discrete symmetry : }$ $\mathbbm{Z}_{\me_\gamma}$ is completely redundant with a transformation in $\txn{U}_\alpha(1) \times \txn{U}_\beta(1) \times \mathbbm{Z}_{N_3}$. For instance 

\begin{align}
\alpha = 2\pi \esp \frac{ q_1^\gamma \me_\alpha }{ N_3 \me_\gamma }
\spa\spa,\spa\spa
\beta = 2\pi \esp \frac{ q_2^\gamma \me_\beta }{ N_3 \me_\gamma }
\spa\spa,\spa\spa
m_3 = 1
\spa\spa,
\end{align}

has the same action on the fermions as a unit $\mathbbm{Z}_{\me_\gamma}$ transformation. Therefore, $\mathbbm{Z}_{\me_\gamma}$ can be completely removed from $G_i$.\\


$\textbf{II. Flavor center:}$ The center $\mathbbm{Z}_{N_1} \times \mathbbm{Z}_{N_2}$ is redundant with $\txn{U}_\alpha(1) \times \txn{U}_\beta(1) \times \mathbbm{Z}_{N_3}$ transformation with

\begin{align*}
\alpha =
&\esp 2\pi \frac{\me_\alpha}{N_1 N_3} \big( N_1 p - m_1 \big)
\spa\spa,\spa\spa
\beta =
2\pi \frac{\me_\beta}{N_2 N_3} \big( N_2 q - m_2 \big)
\spa\spa,
\\
m_3 &= 
- m_1 2T_1 - m_2 2T_2 
+  2N_1T_1 p + 2N_2T_2 q + 2N_3T_3 r
\spa\spa,
\numberthis
\label{integers_sol_flavorcenter_N1_N2}
\end{align*}

for any set of integers\hlfootnote{Since $2T_3 = 1$, both $\me_\alpha$ and $\me_\beta$ divide $N_3$, and thus $m_3$ is an integer.} $p,q$ and $r$. We could choose $p = q = r = 0$, in which case $m_3 = - m_1(N+2) - m_2(N-2) \esp \txn{mod} \esp N_3$, therefore the $\mathbbm{Z}_{N_1} \times \mathbbm{Z}_{N_2}$ symmetry is completely redundant. Regardless of the specific choice of redundant transformation, we conclude that we must quotient the symmetry group by the $\mathbbm{Z}_{N_1} \times \mathbbm{Z}_{N_2}$ center transformation.\\

$\textbf{III. Flavor center:}$ A subgroup of the $\mathbbm{Z}_{N_3}$ center is redundant with $\txn{U}_\alpha(1)\times \txn{U}_\beta(1)$ transformations. These transformations are of the form
\begin{align}
\alpha = 2\pi \esp \frac{p_3 \me_\alpha}{N_3}
\spa\spa,\spa\spa
\beta = 2\pi \esp \frac{q_3 \me_\beta}{N_3}
\spa\spa,\spa\spa
m_3 = 2N_1T_1 p_3 + 2N_2T_2 q_3 + 2N_3T_3 r_3
\spa\spa,
\label{integers_sol_flavorcenter_N3}
\end{align}

for any set of integers $p_3, q_3$ and $r_3$. By virtue of Bézout's identity, the solution for $m_3$ that can be written in this form must be an integer multiple of $\me_\gamma$. Thus, the center $\mathbbm{Z}_{N_3}$ contains a redundant $\mathbbm{Z}_{\hat{N}_3}$ where $\hat{N}_3 = N_3 / \me_\gamma$. We then must quotient the symmetry group by $\mathbbm{Z}_{\hat{N}_3}$.\\

$\textbf{IV. Axial center:}$ There are redundant transformations within $\txn{U}_\alpha(1)\times \txn{U}_\beta(1)$ alone. These are of the form 
\begin{align}
\alpha = 
- 2\pi \esp \frac{2N_2 T_2}{N_3} \frac{\me_\alpha}{\me_\delta} \esp t
\spa\spa,\spa\spa
\beta  = 
2\pi \esp \frac{2N_1 T_1}{N_3} \frac{\me_\beta}{\me_\delta} \esp t
\spa\spa,\spa\spa
\me_\delta = \txn{gcd}\big(2N_1 T_1, 2N_2 T_2\big)
\spa\spa,
\label{integers_sol_axialcenter}
\end{align}

with $t$ an integer. These are $\mathbbm{Z}_{\me_u}$ transformations where $\me_u$ is determined as follows. We look for the smallest $t$ such that $\alpha$ and $\beta$ are integer multiples of $2\pi$. This is 
\begin{align}
t_{\txn{min}} = 
\frac{\me_\delta N_3 }
{ 
\gcd\big( 
2N_1T_1 \me_\beta \esp,\esp 
2N_2T_2 \me_\alpha \esp,\esp
2N_3T_3 \me_\delta 
\big) 
}
\spa\spa.
\label{t_min_Zgu}
\end{align}

Using the following
\begin{align}
\gcd\big( 
2N_1T_1 \me_\beta \esp,\esp 
2N_2T_2 \me_\alpha 
\big) 
=
\gcd\big( 
2N_1T_1 \me_\beta \esp,\esp 
2N_2T_2 \me_\alpha \esp,\esp
2N_3T_3 \me_\delta 
\big) 
=
\frac{\me_\alpha \me_\beta \me_\delta}{\me_\gamma}
\spa\spa,
\end{align}

we conclude that
\begin{align}
\me_u = 
N_3 \me_\delta / \tfrac{\me_\alpha \me_\beta \me_\delta}{\me_\gamma}
=
N_3 \esp \frac{\me_\gamma }{\me_\alpha \me_\beta}
\spa\spa.
\end{align}

Therefore, we must quotient the symmetry group by a $\mathbbm{Z}_{\me_u}$ center symmetry.\\

$\textbf{V. Color center:}$ A $\mathbbm{Z}_{\me_c}$ subgroup of the SU$(N)$ gauge symmetry, with
\begin{align}
\me_c = 
\gcd( N \esp,\esp 2 n_1^c N_1 T_1 +  2 n_2^c N_2 T_2 + 2 n_3^c N_3 T_3 )
\spa\spa,
\end{align}

is redundant with $\txn{U}_\alpha(1)\times \txn{U}_\beta(1) \times \mathbbm{Z}_{N_3}$ transformations of the form
\setlength{\jot}{8pt}
\begin{gather*}
\alpha = 
2\pi \esp \frac{p_c \me_\alpha}{N_3}
-
2\pi n_1^c \esp \frac{m \me_\alpha}{ N_3 \me_c}
\spa\spa,\spa\spa
\beta = 
2\pi \esp \frac{q_c \me_\beta}{N_3}
-
2\pi n_2^c \esp \frac{m \me_\beta}{ N_3 \me_c}
\spa\spa,
\\
m_3 = 
- \frac{m}{\me_c} \sum_i 2 n_i^c N_i T_i
+
\me_c \big( 2 N_1 T_1  p_c +  2 N_2 T_2 q_c + 2 N_3 T_3 r_c \big) 
\spa\spa,
\numberthis
\label{integers_sol_colorcenter}
\end{gather*}
\setlength{\jot}{13pt}
where $p_c, q_c$ and $r_c$ are arbitrary integers. For all theories in Table \tcolor{\ref{table_N1_N2_N3_values}}, using equation (\ref{N3_as_function_of_N1_N2_N}) gives $\me_c = \txn{gcd}(N, N(N_1 + N_2)) = N$. Therefore, we must quotient the symmetry group by the $\mathbbm{Z}_{N}$ color center symmetry.

The full zero-form symmetry group that acts faithfully on the fermions is 
\begin{align}
G
=
\frac{
\txn{SU}(N) \times \txn{SU}(N_1) \times \txn{SU}(N_2) \times \txn{SU}(N_3) \times 
\txn{U}_\alpha(1) \times \txn{U}_\beta(1) 
}
{
\mathbbm{Z}_N \times 
\mathbbm{Z}_{N_1} \times \mathbbm{Z}_{N_2} \times \mathbbm{Z}_{\hat{N}_3}
\times \mathbbm{Z}_{\me_u}
}
\spa\spa.
\label{full_symmetry_group}
\end{align}

When we gauge all of $G$ via background fields\hlfootnote{Background fields are not dynamical, i.e. not integrated in the partition function.}, the denominator indicates gauged one-form symmetries. There are two possible general cases for the value of $\me_\gamma$ which are shown in Table~\tcolor{\ref{table_3flavors_Z_gamma}}. As shown in Table~\tcolor{\ref{table_N1_N2_N3_values}}, 20 theories belong to case A and 10 theories to case B.\\

\renewcommand{\extable}{0.2ex}
\begin{center}
\begin{tabular}[H]{||c | c c||}
\hline\hline
\rule{0pt}{3ex}  
Case \spa& \phantom{--}$N$\phantom{--}  & 
\phantom{--}$\me_\gamma$\phantom{--} \\ [\extable]
\hline\hline
& \phantom{-}  &   \\ [-2ex]
& odd  &  1  \\ [-0.65ex]
$\upsymb{10pt}{A}$ & even &  2  \\ [\extable]
\hline
& \phantom{-}  &   \\ [-2ex]
& odd  &  1  \\ [\extable]
B & $4n$ &  2  \\ [\extable]
& $4n + 2$ &  4  \\ [\extable]
\hline
\end{tabular}
\captionof{table}{ \label{table_3flavors_Z_gamma}
Possible values for $\mathrm{g}_\gamma$ for different $N$.
}
\end{center}

%% file: 04_anomalies.tex
\setlength{\jot}{13pt}

\section{Mixed 't Hooft anomalies by the descent prescription} \label{section_descent_anomalies}

We are interested in anomalies that appear in backgrounds with nontrivial flux in the centers of the color, flavor and axial symmetries, or CFU fluxes for short. Coupling the full global symmetry group to background gauge fields and allowing those fields to transform as spurions introduces two-form background fields $B_i^{(2)}$ for the centers.  These enter in the same way as background fields for one-form center symmetries $\mathbbm{Z}^{(1)}$, so we use that notation.\\

The phase generated by a U$(1)_{\alpha,\beta}$ transformation in the background of a CFU flux is the mixed anomaly between the zero-form U$(1)_{\alpha,\beta}$ and the one-form $\mathbbm{Z}^{(1)}$. The Stora-Zumino descent procedure yields the ordinary mixed 't Hooft anomalies together with those generated by the CFU fluxes.  In Section \ref{section_twisted_gauge_field} they are recomputed by the method of twisted fluxes, where the fractional fluxes appear in the boundary conditions.\\

Following the descent prescription, we start from the anomaly polynomial  in $6d$ given by 
\begin{align}
\mathcal{A}^{6d}
=
\frac{1}{24 \pi^2}
\sum_{i}
\txn{Tr}_{R_i \times F^f_i}
\Big( \mathcal{F}_i \wedge \mathcal{F}_i \wedge \mathcal{F}_i \Big)
\spa\spa.
\label{6d_anomaly_polynomial}
\end{align}

By integrating this expression, we obtain the anomaly in $4d$, or more precisely, the phase shift induced in the partition function under the different gauge transformations. The label $R_i \times F^f_i$ indicates the different traces that have to be computed, both over the SU($N$) color representations $R_i$ and the fundamental flavor SU($N_i$) representations $F^f_i$. The total field strength coupling to each fermion is 

\begin{align}
\mathcal{F}_i = 
F_c \otimes \mathbbm{1}_{N_i} 
+ F_i \otimes \mathbbm{1}_{\txn{dim}\Rep_i}  
+ 
\Big( q^\alpha_i \esp F_\alpha  + q^\beta_i \esp F_\beta \Big)
\otimes \mathbbm{1}_{\txn{dim}\Rep_i} \otimes \mathbbm{1}_{N_i}
\spa\spa,
\end{align}

where $F_c, F_i$ and $F_{\alpha,\beta}$ are the two-form field strengths for the gauge SU($N$), flavor, SU($N_i$) and U(1)$_{\alpha,\beta}$ groups respectively. We also need to include the CFU fluxes by coupling to gauge fields for the color, flavor and axial center transformations. Below we briefly review the procedure  to gauge the center symmetries of interest.\\

In general, the fields introduced to gauge a $\mathbbm{Z}_M$ center symmetry of an SU($M$) group, which we denote as $B^{(p)}$ where $p$ denotes the degree of the form, satisfy the constraint \tcolor{\cite{Gaiotto:Kapustin:Seiberg:Willett:2015,Gaiotto:Kapustin:Komargodski:Seiberg:2017}}

\begin{align}
M B^{(2)} = \de B^{(1)}
\spa\spa.
\label{constrain_B_Z_M_center}
\end{align}

The two-form gauge field can carry fractional ’t Hooft flux given by

\begin{align}
\frac{1}{8\pi^2} \bigintssss_{\Sigma_4} \de B^{(1)} \wedge \de B^{(1)}
\in
\mathbbm{Z}_M
\spa\spa \longrightarrow \spa\spa
\frac{1}{8\pi^2} \bigintssss_{\Sigma_4} B^{(2)} \wedge B^{(2)} 
=
\frac{ k }{M^2} \spa,\spa k \in \mathbbm{Z}_M
\spa\spa,
\label{center_symm_fluxes}
\end{align}

where $\Sigma_4 = \Sigma_2 \times \Sigma_2$ is a topologically non-trivial 4-dimensional manifold. Equivalently, over the closed 2-dimensional manifold $\Sigma_2$ the fractional ’t Hooft flux is 

\begin{align}
\frac{1}{2\pi} \bigintssss_{\Sigma_2} \de B^{(1)} 
\in \mathbbm{Z}_M
\spa\spa \longrightarrow \spa\spa
\frac{1}{2\pi} \bigintssss_{\Sigma_2} B^{(2)} 
= \frac{\ell}{M} \spa,\spa \ell \in \mathbbm{Z}_M
\spa\spa.
\end{align}

The constraint in (\tcolor{\ref{constrain_B_Z_M_center}}) is invariant under the one-form gauge transformation 

\begin{align}
B^{(2)} \esp \longrightarrow \esp 
B^{(2)} + \de \lambda^{(1)}
\spa\spa\spa &, \spa\spa\spa
B^{(1)} \esp \longrightarrow \esp 
B^{(1)} + M \lambda^{(1)}
\spa\spa.
\end{align}

If we were to gauge only a subgroup $\mathbbm{Z}_{\hat{M}}$ of the center symmetry where $\hat{M} = M/m$ for some integer $m$, then we need to introduce a 2-form gauge field which is related to the full $\mathbbm{Z}_{M}$ gauge field as

\begin{align}
\hat{B}^{(2)} = m B^{(2)}
\spa\spa,
\end{align}

with fractional flux given by 

\begin{align}
\frac{1}{8\pi^2} \bigintssss_{\Sigma_4} 
\hat{B}^{(2)} \wedge \hat{B}^{(2)} 
=
\frac{ k }{\hat{M}^2} = k \esp \frac{m^2}{ M^2 }
\spa,\spa k \in \mathbbm{Z}_{\hat{M}}
\spa\spa.
\label{b_flux_subgroup}
\end{align}

Now we promote the SU($M$) gauge fields to U($M$) gauge fields and embed $B^{(1)}$ inside the U(1) factor for U($M$),
\begin{align}
\widetilde{A}^\Rep =
A^\Rep 
+ \frac{1}{M} B^{(1)} \otimes \mathbbm{1}_{M}
\spa\spa \Longrightarrow \spa\spa
\widetilde{F} = F + B^{(2)} \otimes \mathbbm{1}_{M}
\spa\spa.
\end{align}

Note that when tracing over the fundamental representation

\setlength{\jot}{10pt}
\begin{align*}
\txn{Tr}\big( F \big) &= \txn{Tr}\big( \widetilde{F} - B^{(2)} \big) = 0
\spa\spa,
\\
\txn{Tr}\big( \widetilde{F} \big) &= M B^{(2)}  
\spa\spa,
\\
\txn{Tr}\big( \widetilde{F} \wedge \widetilde{F} \big) 
&= 
\txn{Tr}\big( F^2 + 2 F \wedge B^{(2)} +  (B^{(2)})^2 \big)
=
\txn{Tr}\big( F^2 \big) + M B^{(2)} \wedge B^{(2)}
\spa\spa.
\numberthis
\label{traces_F_tilde}
\end{align*}
\setlength{\jot}{13pt}

As we enlarge the gauge group, we introduce new gauge degrees of freedom. These can be eliminated by demanding invariance under the one-form gauge transformation 
\begin{align}
\widetilde{F}  
\longrightarrow 
\widetilde{F} + \de \lambda^{(1)}
\spa\spa,
\end{align}

which renders the combination $F = \widetilde{F} - B^{(2)}$ one-form gauge invariant.\\

Using the above technology, after gauging the center symmetries in (\tcolor{\ref{full_symmetry_group}}), we can write the proper field strengths felt by the fermions as

\begin{align*}
\mathcal{F}_i
&=
\Big( 
\widetilde{F}_c - B_c^{(2)} \esp \mathbbm{1}_{\txn{dim}\Rep_i}
\Big) 
\otimes \mathbbm{1}_{N_i}
+
\Big( \widetilde{F}_i - B_i^{(2)}  \esp \mathbbm{1}_{N_i}\Big) 
\otimes \mathbbm{1}_{\txn{dim}\Rep_i}
\\
&+
\bigg(
q^\alpha_i \Big( F_\alpha - \kappa_\alpha B_u^{(2)} \Big) 
+
q^\beta_i \Big( F_\beta - \kappa_\beta B_u^{(2)} \Big) 
\bigg)
\otimes \mathbbm{1}_{\txn{dim}\Rep_i}
\otimes \mathbbm{1}_{N_i}
\spa\spa,
\numberthis
\end{align*}

with parameters $\kappa_{\alpha,\beta}$ to be defined shortly. We stress that we are gauging only a subgroup $\mathbbm{Z}_{N_3/\me_\gamma}$ of the full center  $\mathbbm{Z}_{N_3}$ of $SU(N_3)$, due to the structure of the faithful symmetries. This has the effect of modifying the flux sourced by $B^{(2)}_3$ as in (\tcolor{\ref{b_flux_subgroup}}).\\

The background field $B^{(2)}_u$ is introduced to gauge the center symmetry $\mathbbm{Z}_{\me_u}$ of the axial combination $\txn{U}_\alpha(1) \times \txn{U}_\beta(1)$. It obeys the condition

\begin{align}
\me_u \esp B^{(2)}_u = \de B^{(1)}_u
\spa\spa,
\end{align}

and the flux induced by this field is 

\begin{align}
\frac{1}{8\pi^2} \bigintssss_{\Sigma_4} B^{(2)}_u \wedge B^{(2)}_u 
=
\frac{ d }{\me_u^2} \spa,\spa d \in \mathbbm{Z}_{\me_u}
\spa\spa.
\end{align}

It is convenient to write the per-factor U(1) fluxes associated with a $B^{(2)}_u$ flux given by 

\begin{align}
\frac{1}{2\pi} 
\bigintssss_{\Sigma_2} B^{(2)}_\alpha \bigg|_{\mathbbm{Z}_{\me_u}}
=
- \frac{2 N_2 T_2}{N_3} \frac{\me_\alpha}{\me_\delta} \esp d
\spa\spa,\spa\spa
\frac{1}{2\pi} 
\bigintssss_{\Sigma_2} B^{(2)}_\beta \bigg|_{\mathbbm{Z}_{\me_u}}
=
\frac{2 N_1 T_1}{N_3} \frac{\me_\beta}{\me_\delta} \esp d
\spa\spa.
\end{align}

These fluxes have to be turned on simultaneously. From these expressions we see that the coefficients $\kappa_\alpha$ and $\kappa_\beta$ that yield appropriate fluxes are given by 

\begin{align}
\kappa_\alpha =
- 2 N_2 T_2 \frac{\me_\gamma}{ \me_\beta \me_\delta }
\spa\spa,\spa\spa
\kappa_\beta =
2 N_1 T_1 \frac{\me_\gamma}{ \me_\alpha \me_\delta }
\spa\spa.
\end{align}

In principle, we could condense the fluxes for $B^{(2)}_\alpha$ and $B^{(2)}_\beta$ into a single flux for $B^{(2)}_u$. Doing this would make the relation with the twisted flux method used in Section \tcolor{\ref{section_twisted_gauge_field}} somewhat more transparent.

\subsection{Ordinary zero-form anomalies}

To compute the anomalies involving only 0-form continuous symmetries, we first turn off the CFU fluxes. By pulling out the exterior derivative on various terms, we can integrate the $6d$ anomaly polynomial  down to $5d$. Then we perform a gauge transformation for any of the symmetries participating in the anomaly, integrate again, and extract the phase induced in the $4d$ partition function.\\ 

For the anomalies involving the U(1)$_{\alpha,\beta}$ symmetry, we can perform a gauge transformation for the $A_{\alpha,\beta}$ gauge field in the $5d$ anomaly polynomial as

\begin{align}
A_{\alpha,\beta}
\spa &\longrightarrow \spa
A_{\alpha,\beta} + \esp \de\lambda_{\alpha,\beta}
\spa\spa.
\end{align}

Evaluating the surface term $\lambda_{\alpha,\beta} = \alpha,\beta$, where $\alpha$ and $\beta$  parametrize the U(1)$_{\alpha,\beta}$ rotations, we obtain the variation of the anomaly polynomial  in $4d$

\begin{align}
S^{5d}_i = \mathcal{A}^{5d}_i \esp
\frac{1}{8\pi^2} \bigintsss_{\Sigma_5} 
\Omega \wedge A_{\alpha,\beta}
\spa\spa \longrightarrow \spa\spa
S^{4d}_i
=
\lambda_{\alpha,\beta} \esp  \mathcal{A}^{5d}_i\esp
\frac{1}{8\pi^2} \bigintsss_{\Sigma_4} \Omega 
\spa\spa,
\label{anomaly_flow_down_to_4d}
\end{align}

and $\Omega$ is a general product of 2-forms related to the symmetries participating in the anomaly. By integrating these 2-forms over the topologically non-trivial 4-dimensional manifold $\Sigma_4 \equiv \Sigma_2 \times \Sigma_2$, where $\Sigma_2$ is a closed 2-dimensional manifold, we obtain fluxes which yield a final result for the anomaly in $4d$. The induced phase in the partition function under the different mixed transformations is  

\begin{align}
\mathcal{Z} 
\spa \longrightarrow \spa
\eu{ 2\pi i  S^{4d}_i} \esp \mathcal{Z}
\spa\spa.
\end{align}

As long as $\mathcal{A}^{5d}_i$ is non-zero there is an anomaly. There might be values for $\alpha,\beta$ that make the induced phase in the partition function trivial, which correspond to non-anomalous subgroups of the U(1)$_{\alpha,\beta}$ transformations. Below we list the 15 zero-form anomalies present in the theory.\\

There are 10 mixed anomalies involving the U(1)$_{\alpha,\beta}$ symmetries. These are given by 
\setlength{\jot}{6pt}
\begin{align*}
\Big( \txn{U(1)}_{\alpha,\beta} \Big)^3
\spa \txn{:} \spa\spa
\mathcal{A}^{5d} 
&= 
\sum_i \big( q_i^{\alpha,\beta }\big)^3 \esp N_i \esp \txn{dim}(\Rep_i)
\spa\spa,
\numberthis
\\
\txn{U(1)}_{\alpha}  \times \Big( \txn{U(1)}_{\beta} \Big)^2
\spa \txn{:} \spa\spa
\mathcal{A}^{5d} 
&= 
\sum_i q_i^{\alpha} \big( q_i^{\beta} \big)^2 \esp N_i \esp \txn{dim}(\Rep_i)
\spa\spa,
\numberthis
\\
\txn{U(1)}_{\beta}  \times \Big( \txn{U(1)}_{\alpha} \Big)^2
\spa \txn{:} \spa\spa
\mathcal{A}^{5d} 
&= 
\sum_i q_i^{\beta} \big( q_i^{\alpha} \big)^2 \esp N_i \esp \txn{dim}(\Rep_i)
\spa\spa,
\numberthis
\\
\txn{U(1)}_{\alpha,\beta} \times \Big( \txn{SU}(N_i) \Big)^2
\spa \txn{:} \spa\spa
\mathcal{A}^{5d}  
&= 
q_i^{\alpha,\beta} \esp \txn{dim}(\Rep_i)
\spa\spa.
\numberthis
\end{align*}
\setlength{\jot}{13pt}
Note that the mixed anomalies involving SU($N_i)^2$ symmetries do not explicitly involve the Dynkin indices $T_{N_i} = 1/2$. These are contained inside the trace of field strengths, denoted by $\Omega$ for instance in \tcolor{(\ref{anomaly_flow_down_to_4d})}. Similarly, for the SU($N_i)^3$ anomalies there is a factor of $A_{N_i} = 1$ inside $\Omega$, which is the cubic anomaly coefficient.\\ 

The gauge anomaly is 
\begin{align}
\Big( \txn{SU}(N) \Big)^3
\spa \txn{:} \spa
\mathcal{A}^{5d}_{abc} \sim 
\sum_i N_i \esp \txn{Tr}\Big(T^a T^b T^c\Big)_{\Rep_i}
=
\sum_i N_i \bigg(
\frac{i}{2} \esp T_{\Rep_i} f^{abc} + \frac{1}{4} \esp A_{\Rep_i} d^{abc}
\bigg)
\spa\spa,
\end{align}
it is well known that a priori the gauge anomaly is proportional to both $T_\Rep$ and $A_\Rep$. The contribution proportional to $T_\Rep$ is UV divergent, but it can be renormalized while preserving gauge invariance. In fact, it contributes to the renormalization of the operator $f^{abc}A^a_\mu A^b_\nu \del_\mu A^c_\nu$ in the Yang-Mills Lagrangian. We then consider only the term proportional to $A_\Rep$, and the anomaly vanishes by construction
\begin{align}
\mathcal{A}^{5d} 
\sim
\sum_i N_i A_{\Rep_i} = 0
\spa\spa.
\end{align}
Since the generators of the gauge group are traceless, there is no mixed gauge-gravitational anomaly. The U$_{\alpha,\beta}$(1) fermion charges are chosen to cancel the ABJ anomaly with SU($N$),
\begin{align}
\txn{U(1)}_{\alpha,\beta} \times \Big( \txn{SU}(N) \Big)^2
\spa \txn{:} \spa\spa
\mathcal{A}^{5d}  \sim  \sum_{i}  q_i^{\alpha,\beta} |N_i|  T_{\Rep_i} = 0
\spa\spa.
\label{U(1)_color_mixed_anomaly}
\end{align}
There are 3 more anomalies between the non-abelian flavor symmetries only
\begin{align}
\Big( \txn{SU}(N_i) \Big)^3
\spa \txn{:} \spa\spa
\mathcal{A}^{5d}  = \txn{dim}(\mathcal{R}_i)
\spa\spa.
\end{align}

Note also that there are no mixed anomalies among non-abelian flavor symmetries because the generators for SU($N_i$) are traceless 
\begin{align}
\txn{SU}(N_i) \times \Big( \txn{SU}(N_j) \Big)^2
\spa \txn{:} \spa\spa
\mathcal{A}^{5d}  \sim \txn{Tr}\big( T^a_i ) = 0
\spa\spa.
\end{align}
Finally, we may also consider mixed U(1)$_{\alpha,\beta}$-gravitational anomalies. The $5d$ anomaly polynomial  reads
\begin{align}
-\frac{1}{8\pi^2} \bigintsss_{\mathbbm{M}^5}
\frac{ \mathcal{A}^{5d} }{ 24 } \esp 
A_{\alpha,\beta} \wedge \txn{Tr}\big( R \wedge R \big)
\spa\spa,
\end{align}

where $R$ is the space-time curvature 2-form. The anomaly is 

\begin{align}
\txn{U(1)}_{\alpha,\beta} \times \big( \txn{Gravity} \big)^2
\spa \txn{:} \spa\spa
\mathcal{A}^{5d} = \sum_{i} q^{\alpha,\beta}_i N_i \dim(R_i)
\spa\spa.
\end{align}

\subsection{Generalized anomalies}

Now we turn on the CFU fluxes to extract the mixed anomalies induced by a $\txn{U}(1)_{\alpha,\beta}$ transformation in the background of the CFU fluxes. See \cite{Anber:Hong:Son:2022} where the authors follow this method to study the 2-index chiral theories in Table \ref{two_species_table_appendix}. The relevant terms in the $6d$ anomaly polynomial  are 

\begin{align*}
\mathcal{A}^{6d}
\supset
\frac{1}{24\pi^2} \sum_i 
\txn{Tr}_{\Rep_i \times F_f}
\Bigg[
&3 \esp    
\Big( \widetilde{F}_i - B_i^{(2)} \mathbbm{1}_{N_i} \Big)^2
\wedge 
\esp  q^\alpha_i \Big( F_\alpha - \kappa_\alpha B_u^{(2)} \Big) 
\otimes \mathbbm{1}_{\txn{dim}\Rep_i} \otimes \mathbbm{1}_{N_i}
\\
+ \esp & 3 \esp    
\Big( \widetilde{F}_i - B_i^{(2)} \mathbbm{1}_{N_i} \Big)^2
\wedge 
\esp  q^\beta_i \Big( F_\beta - \kappa_\beta B_u^{(2)} \Big) 
\otimes \mathbbm{1}_{\txn{dim}\Rep_i} \otimes \mathbbm{1}_{N_i}
\Bigg]
\\
+ \esp 
\frac{1}{24\pi^2} \esp N_i& \esp \txn{dim}(\Rep_i) 
\Bigg[
q^\alpha_i  \Big( F_\alpha - \kappa_\alpha B_u^{(2)} \Big)
+
q^\beta_i  \Big( F_\beta - \kappa_\beta B_u^{(2)} \Big) 
\Bigg]^3 
\esp.
\numberthis
\end{align*}

By evaluating the traces and defining the following topological charge densities  

\begin{align*}
\rho_i 
&=
\frac{1}{8\pi^2}
\bigg(
\txn{Tr} \big[ \widetilde{F}_i  \wedge \widetilde{F}_i \big]
-  N_i \esp  B_i^{(2)}  \wedge B_i^{(2)} 
\bigg)
\\
\rho_{\alpha,\beta} 
&= 
\frac{1}{8\pi^2}
\big( F_{\alpha,\beta} - \kappa_{\alpha,\beta} B_{u}^{(2)} \big)
\wedge
\big( F_{\alpha,\beta} - \kappa_{\alpha,\beta} B_{u}^{(2)} \big)
\\
\rho_{\alpha\beta} &= 
\frac{1}{8\pi^2}
\big( F_{\alpha} - \kappa_{\alpha} B_{u}^{(2)} \big)
\wedge
\big( F_{\beta} - \kappa_{\beta} B_{u}^{(2)} \big)
\spa\spa,
\numberthis
\label{topo_charge_densisties_descent}
\end{align*}

and then collecting the terms proportional to $\de A_{\alpha,\beta}$, we write

\begin{align*}
\mathcal{A}^{6d}
\supset
\sum_i 
\big( q_i^\alpha \esp \de A_\alpha 
+ q_i^\beta \esp \de A_\beta 
\big)
\wedge 
\bigg[
\txn{dim}(\Rep_i)  \rho_i
+
\frac{N_i}{3} \txn{dim}(\Rep_i) 
\Big(
(q^\alpha_i)^2  \rho_\alpha
+ 2 q^\alpha_i q^\beta_i  \rho_{\alpha\beta}
+(q^\beta_i)^2 \rho_\beta
\Big)
\bigg]
\esp.
\\
\numberthis
\end{align*}

The $5d$ anomaly polynomial reads
\begin{align*}
\mathcal{A}^{5d}
&\supset
\sum_{i} \bigintsss_{\Sigma^5}
\big( q_i^\alpha  A_\alpha 
+ q_i^\beta   A_\beta 
\big)
\wedge 
\bigg[
\txn{dim}(\Rep_i)  \rho_i
+
\frac{N_i}{3} \txn{dim}(\Rep_i) 
\Big(
(q^\alpha_i)^2  \rho_\alpha
+ 2 q^\alpha_i q^\beta_i  \rho_{\alpha\beta}
+(q^\beta_i)^2 \rho_\beta
\Big)
\bigg]
\esp.
\numberthis
\end{align*}

By evaluating a gauge transformation $A_{\alpha,\beta} \longrightarrow A_{\alpha,\beta}  + \de \lambda_{\alpha,\beta}$, where at the boundary $\lambda_{\alpha,\beta} = \alpha,\beta$ is the parameter for the U(1)$_{\alpha,\beta}$ transformation, and integrating the topological charge densities which yield the non-integer topological charges $Q_i$ and $Q_{\alpha,\beta}$

\begin{align}
Q_c
=
\bigintsss_{\Sigma_4} \rho_c
\spa,\spa
Q_i
=
\bigintsss_{\Sigma_4} \rho_i
\spa,\spa
Q_\alpha
=
\bigintsss_{\Sigma_4} \rho_\alpha
\spa,\spa
Q_\beta
=
\bigintsss_{\Sigma_4} \rho_\beta
\spa,\spa
Q_{\alpha\beta}
=
\bigintsss_{\Sigma_4} \rho_{\alpha\beta}
\spa\spa.
\end{align}

Here $\rho_c$ is the equivalent density as $\rho_i$ in (\ref{topo_charge_densisties_descent}), but given in terms of $\widetilde{F}_c$ and $B_c^{(2)}$. Since we are gauging the subgroup  $\mathbbm{Z}_{N_3 / \me_\gamma}$  the $Q_3$ charge is proportional to $\me_\gamma^2$ 

\begin{align*}
Q_3 
\supset 
- \frac{1}{8\pi^2} \esp N_3 
\bigintsss_{\Sigma_4} 
B^{(2)}_3 \wedge B^{(2)}_3 
=
- \frac{ \me_\gamma^2  }{N_3} 
\spa\spa.
\numberthis
\end{align*}

The $4d$ anomaly reads
\begin{align*}
\mathcal{A}^{4d}
&\supset
\lambda_{\alpha,\beta} 
\sum_{i} 
\bigintsss_{\Sigma^4}
\bigg(
\txn{dim}(\Rep_i)  \rho_i
+
\frac{1}{3} \esp N_i \txn{dim}(\Rep_i) 
\Big(
(q^\alpha_i)^2  \rho_\alpha
+ 2 q^\alpha_i q^\beta_i  \rho_{\alpha\beta}
+(q^\beta_i)^2 \rho_\beta
\Big)
\bigg)
\spa\spa.
\numberthis
\label{4d_U(1)_CFU_anomaly_descent}
\end{align*}

The Stora-Zumino descent formalism generates a factor of $1/3$ for the U$(1)^3$ anomaly, which appears to be a limitation of the abelian descent procedure; for U$(1)^3$ there is no non-vanishing counterterm available to convert it into the covariant anomaly. The index theorem yields the  covariant anomaly coefficient of 1 (see Section \tcolor{\ref{section_twisted_gauge_field}}). To establish a relation between these methods (as discussed in Section \ref{section_relations_between_fluxes}), we will ignore this 1/3 factor. The induced shift in the partition function under a U(1)$_{\alpha,\beta}$ transformation in the background of the CFU center fluxes  is 
\begin{align}
\mathcal{Z} 
\longrightarrow 
\eu{2\pi i \alpha \sum_i q^\alpha_i \Theta^i}
\esp
\mathcal{Z} 
\spa\spa,\spa\spa
\mathcal{Z} 
\longrightarrow 
\eu{2\pi i \beta \sum_i q^\beta_i \Theta^i}
\esp
\mathcal{Z} 
\spa\spa.
\end{align} 

Here $\alpha$ and $\beta$ are real-valued U$_\alpha$(1) and U$_\beta$(1) parameters and we have defined the index 
\begin{align}
\Theta^i 
=
2N_i T_i Q_c
+
\txn{dim}(\Rep_i) Q_i 
+ 
N_i \esp \txn{dim}(\Rep_i)  
\Big(
\big( q^{\alpha}_i \big)^2 Q_{\alpha}
+ 2  q^{\alpha}_i q^{\beta}_i Q_{\alpha\beta}
+ \big( q^{\beta}_i \big)^2 Q_{\beta}
\Big)
\spa\spa.
\label{CFU_anomalies_stora_zumino}
\end{align}

Note that we have included for generality the contribution coming from the color center fluxes proportional to the Dynkin indices $T_i$ in equation (\tcolor{\ref{CFU_anomalies_stora_zumino}}). However, as discussed around equation (\tcolor{\ref{U(1)_color_mixed_anomaly}}), this contribution vanishes by construction regardless of the value of $Q_c$ when summed over the matter content. If there were an extra genuine discrete symmetry that acts faithfully on the fermions, which can come from a disconnected component of the U(1) symmetries for instance, then the anomaly induced by this 0-form discrete symmetry transformation in the background of the CFU fluxes would have an extra piece proportional to the Dynkin indices $T_i$. For the theories studied here however, we have shown explicitly that there is no such symmetry. This putative faithful transformation can be absorbed in the center of the flavor SU($N_3$) symmetry, i.e. the one acting on the fermions transforming under the anti-fundamental representation of the gauge group SU($N$). This is an important difference from other chiral gauge theories with no fermions in this representation like those studied in \tcolor{\cite{Anber:Chan:2023}}.\\

Other studies have defined the mixed anomaly in a slightly different way. For instance, in \tcolor{\cite{Bolognesi:Konishi:Luzio:2020}} the anomaly is computed by collecting the following terms in the $6d$ functional 
\begin{align}
\frac{\mathcal{A} }{8 \pi^2 }  
\bigintsss_{\Sigma_6}
\de A_\alpha \wedge B^{(2)}_i \wedge B^{(2)}_i 
\spa\spa.
\end{align}
This yields the $4d$ anomaly
\begin{align}
\alpha \esp 
\frac{\mathcal{A}}{8 \pi^2 }  
\bigintsss_{\Sigma_4}
B^{(2)}_i \wedge B^{(2)}_i
=
\alpha \mathcal{A} \esp \frac{ k }{N_i^2} \spa,\spa k \in \mathbbm{Z}_{N_i}
\spa\spa,
\end{align}

which is in general fractional. This contribution to the anomaly is contained in the more general topological charge densities (\tcolor{\ref{topo_charge_densisties_descent}}).

%% file: 05_anomalies_2.tex
\setlength{\jot}{13pt}

\section{Anomalies from twisted gauge field configurations}
\label{section_twisted_gauge_field}

While the descent procedure systematically yields all the anomalies in a theory with proper anomaly coefficients, we can also determine the anomalies induced by fractional fluxes by taking full advantage of the solutions to the cocycle conditions in (\tcolor{\ref{consistency_conditions_global_symm_section}}) given in equations (\tcolor{\ref{integers_sol_flavorcenter_N1_N2}}), (\tcolor{\ref{integers_sol_flavorcenter_N3}}), (\tcolor{\ref{integers_sol_axialcenter}}) and (\tcolor{\ref{integers_sol_colorcenter}}). These 5 sets of redundancies define the discrete structure of the full symmetry group in (\ref{full_symmetry_group}) that acts faithfully on the fermions. In this approach the mixed anomalies between zero- and one-form center symmetries are sourced by twisted fluxes for one-form gauge fields, instead of those of two-form gauge fields. Here we follow e.g. \cite{Anber:Poppitz:2019,Anber:Hong:Son:2022}, where this method was used to study some strongly coupled gauge theories.\\

The method proceeds as follows. On a 4-torus $\mathbbm{T}^4$ with period length $\ell$, the twisted gauge field configurations compatible with the cocycle conditions are
\setlength{\jot}{8pt}
\begin{align*}
a_\mu &= 
\bigg(
0 \esp,\esp  
C^{c}_1 \esp y  \esp,\esp 
0  \esp,\esp  
C^{c}_2  \esp t  
\bigg) 
\frac{ 2\pi }{\ell^2} \esp \bm{H}^c \cdot \bm{\nu}^c
\spa \txn{: color SU(\textit{N}) dynamical gauge field}
\\
\Gamma^i_\mu &= 
\bigg(
0 \esp,\esp  
C^{i}_1 \esp y  \esp,\esp 
0  \esp,\esp  
C^{i}_2  \esp t  
\bigg) 
\frac{ 2\pi }{\ell^2} \esp \bm{H}^i \cdot \bm{\nu}^i
\spa \txn{: flavor SU}(N_i) \txn{ background gauge fields}
\\
A^{\alpha,\beta}_\mu &= 
\bigg(
0 \esp,\esp  
C^{\alpha,\beta}_1 \esp y  \esp,\esp 
0  \esp,\esp  
C^{\alpha,\beta}_2  \esp t
\bigg) 
\frac{2\pi}{\ell^2}
\spa \txn{ : } \txn{U}(1)_{\alpha,\beta}\txn{ background gauge fields}
\spa\spa.
\numberthis
\label{CFU_fluxes}
\end{align*}
\setlength{\jot}{13pt}

Here $\bm{H} = \big( H_1,...,H_{N-1} \big)$ are the Cartan generators of SU($N$), and $\bm{\nu} = \big( \nu_1,...,\nu_{N-1} \big)$ are the weights in the particular representation for the group. It must be true that at $x = \ell$ and $m_i = 1$, the combination $\bm{H}\cdot \bm{\nu}$ takes the form $\bm{H}\cdot \bm{\nu} =  \txn{diag}\big( n_1,...,n_N \big)$ where $n_i$ are integers and $\txn{Tr}(\bm{H}\cdot \bm{\nu}) = 0$, and we choose $\bm{H}\cdot \bm{\nu} = \txn{diag}(1,1,...,1,1-N)$. All the gauge fields, denoted in general by $\Omega^i_\mu$ below, must obey the torus boundary conditions
\setlength{\jot}{8pt}
\begin{align*}
\Omega^i_x(y + \ell) 
&= 
\Omega^i_x(y) 
+ 
i \esp U^i_1(x) \del_{x} U_1^{i\dagger}(x)
\spa\spa,
\\
\Omega^i_z(t + \ell) 
&= 
\Omega^i_z(t) 
+ 
i \esp U^i_2(z) \del_{z} U_2^{i\dagger}(z)
\spa\spa.
\numberthis
\end{align*}
\setlength{\jot}{13pt}

The transition functions come in pairs, one for each twisted plane. Here $U_1^i(x)$ implements the shift around the $y$ cycle, and $U_2^i(z)$ the shift around the $t$ cycle. By choosing suitable transformations 
\begin{align*}
U_1^c(x) &= \eu{
2\pi i \frac{x}{\ell} \frac{m}{N} \bm{H}^c \cdot \bm{\nu}^c
}
\spa\spa,\spa\spa
U_1^i(x) = \eu{
2\pi i \frac{x}{\ell} \frac{m_i}{N_i} \bm{H}^i \cdot \bm{\nu}^i
}
\spa\spa,\spa\spa
U_1^{\delta_i}(x) = \eu{ i \delta_i x/\ell} 
\spa\spa,
\numberthis
\end{align*}

with $\delta_i = \alpha_i,\beta_i$. To obtain $U_2^i(z)$ take $x \to z$ together with $(m,m_i,\delta_1) \to (m',m_i',\delta_2)$. The coefficients compatible with the boundary conditions are given by
\setlength{\jot}{8pt}
\begin{gather*}
C^c_1 = \frac{m}{N}
\spa,\spa
C^c_2 = \frac{m'}{N}
\spa,\spa
C^i_1 = \frac{m_i}{N_i}
\spa,\spa
C^i_2 = \frac{m'_i}{N_i}
\spa,\spa
C^{\alpha}_{i} =  \frac{\alpha_i}{2\pi}
\spa,\spa
C^{\beta}_{i} =  \frac{\beta_i}{2\pi}
\spa\spa.
\numberthis
\label{C_coeff_twisted_f}
\end{gather*}
\setlength{\jot}{13pt}

The values of $\{m,m',m_i,m'_i, \alpha_i, \beta_i\}$ are fixed. Each set corresponds to a solution to the consistency conditions in equation (\tcolor{\ref{consistency_conditions_global_symm_section}}) that define the redundancies within the symmetry group, i.e. the possible discrete center transformations that act trivially on the fermions. These are given by equations (\tcolor{\ref{integers_sol_flavorcenter_N1_N2}}), (\tcolor{\ref{integers_sol_flavorcenter_N3}}), (\tcolor{\ref{integers_sol_axialcenter}}) and (\tcolor{\ref{integers_sol_colorcenter}}). Each set defines a twisted sector that we label as $s$. Given the consistent set of gauge field configurations, we can compute the following topological charges  
\setlength{\jot}{5pt}
\begin{gather*}
\mathcal{Q}_c^{(s)} 
= 
\frac{1}{8\pi^2 } \bigintsss_{\mathbbm{T}^4}
\txn{Tr}( F \wedge F ) 
\spa\spa,\spa\spa
\mathcal{Q}_i^{(s)}
= 
\frac{1}{8\pi^2 } \bigintsss_{\mathbbm{T}^4}
\txn{Tr}( F_i \wedge F_i ) 
\spa\spa,
\\
\mathcal{Q}_{\alpha,\beta}^{(s)}
= 
\frac{1}{8\pi^2 } \bigintsss_{\mathbbm{T}^4}
F_{\alpha,\beta} \wedge F_{\alpha,\beta}  
\spa\spa,\spa\spa
\mathcal{Q}_{\alpha\beta}^{(s)}
= 
\frac{1}{8\pi^2 } \bigintsss_{\mathbbm{T}^4}
F_{\alpha} \wedge F_{\beta}  
\spa\spa.
\numberthis
\end{gather*}
\setlength{\jot}{13pt}

Since the non-abelian gauge fields are proportional to a linear combination of the Cartan generators, the non-abelian contribution to the field strengths given by the commutator of the gauge fields vanishes. The topological charges take the form\hlfootnote{Recall that $\alpha$ and $\beta$ are $2\pi$-valued and thus $\mathcal{Q}_{\alpha,\beta}$ is rational.}

\begin{gather*}
\mathcal{Q}_c^{(s)} = m m' \bigg( 1 - \frac{1}{N}\bigg)
\spa\spa,\spa\spa
\mathcal{Q}_i^{(s)} = m_i m_i' \bigg( 1 - \frac{1}{N_i}\bigg)
\spa\spa,
\\
\mathcal{Q}_{\alpha}^{(s)} 
= \frac{\alpha_1}{2\pi}\frac{\alpha_2}{2\pi}
\spa\spa,\spa\spa 
\mathcal{Q}_{\beta}^{(s)} 
= \frac{\beta_1}{2\pi}\frac{\beta_2}{2\pi}
\spa\spa,\spa\spa 
\mathcal{Q}_{\alpha\beta}^{(s)} 
=
\frac{\alpha_1\beta_2 + \alpha_2\beta_1}{8\pi^2}
\spa\spa,
\numberthis
\label{topo_charges}
\end{gather*}

where $m_3$ must be a multiple of $\me_\gamma$ because we are gauging the subgroup $\mathbbm{Z}_{\Hat{N}_3}$ only. The fluxes in (\tcolor{\ref{CFU_fluxes}}) admit fermion zero modes with Dirac indices given by 

\begin{align*}
\mathcal{I}^{(s)}_i
=
2 N_i T_i \mathcal{Q}_c^{(s)}
+
\txn{dim}(\Rep_i) \mathcal{Q}_i ^{(s)}
+ 
\esp  N_i \esp \txn{dim}(\Rep_i)
\bigg(
\big( q^{\alpha}_i \big)^2 \mathcal{Q}_{\alpha}^{(s)}
+  2 q^{\alpha}_i q^{\beta}_i \mathcal{Q}_{\alpha\beta}^{(s)}
+\big( q^{\beta}_i \big)^2 \mathcal{Q}_{\beta}^{(s)}
\bigg)
\spa\spa.
\numberthis
\label{CFU_anomalies_twisted_fields}
\end{align*}

Note that we have included for generality the contribution coming from the color center fluxes proportional to the Dynkin indices $T_i$. However, as discussed around equation (\tcolor{\ref{U(1)_color_mixed_anomaly}}), this contribution vanishes by construction regardless of the value of $\mathcal{Q}_c$ when summed over the matter content. The anomalous action of a U$(1)_{\alpha,\beta}$ transformation in the background of these fluxes induces a phase in the partition function for each twisted sector $s$ of the form
\begin{align}
\mathcal{Z} 
\longrightarrow 
\eu{ 
2\pi i \alpha \sum_i q^\alpha_i \mathcal{I}^{(s)}_i
} 
\esp
\mathcal{Z} 
\spa\spa,\spa\spa
\mathcal{Z}
\longrightarrow 
\eu{
2\pi i \beta\sum_i q^\beta_i \mathcal{I}^{(s)}_i
} 
\esp
\mathcal{Z} 
\spa\spa,
\label{partition_function_shift_twisted_fluxes}
\end{align} 

where $\alpha$ and $\beta$ are real-valued U(1)$_\alpha$ and U(1)$_\beta$ parameters.

%% file: 06_fluxes.tex
\section{Evaluating the background fluxes}
\label{fluxes_section}

The total index for each U(1)$_\delta$ rotation in a twisted sector $s$ is given by

\begin{align}
\mathcal{I}_{\delta}^{(s)} 
= \sum_i 
q^\delta_i \esp \mathcal{I}^{(s)}_i
\spa\spa,\spa\spa
\delta \equiv \alpha,\beta
\spa\spa.
\label{total_index_twisted}
\end{align}

As we explain below, it is useful to consider the individual indices $\Phi^i_\delta$ induced by independent $B^{(2)}_i$ background fluxes
\begin{align}
\Phi^j_\delta  = 
q^\delta_j  \Theta^j = 
q_j^\delta  \dim(\mathcal{R}_j) Q_j
\spa\spa,\spa\spa 
j \equiv 1,2,3
\spa\spa.
\label{partial_index_B_flux}
\end{align}

For $j \equiv \alpha,\beta, \alpha\beta$, given that all the fermions are charged under both U$(1)_\alpha$ and U$(1)_\beta$, we must sum over the matter content as

\setlength{\jot}{8pt}
\begin{gather*}
\Phi^\alpha_\delta
=
\sum_i q_i^\delta N_i \dim(\mathcal{R}_i) \esp  
(q_i^\alpha)^2 Q_\alpha 
\spa\spa,\spa\spa
\Phi^\beta_\delta
=
\sum_i q_i^\delta N_i \dim(\mathcal{R}_i) \esp  
(q_i^\beta)^2 Q_\beta 
\\
\Phi^{\alpha\beta}_\delta
=
2 \sum_i q_i^\delta N_i \dim(\mathcal{R}_i) \esp  
q_i^\alpha q_i^\beta  Q_{\alpha\beta} 
\spa\spa.
\numberthis
\label{partial_index_gu_B_flux}
\end{gather*}
\setlength{\jot}{13pt}

In what follows we evaluate the anomalies for particular sets of minimal background fluxes. We show this for both methods used to compute the anomalies. The first approach follows the descent prescription for the anomaly polynomial in (\tcolor{\ref{6d_anomaly_polynomial}}) from $6d$ to $4d$, where two-form background fields are introduced to gauge the center symmetries. In the second approach the theory is put on a 4-torus, which allows us to define twisted gauge field configurations for ordinary one-form gauge fields (\tcolor{\ref{CFU_fluxes}}). These configurations are such that the consistency conditions in (\ref{consistency_conditions_global_symm_section}) that define the complete global symmetry group are satisfied. Explicit values for the total Dirac indices are given in Appendix \tcolor{\ref{appendix_fluxes}}.

\subsection{Two-form gauge fields}

By turning on only the fluxes induced by the $B^{(2)}_i$ and $B^{(2)}_{u}$ fields, the integrated topological charge densities in (\ref{topo_charge_densisties_descent}) take the form 
\setlength{\jot}{10pt}
\begin{gather*}
Q_c = - \frac{1}{N}
\spa\spa,\spa\spa
Q_1 = - \frac{1}{N_1}
\spa\spa,\spa\spa
Q_2 = - \frac{1}{N_2}
\spa\spa,\spa\spa
Q_3 = - \frac{\me_\gamma^2}{N_3}
\spa\spa,\\
Q_\alpha 
= 
\bigg(
\frac{2 N_2 T_2}{N_3} \frac{\me_\alpha}{\me_\delta}
\bigg)^2
\spa\spa,\spa\spa
Q_\beta 
= 
\bigg(
\frac{2 N_1 T_1}{N_3} \frac{\me_\beta}{\me_\delta}
\bigg)^2
\spa\spa,\spa\spa
Q_{\alpha\beta} = 
- \frac{(2N_1 T_1)(2N_2 T_2)}{N_3^2}
\esp
\frac{\me_\alpha \me_\beta}{\me_\delta^2}
\spa\spa.
\label{B_fields_topo_charges}
\numberthis
\end{gather*}
\setlength{\jot}{13pt}

Each independent flux defines a different anomaly given by $\Phi^i_{\alpha,\beta}$. Recall however that $Q_\alpha$, $Q_\beta$ and $Q_{\alpha\beta}$ have to be turned on simultaneously. In Appendix \tcolor{\ref{appendix_fluxes_B_fields}} we show these anomalies for each independent flux and different values of $N$.

\subsection{Twisted one-form gauge fields}
\label{section_evaluating_fluxes_twisted_one_form}

The fluxes induced by the twisted gauge field configurations are given by
\setlength{\jot}{10pt}
\begin{gather*}
\mathcal{Q}^{(s)}_1 = m_1^2 \bigg( 1 - \frac{1}{N_1}\bigg)
\spa\spa,\spa\spa
\mathcal{Q}^{(s)}_2 = m_2^2 \bigg( 1 - \frac{1}{N_2}\bigg)
\spa\spa,\spa\spa
\mathcal{Q}^{(s)}_3 = m_3^2 \bigg( 1 - \frac{1}{N_3}\bigg)
\spa\spa,
\\
\mathcal{Q}^{(s)}_c = m^2 \bigg( 1 - \frac{1}{N}\bigg)
\spa\spa,\spa\spa
\mathcal{Q}^{(s)}_\alpha = 
\frac{\alpha_1}{2\pi}\frac{\alpha_2}{2\pi}
\spa\spa,\spa\spa
\mathcal{Q}^{(s)}_\beta =
\frac{\beta_1}{2\pi}\frac{\beta_2}
{2\pi}
\spa\spa,\spa\spa
\mathcal{Q}^{(s)}_{\alpha\beta} = \frac{\alpha_1\beta_2 + \alpha_2\beta_1}{8\pi^2}
\spa\spa.
\label{twisted_topo_charges}
\numberthis
\end{gather*}
\setlength{\jot}{13pt}

There are five independent sets of integers $\{m,m_i,\alpha_i,\beta_i\}$ that satisfy the consistency conditions in (\ref{consistency_conditions_global_symm_section}), and each of these defines an anomaly given by $\mathcal{I}^{(s)}_{\alpha,\beta}$ (indexed here by the superscript $s$). The solutions are shown in equations (\tcolor{\ref{integers_sol_flavorcenter_N1_N2}}), (\tcolor{\ref{integers_sol_flavorcenter_N3}}), (\tcolor{\ref{integers_sol_axialcenter}}) and (\tcolor{\ref{integers_sol_colorcenter}}), where we used the following integer parameters to evaluate the topological charges

\begin{align}
&p = q = r = 0
\spa\spa,\spa\spa
p_3 = q_3 = r_3 = 1
\spa\spa,\spa\spa
t = 1
\spa\spa,\spa\spa
p_c = q_c = r_c = 0
\spa\spa.
\end{align}

In Appendix \tcolor{\ref{appendix_fluxes_twisted_fields}} we show these anomalies for each twisted sector and some values of $N$.


\subsection{Relation between fluxes}
\label{section_relations_between_fluxes}

We can regard the anomalies induced in the five twisted sectors as a linear combination of those sourced by $B^{(2)}_i$ fields. The coefficient in the expansion of anomalies computed for each twisted sector can be written in terms of those computed from $B^{(2)}_i$ fields by taking the ratio between fluxes. We write these coefficients as $\mathcal{K}^{(s)}_{j}$ where $s$ labels each sector and $j \equiv c,1,2,3,\alpha,\beta,{\alpha\beta}$ stands for each of the center symmetries participating in the anomaly
\begin{align}
\mathcal{I}^{(s)}_{\delta}
=
\sum_j \mathcal{K}^{(s)}_{j}
\Phi^{j}_{\delta} 
\spa\spa,\spa\spa
\mathcal{K}^{(s)}_{j}
=
\frac{ \mathcal{Q}^{(s)}_j }{Q_j}
\spa\spa.
\label{fluxes_linear_comb}
\end{align}

Below we show the non-zero coefficients for each twisted sector.\\

$\bullet$ \textbf{Flavor center} $\mathbbm{Z}_{N_1}$\textbf{: \textit{s} = 1 twisted sector}

\begin{gather*}
\mathcal{K}^{(1)}_{1}
=
1 - N_1
\spa\spa,\spa\spa
\mathcal{K}^{(1)}_{3}
= (1 - N_3) \bigg(\frac{2 T_1}{\me_\gamma}\bigg)^2
\spa\spa,\spa\spa
\mathcal{K}^{(1)}_{\alpha}
= \bigg( \frac{\me_\delta  }{ 2 T_2 N_2 N_1} \bigg)^2
\spa\spa.
\numberthis
\end{gather*}

$\bullet$ \textbf{Flavor center} $\mathbbm{Z}_{N_2}$\textbf{: \textit{s} = 2 twisted sector}

\begin{gather*}
\mathcal{K}^{(2)}_{2}
=
1 - N_2
\spa\spa,\spa\spa
\mathcal{K}^{(2)}_{3}
= (1 - N_3) \bigg(\frac{2 T_2}{\me_\gamma}\bigg)^2
\spa\spa,\spa\spa
\mathcal{K}^{(2)}_{\beta}
= \bigg( \frac{\me_\delta  }{ 2 T_1 N_1 N_2} \bigg)^2
\spa\spa.
\numberthis
\end{gather*}

$\bullet$ \textbf{Flavor center} $\mathbbm{Z}_{N_3}$\textbf{: \textit{s} = 3 twisted sector}

\setlength{\jot}{5pt}
\begin{gather*}
\mathcal{K}^{(3)}_{3}
= \frac{1 - N_3}{ \me_\gamma^2 }
\bigg( \sum_{i} 2 T_i N_i \bigg)^2
\spa\spa,\spa\spa
\mathcal{K}^{(3)}_{\alpha}
=
\bigg( \frac{\me_\delta  }{ 2 T_2 N_2 } \bigg)^2
\spa\spa,\spa\spa
\mathcal{K}^{(3)}_{\beta}
=
\bigg( \frac{\me_\delta }{ 2 T_1 N_1  } \bigg)^2
\spa\spa,\\
\mathcal{K}^{(3)}_{\alpha\beta} = 
-\frac{\me_\delta^2}{(2N_1 T_1)(2N_2 T_2)}
\spa\spa.
\numberthis
\end{gather*}
\setlength{\jot}{13pt}
$\bullet$ \textbf{Color center} $\mathbbm{Z}_{N_c}$\textbf{: \textit{s} = 4 twisted sector}

\setlength{\jot}{5pt}
\begin{gather*}
\mathcal{K}^{(4)}_{c}
= 1 - N
\spa\spa,\spa\spa
\mathcal{K}^{(4)}_{3}
= \frac{1 - N_3}{ \me_\gamma^2 N^2 }
\bigg( \sum_{i} 2 T_i N_i n_i^c \bigg)^2
\spa\spa,\spa\spa
\mathcal{K}^{(4)}_{\alpha}
=
\bigg( \frac{\me_\delta \esp n_1^c }{ 2 T_2 N_2 N} \bigg)^2
\spa\spa,
\\
\mathcal{K}^{(4)}_{\beta}
=
\bigg( \frac{\me_\delta \esp n_2^c }{ 2 T_1 N_1  N} \bigg)^2
\spa\spa,\spa\spa
\mathcal{K}^{(4)}_{\alpha\beta} = 
-\frac{n_1^c n_2^c \me_\delta^2}{(2N_1 T_1)(2N_2 T_2)N^2}
\spa\spa.
\numberthis
\end{gather*}
\setlength{\jot}{13pt}

$\bullet$ \textbf{Axial} $\txn{U}_\alpha(1) \times \txn{U}_\beta(1)$ \textbf{center:} \textbf{\textit{s} = 5 twisted sector}
\begin{align*}
\mathcal{K}^{(5)}_{\alpha}
= 1
\spa\spa,\spa\spa
\mathcal{K}^{(5)}_{\beta}
= 1
\spa\spa,\spa\spa
\mathcal{K}^{(5)}_{\alpha\beta} = 1
\spa\spa.
\numberthis
\end{align*}

See Appendix \ref{appendix_fluxes} for explicit evaluations of equation (\ref{fluxes_linear_comb}).

%% file: 05_02_witten_anomalies.tex
\setlength{\jot}{13pt}

\section{Witten anomalies}
\label{section_witten_anomalies}

As is well known, SU(2) gauge theory with an odd number of Weyl fermions in the fundamental representation is an inconsistent theory, as it suffers from a gauge anomaly \tcolor{\cite{Witten:1982,Wang:Weng:Witten:2019}}. Since all the representations of SU(2) are real or pseudoreal, the perturbative anomaly coefficient $A$ given by the symmetric invariant $ d^{abc} = \tfrac{1}{2}\txn{Tr}(T^a \{ T^b,T^c\})$ vanishes. Anomalies given by $d^{abc}$ are local in the sense that they are perturbative and related to \textit{small} gauge transformations. The anomaly of SU(2) discovered in \tcolor{\cite{Witten:1982}} is global in the sense that it is not perturbative and is sourced by \textit{large} gauge transformations (those not continuously connected to the identity). Whether SU(2) is a gauge or a flavor symmetry, its anomaly is said to be global, as defined above. Its origin is topological as it is classified by the fourth homotopy group $\pi_4(\txn{SU}(2)) = \mathbbm{Z}_2$. The upshot is that a theory with an odd number of Weyl doublets\hlfootnote{More generally, the anomaly is present for an odd number of multiplets in  representations of odd Dynkin index \tcolor{\cite{Wang:Weng:Witten:2019}.}} under SU(2) suffers from an anomaly; the partition function acquires a minus sign induced by a nontrivial large gauge transformation.\\

For the theories studied here, the gauge group is SU($N$) with $N \geq 5$, so there is no gauge Witten anomaly. For some of these theories it is the case that $|N_i| = 2$, and thus the flavor symmetry group contains a flavor SU(2) factor. If we gauge this symmetry to compute the 't Hooft anomalies, we must check whether the total number of doublets, given by the dimension of the color representation, is odd, indicating that there is a 't Hooft pure Witten anomaly. It is also possible to consider mixed 't Hooft anomalies given by SU(2) large gauge transformations in the background of the CFU fluxes. To properly account for this type of mixed anomaly, we need to employ the machinery of the $\eta$-invariant \tcolor{\cite{Atiyah:Patodi:Singer:1975,Atiyah:Singer:1971,Callan:Harvey:1984,Dai:Freed:1994,Witten:2015,Witten:Yonekura:2019}}. Here we only indicate for which cases there is a pure Witten anomaly. Note, however, that theories lacking a pure Witten anomaly may still contain a mixed Witten–CFU anomaly. Given the dimensions of the color representations, in Table \tcolor{\ref{table_Ni_N_values_witten_anomaly}} we show the values of $N$ that yield a Witten anomaly when $|N_i| = 2$ for each of the theories in Table \tcolor{\ref{table_N1_N2_N3_values}}.

\renewcommand{\extable}{0.7ex}
\begin{center}
\begin{tabular}{||c | c | c   ||} 
\hline\hline
\rule{0pt}{3ex}  
$N_i$ & $N$ & $R_i$ \\ [0.5ex]
\hline\hline
\phantom{-} &  &  \\ [-2ex]
$N_1$  & $N \equiv 1,2 \esp (\txn{mod } 4) = 5,6,9,10,13,14,...$ &
$R_{12}, \esp R_{13}, \esp R_{14}, \esp R_{15}, \esp R_{16}$
\\ [\extable]
$N_2$  & $N \equiv 2,3 \esp (\txn{mod } 4) = 6,7,10,11,14,15,...$ &
$R_{4}, \esp R_{8}, \esp R_{18}, \esp R_{20}, \esp R_{28}$
\\ [\extable]
$N_3$  & $N \esp:\esp \txn{odd}$ & 
$R_9 \txn{ for } N = 7,9$ \spa\spa,\spa\spa $R_{22} \txn{ for } N = 15,17$
\\ [\extable]
\hline
\end{tabular}
\end{center}
\captionof{table}{ \label{table_Ni_N_values_witten_anomaly}
Theories for which there is a Witten anomaly at particular $N$.\\
}

Note that since $N_3$ depends on $N$, there is only a finite set of values for $N$ that yield an anomaly. The value $|N_3| = 2$ takes place at odd $N$ only for $R_9$ and $R_{22}$. For both the $N_1$ and $N_2$ factors, the anomaly is present for 2 out of every 4 consecutive integers $N$. For theories with two different flavor SU(2) factors, it is always the case that only one of these yields a pure Witten anomaly\hlfootnote{E.g. $R_{15}$ at $N = 18$ has $N_1 = 2$ and $N_3 = 2$, but $\txn{dim}(\txn{\textbf{S}}_2) = 171$ is odd, while $\txn{dim}(\txn{\textbf{F}}) = 18$ is even.}. This anomaly must be matched in the IR: a trivially gapped symmetric phase is excluded, and if the theory confines while preserving the flavor SU(2), its massless composites must contain an odd number of doublets.

%% file: 07_Large_N.tex
\setlength{\jot}{13pt}

\section{The large $N$ limit} \label{section_large_N}

The theories in Table \tcolor{\ref{table_N1_N2_N3_values}} satisfy gauge anomaly cancellation and are asymptotically free for arbitrarily large $N$, and thus we can apply large $N$ techniques to infer consistent infrared behaviors. In this section we discuss possible patterns of symmetry breaking, putting together large $N$ with ordinary zero-form anomaly matching; the case of anomaly matching with generalized symmetries will be taken up in the next section. In \cite{Eichten:Peccei:Preskill:Zeppenfeld:1986} a large $N$ analysis for theories $R_3$ and $R_7$ is discussed. See \cite{Karasik:Onder:Tong:2022} for a discussion of $R_1$ also based on this technique.\\

We start by reviewing the results in \cite{Eichten:Peccei:Preskill:Zeppenfeld:1986} for $R_3$, and then extend them for the rest of the EKK models. In particular, in Section \tcolor{\ref{sub_section_large_N_N2_in_A}} we present a solution to the constraints from the ’t Hooft anomalies for $R_5$ and $R_{17}$. This IR phase is realized by a partial chiral symmetry breaking pattern and a set of massless composite fermions. We also briefly comment on the possibility that the gauge symmetry could break at low energies for the theories where the $N_2$ fermions $\lambda$ transform in the complex-conjugated two-index anti-symmetric representation.\\ 

In theories where the number of flavors $N_f \ll N$  in the large $N$ limit, fundamental fermion loops are suppressed; this scenario is usually referred to as the 't Hooft limit \tcolor{\cite{tHooft:1974}}. In this case the baryon masses scale as $m \sim N \Lambda_{\txn{QCD}}$, and thus the baryons are decoupled\hlfootnote{Arguing that baryons decouple in the large $N$ limit assumes that chiral symmetry breaking takes place. If a chiral symmetry is unbroken, it can protect some baryons, keeping them light.} \cite{Witten:1979}. For theories with $N_f \sim N$ fundamentals, like the chiral gauge theories studied here where $N_3 \sim N$, fermion loops are not suppressed. This scenario is usually referred to as the Veneziano limit \tcolor{\cite{Veneziano:1976}}. Moreover, matter loops involving ${\cal{O}}(1)$ flavors of two-index tensor irreps are also  unsuppressed at large $N$. \\


At large $N$ the theory is dominated by planar diagrams, and the physical spectrum can be read off by cutting them. A cut planar diagram is a planar vacuum diagram severed along a line, so that the lines crossing the cut specify the field content of a gauge-invariant operator together with its $1/N$ scaling. Operators appearing on such a cut are those that survive the planar limit, while any requiring a non-planar contraction is suppressed by powers of $1/N$. Here $N_3$ scales with $N$, so the relevant counting is that of the Veneziano rather than the 't Hooft limit. Fermion loops are unsuppressed and quark lines are retained in the planar class. This is what makes the limit useful for constraining symmetry breaking, since the operators appearing in cut planar diagrams delimit the set of composite fermions and condensates available\hlfootnote{Whether a given composite remains massless is a further dynamical question.} at large $N$. A condensate appears in a cut planar diagram in the sense that the corresponding fermion bilinear interpolates for the state exposed by the cut; a nonzero expectation value is then an unsuppressed effect at large $N$, and the charges carried by that operator determine which symmetry generators break and which survive. See \tcolor{\cite{Eichten:Kang:Koh:1982,Karasik:Onder:Tong:2022}} for a brief review of the large $N$ technique.

\subsection{The case of $R_3$}

For the $R_3$ theory $N_1=1$, $N_2 =1$, and $N_3 = 2N$. Ignoring redundant center transformations for the moment, the symmetry group is 

\begin{align}
G_{\txn{UV}}
=
\txn{SU}(N) \times 
\txn{SU}(2N) \times 
\txn{U}_\alpha(1) \times \txn{U}_\beta(1) 
\spa\spa.
\end{align}

This theory is the third simplest among those in Table \tcolor{\ref{table_N1_N2_N3_values}}. Before discussing it let us briefly recall what is known about the simplest theories, $R_1$ and $R_2$.

Famously the authors of \tcolor{\cite{Bars:Yankielowicz:1981}} noted that the IR of $R_1 = \txn{\textbf{S}}_2 \oplus (N+4) \overline{\txn{\textbf{F}}}$ admits a non-SUSY symmetric confining phase. The massless composite fermions given by 

\begin{align}
\eta^a_i \psi^{ij} \eta^b_j
\spa\spa \txn{in} \spa\spa 
\txn{\textbf{A}}_2 
\spa\spa \txn{of} \spa\spa \txn{SU}(N+4)
\spa\spa,
\label{large_N_R1_baryons}
\end{align}

alone match all the 't Hooft anomalies for all $N$ preserving the global symmetry group at low energies. The story is analogous for $R_2 = \txn{\textbf{A}}_2 \oplus (N-4) \overline{\txn{\textbf{F}}}$. Given these solutions, a natural expectation is that the following set of massless composite fermions 
\begin{align}
\mathcal{B}_S^{a,b} &= \eta^a_i \psi^{ij} \eta^b_j
\spa\spa \txn{in} \spa\spa 
\txn{\textbf{A}}_2 
\spa\spa \txn{of} \spa\spa \txn{SU}(2N)
\spa\spa,
\\
\mathcal{B}_A^{a,b} &= \eta^a_i \lambda^{ij} \eta^b_j
\spa\spa \txn{in} \spa\spa 
\txn{\textbf{S}}_2 
\spa\spa \txn{of} \spa\spa \txn{SU}(2N)
\spa\spa,
\label{large_N_R3_baryons}
\end{align}

might match the complete set of 't Hooft anomalies in a symmetric confining phase of $R_3$. This is not the case, however.  If a symmetric confining phase can take place, more exotic composite color singlet fermionic operators are needed (see Section \tcolor{\ref{section_massless_general_commposite_operators}} for a brief discussion on this possibility).\\

Based on the large $N$ analysis, the authors of \tcolor{\cite{Eichten:Peccei:Preskill:Zeppenfeld:1986}} suggested a symmetry breaking pattern for $R_3$ that matches all the traditional zero-form 't Hooft anomalies. In their study they used the following fermion charge assignment 

\begin{align*}
\widetilde{q_i}^\alpha = 
\Big(
2N_3 T_3  \esp,\esp 2N_3 T_3 \esp,\esp - 2N_1 T_1  - 2N_2 T_2
\Big)
\spa\spa,\spa\spa
\widetilde{q_i}^\beta = 
\Big(
2 N_2 T_2 \esp,\esp - 2N_1 T_1 \esp,\esp 0
\Big)
\spa\spa,
\numberthis
\label{EPPZ_tilde_charge_assigment}
\end{align*}

which differs from the charge assignment used in this work
\begin{align*}
q^\alpha_i = 
\Big(
2N_3 T_3  \esp,\esp 0 \esp,\esp - 2N_1 T_1 
\Big)
\spa\spa,\spa\spa
q^\beta_i = 
\Big(
0 \esp,\esp 2N_3 T_3 \esp,\esp - 2N_2 T_2 
\Big)
\spa\spa.
\numberthis
\end{align*}

Here we are ignoring the overall $1/\me_{\alpha}$ and $1/\me_{\beta}$ factors as these are not relevant for the matching of anomalies for zero-form symmetries. The relation between these bases is

\begin{align}
\widetilde{q_i}^\alpha = 
q^\alpha_i + q^\beta_i
\spa\spa,\spa\spa
\widetilde{q_i}^\beta =
\frac{N_2 T_2}{N_3 T_3} q^\alpha_i - \frac{N_1 T_1}{N_3 T_3} q^\beta_i
\spa\spa,
\end{align}

or equivalently
\begin{align}
q^\alpha_i = 
\frac{
N_1 T_1 \esp \widetilde{q_i}^\alpha
+ N_3 T_3 \esp \widetilde{q_i}^\beta
}
{N_1 T_1 + N_2 T_2}
\spa\spa,\spa\spa
q^\beta_i = 
\frac{
N_2 T_2 \esp \widetilde{q_i}^\alpha
- N_3 T_3 \esp \widetilde{q_i}^\beta
}
{N_1 T_1 + N_2 T_2}
\spa\spa.
\label{U(1)_change_of_basis}
\end{align}

For $R_3$ the fermion charges are 

\begin{align*}
\widetilde{q}^{\esp\alpha}
&= \big(1,1,-1\big)
\spa\spa,\spa\spa
\widetilde{q}^{\esp\beta}
= \big(N-2,-(N+2),0 \big)
\spa\spa,
\numberthis
\\
q^{\alpha} 
&= \big( 2N, 0 , -(N+2) \big)
\spa\spa,\spa\spa
q^{\beta}
= \big( 0 , 2N , -(N-2) \big)
\spa\spa.
\numberthis
\end{align*}

In the IR, the only baryons appearing in the large $N$ cut planar diagrams are

\begin{align}
\mathcal{B}_S^{a,b} = \eta^a_i \psi^{ij} \eta^b_j
\spa\spa \txn{and} \spa\spa
\mathcal{B}_A^{a,b} = \eta^a_i \lambda^{ij} \eta^b_j
\spa\spa,
\label{R3_large_N_baryons}
\end{align}

with charges
\begin{align*}
\widetilde{q_S} &= \big(-1,N-2 \big) 
\spa\spa,\spa\spa
\widetilde{q_A} = \big(-1,-(N+2) \big) 
\spa\spa,
\numberthis
\\
q_S &= \big(-4,-2(N-2) \big) 
\spa\spa,\spa\spa
q_A = \big(-2(N+2),4 \big) 
\spa\spa.
\numberthis
\end{align*}

The only charged condensate appearing in the large $N$ cut 
planar diagrams is

\begin{align*}
\mathcal{C}^{a,b}  
=
\eta_i^a \psi^{ij} 
\lambda^\dagger_{jk}
\eta^{b, \dagger k}
\equiv
\overline{\txn{\textbf{F}}} \otimes 
\txn{\textbf{S}}_2 \otimes
\overline{\txn{\textbf{A}}}_2 \otimes
\txn{\textbf{F}}
\spa\spa,
\numberthis
\label{R3_condensate}
\end{align*}

and its charges are 

\begin{align*}
q_\mathcal{C} = \big( 2N, -2N \big)
\spa\spa,\spa\spa
\widetilde{q_\mathcal{C}} = \big( 0 ,2N \big)
\spa\spa.
\numberthis
\end{align*}

Let us briefly work in the $\widetilde{q}^{\esp\alpha,\beta}$ basis to discuss the 't Hooft anomaly matching of ordinary zero-form continuous symmetries as presented in \tcolor{\cite{Eichten:Peccei:Preskill:Zeppenfeld:1986}}. In this basis the condensate in (\tcolor{\ref{R3_condensate}}) breaks U$_\beta(1)$ leaving U$_\alpha(1)$ intact. We start by matching the U$^3_\alpha(1) \to - N^2$ anomaly, which is the same as the U$_\alpha(1) \times \txn{Grav}^2 \to -N^2$ anomaly. In the IR these anomalies are respectively  $\sum \mathcal{N}_{\mathcal{B}_i} \big(q^\alpha_{\mathcal{B}_i}\big)^3  =  - \mathcal{N}_\mathcal{B}$  and  $\sum \mathcal{N}_{\mathcal{B}_i}  q^\alpha_{\mathcal{B}_i} = - \mathcal{N}_\mathcal{B}$, where $\mathcal{N}_\mathcal{B} = \sum_i \mathcal{N}_{\mathcal{B}_i} $ is the total number of baryons including the flavor multiplicity. If there are $\mathcal{N}_\mathcal{B} = N^2$ massless baryons then these anomalies are matched. For an unbroken flavor SU$(2N)$, there would be $\txn{dim}(\txn{\textbf{S}}_2) + \txn{dim}(\txn{\textbf{A}}_2) = 4N^2$ massless baryons in the IR, and this yields anomalies $\txn{U}^3_\alpha(1) \to -4N^2$ and $\txn{U}_\alpha(1) \times \txn{Grav}^2 \to -4N^2$. Therefore this flavor symmetry must be broken down to SU$(N)$ giving $\txn{dim}(\txn{\textbf{S}}_2) + \txn{dim}(\txn{\textbf{A}}_2) = N^2$ baryons and thus matching the anomalies. We are left to match the $\txn{U}_\alpha(1) \times \txn{SU}(N)^2 \to q^\alpha_3 T^f_3 \txn{dim}(\mathcal{R}_3)$ and $\txn{SU}(N)^3 \to A^f_3 \txn{dim}(\mathcal{R}_3)$ anomalies. Take the unbroken SU$(N)$ generators $\mathcal{T}^a_{N}$ to be embedded in the SU$(2N)$ Lie algebra as $\mathcal{T}^a_{2N} = \tfrac{1}{\sqrt{2}} \esp \txn{diag}(\mathcal{T}^a_{N},\mathcal{T}^a_{N})$, which have the same fundamental Dynkin index as the IR SU$(N)$, but now with two fundamental \say{flavors} corresponding to $\eta_{i\leq N}$ and $\eta_{i\geq N}$, and thus $\txn{dim}(\mathcal{R}_3) = 2N$. In the UV the first anomaly is $\txn{U}_\alpha(1) \times \txn{SU}(N)^2 \to -N$, which is matched in the IR as $q^\alpha_S T_{\txn{\textbf{A}}_2} + q^\alpha_A T_{\txn{\textbf{S}}_2} = -N$. The UV flavor cubic anomaly is $\txn{SU}(N)^3 \to 2N$, which is matched in the IR as $A_{\txn{\textbf{S}}_2} + A_{\txn{\textbf{A}}_2} = 2N$.\\

In the $q^{\alpha,\beta}$ basis, the condensate is charged under both U$_\alpha(1)$ and U$_\beta(1)$, but $q^\xi_\mathcal{C} = q_{\mathcal{C}}^{\alpha} + q_{\mathcal{C}}^{\beta} = 0$, and thus the linear combination of generators $T_{\alpha} + T_{\beta}$ survives in the IR. We denote this remnant symmetry U$_\xi(1)$ where $q^\xi_i = q^\alpha_i + q^\beta_i$. Since $q^\chi_\mathcal{C} = q_\mathcal{C}^{\alpha} - q_{\mathcal{C}}^{\beta} = 4N$, there is a broken direction $T_{\alpha} - T_{\beta}$. If SU$(2N)$ is broken down to SU$(N)$ such that there are a total  $\txn{dim}(\txn{\textbf{S}}_2) + \txn{dim}(\txn{\textbf{A}}_2) = N^2$ massless baryons, then the $\txn{SU}(N)^3$ anomaly is matched as discussed previously. The UV anomaly $\txn{U}_\xi(1)^3 \to \sum ( q^\alpha_i + q^\beta_i )^3 = - 8N^5$ is matched in the IR by the baryons with charges $q_S^\xi = q_A^\xi = - 2N$ as  $\txn{U}_\xi(1)^3 \to q_S^3 \txn{dim}(\txn{\textbf{S}}_2) + q_A^3 \txn{dim}(\txn{\textbf{A}}_2) = - 8N^5$. The UV anomaly $\txn{U}_\xi(1) \times \txn{SU}(N)^2 \to - 2N^2$ is matched in the IR as $q_S T_{\txn{\textbf{S}}_2} + q_A T_{\txn{\textbf{A}}_2} = - 2N^2$. Finally, the UV anomaly $\txn{U}_\xi(1) \times \txn{Grav}^2 \to - 2N^3$ is matched in the IR as $q_S \txn{dim}(\txn{\textbf{S}}_2) + q_A \txn{dim}(\txn{\textbf{A}}_2) = - 2N^3$.\\

There is a caveat which might become important when discussing the matching of anomalies sourced by fractional fluxes. The charge of the condensate that breaks the $T_\alpha - T_\beta$ direction of the $\txn{U}_\alpha(1) \times \txn{U}_\beta(1)$ combination is in detail 
\begin{align}
q^\chi_\mathcal{C}
&= (q_\mathcal{C}^\alpha \me_\alpha - q_\mathcal{C}^\beta \me_\beta) / \me_\chi
= 4N / \me_\chi
\spa\spa,
\\
\me_\chi &= \txn{gcd}( 2 N_3 T_3, -2 N_1 T_1 + 2 N_2 T_2 )
\spa\spa.
\end{align}

so the U$_\chi(1)$ charges read
\begin{align*}
\me_\chi &= 2 \txn{ for \textit{N} odd} 
\spa \longrightarrow \spa
q^\chi_\mathcal{C} = 2N
\spa\spa,
\\
\me_\chi &= 4 \txn{ for \textit{N} even} 
\spa \longrightarrow  \spa
q^\chi_\mathcal{C} = N
\spa\spa.
\numberthis
\end{align*}

There is a putative symmetry that remains unbroken given by a discrete subgroup $\mathbbm{Z}^\chi_{q^\chi_\mathcal{C}}$ (up to redundancies). Similarly, for the baryons the charge is 

\begin{align*}
q^\xi_{S,A} 
&= (q_{S,A}^\alpha \esp \me_\alpha + q_{S,A}^\beta \esp \me_\beta) / \me_\xi
= -2N / \me_\xi
\spa\spa,
\numberthis
\\
\me_\xi &= \txn{gcd}( 2 N_3 T_3, 2 N_1 T_1 + 2 N_2 T_2 ) 
\spa\spa,
\numberthis
\end{align*}

in the case of $R_3$
\begin{align}
\me_\xi = 2N
\spa \longrightarrow  \spa
q^\xi_S = q^\xi_A = - 1
\spa\spa. 
\end{align}

The symmetry group in the IR is 

\begin{align}
G_{\txn{IR}} = 
\txn{SU}(N) \times 
\txn{U}_\xi(1) \times
\mathbbm{Z}^\chi_{q^\chi_\mathcal{C}} 
\label{R3_large_N_symm_group}
\spa\spa.
\end{align}

Here we are ignoring the redundancies given by center transformations, as these are not relevant for the matching of zero-form anomalies. As discussed in Section \ref{higher_form_matching_R3}, the zero-form $\mathbbm{Z}^\chi_{q^\chi_\mathcal{C}}$ does not act faithfully on the baryons, as it can be removed by center transformations.\\

Before addressing the remaining theories, let us briefly discuss some results of \tcolor{\cite{Karasik:Onder:Tong:2022}}. In order to match the anomalies, the baryons must remain massless. Equivalently, if the candidate set of baryons that match the anomalies acquire a mass, then the symmetry must be spontaneously broken. The fact that some set of baryons match all the 't Hooft anomalies could itself be seen as compelling evidence suggesting that these baryons indeed remain massless in the IR. According to \tcolor{\cite{Karasik:Onder:Tong:2022}}, for the chiral gauge theories of interest, given that $ N_3 \sim N$ it is possible to argue that the baryons in (\tcolor{\ref{R3_large_N_baryons}}) remain massless whether they match the anomalies or not. The argument presented in \tcolor{\cite{Karasik:Onder:Tong:2022}} goes as follows. Suppose the baryons acquire a mass via an eﬀective Yukawa interaction 

\begin{align}
y \Phi_{abc n m \ell } \mathcal{B}^{abc} \mathcal{B}^{n m \ell }  + \txn{h.c.}
\spa\spa,\spa\spa
a,b, n , m = 1, \esp ..., \esp N_3
\spa\spa,\spa\spa
c, \ell = 1, \esp ..., \esp N_1 \txn{ or } N_2
\spa\spa.
\label{large_N_yukawa_operator}
\end{align}

The scalar $\Phi$ can be formed out of multi-meson (i.e. multi-trace) operators or out of contractions involving SU($N$) epsilon symbols. The simplest of these for each case are $\Phi_M \sim \mathcal{B}^\dagger \mathcal{B}^\dagger$ and $\Phi_B \sim \eps \psi^{(N-4)} \eta^{(N-8)}$ or $\Phi_B \sim \eps \lambda^{(N-4)} \eta^{(N-8)}$. If operators like $\Phi_M$ condense with an expectation value of order $v$, given that both $N_1$ and $N_2$ are order unity with $N_3$ order $N$, the only way the operator in (\tcolor{\ref{large_N_yukawa_operator}}) can survive in the large $N$ limit is that $v \sim N$; however this is not allowed since the vacuum energy would scale as $N^3$. If operators like $\Phi_B$ condense it must be that $v \sim 1/\sqrt{N}$, and then the mass $m \sim v y$ vanishes for large $N$ assuming that $y$ does not scale with $N$.

\subsection{The rest of the theories} \label{large_N_general_case}

First we will restrict to the cases where $N_2 > 0$, that is, those in which the $\lambda$ fermions transform in the two-index anti-symmetric representation of the gauge group. For these models we test whether a pattern of symmetry breaking as realized in the $R_3$ theory could match the anomalies for the rest of the theories in Table \tcolor{\ref{table_N1_N2_N3_values}}. Here we make the assumption that the gauge group confines. This analysis indicates that only the theories shown in Table \tcolor{\ref{table_NB_massless_baryons_large_N_N2_rep_A}} could allow for such a pattern of symmetry breaking. From these, we find a set of massless composite fermions that match the complete set of 't Hooft anomalies for $R_{5}$ and $R_{17}$.\\

We also discuss some aspects of the theories with $N_2 < 0$, where the $\lambda$ fermions transform in the two-index complex-conjugated anti-symmetric. For these models $N_3$ can take negative values indicating that the $\eta$ fermions transform in the fundamental, contrary to the case with $N_3$ positive for which the representation is the anti-fundamental. First some earlier results of \tcolor{\cite{Eichten:Peccei:Preskill:Zeppenfeld:1986}} and \tcolor{\cite{Goity:Peccei:Zeppenfeld:1985}} are discussed, in which the simplest of these models given by $R_7$ is studied. We then extend these ideas to the rest of the models.

\subsubsection{$N_2 > 0$}
\label{sub_section_large_N_N2_in_A}

The UV global symmetry group is 
\begin{align}
G_{\txn{UV}} =
\txn{SU}(N_1) \times \txn{SU}(N_2) \times \txn{SU}(N_3) \times \txn{U}_\alpha(1) \times \txn{U}_\beta(1)
\spa\spa.
\end{align}

The only charged operator that could condense appearing in the large $N$ cut planar diagrams is
\begin{align}
\mathcal{C}^{ab}
=
\eta_i^a \psi^{ij} \lambda^\dagger_{jk} \eta^{b, \dagger k}
\spa\spa.
\end{align}

This operator is charged under both U$_\alpha(1)$ and U$_\beta(1)$, but only the U$_\xi(1)$ direction with $q_\xi = q_\alpha + q_\beta$ remains unbroken, while the direction U$_\chi(1)$ with $q_\chi = q_\alpha - q_\beta$ breaks
\begin{align}
q_\mathcal{C}^\chi = 2 N_3 / \me_\chi 
\spa\spa,\spa\spa
q_\mathcal{C}^\xi = 0
\spa\spa.
\end{align}

There is a putative symmetry that remains unbroken given by a discrete subgroup $\mathbbm{Z}^\chi_{q^\chi_\mathcal{C}}$.\\

The only baryons appearing in the large $N$ cut planar diagrams are
\begin{align}
\mathcal{B}_S^{a,b,c} = \eta^a_i \psi^{c, ij} \eta^b_j
\spa\spa \txn{and} \spa\spa
\mathcal{B}_A^{a,b,c} = \eta^a_i \lambda^{c, ij} \eta^b_j
\spa\spa.
\label{large_N_baryons_general_Ri}
\end{align}

The number of them will depend on how the flavor symmetries SU$(N_i)$ are broken (or not). Their charges are
\setlength{\jot}{8pt}
\begin{align}
q_S^\xi &= q_A^\xi = - N (N_1 + N_2) / \me_\xi
\spa\spa,
\\
q_S^\chi &= - N N_1 + N_2(3N-8)
\spa\spa ,\spa\spa
q_A^\chi = N N_2 - N_1(3N+8)
\spa\spa.
\end{align}
\setlength{\jot}{13pt}

With the U$_\chi$(1) direction broken, we are left to match 't Hooft anomalies involving U$_\xi$(1) only (in addition to the pure non-abelian flavor cubic anomalies of course). We start by matching the cubic U$_\xi^3$(1) and gravitational $\txn{U}_\xi(1) \times \txn{Grav}^2$ anomalies given by 

\begin{align}
\mathcal{A_\xi}
&=
\sum_i \big( q_i^\xi \big)^3  N_i \esp \txn{dim}(\mathcal{R}_i)
=
\big(q_S^\xi)^3 \mathcal{N}_S 
+ 
\big(q_A^\xi)^3 \mathcal{N}_A
\spa\spa,
\\
\mathcal{A_{\xi \times \txn{Grav}}}
&=
\sum_i q_i^\xi   N_i \esp  \txn{dim}(\mathcal{R}_i)
=
q_S^\xi \mathcal{N}_S 
+ 
q_A^\xi \mathcal{N}_A
\spa\spa.
\end{align}

Given that the charges for the baryons are equal, and defining the total number of baryons $\mathcal{N}_\mathcal{B} = \mathcal{N}_S +\mathcal{N}_A$, we can write 

\begin{align}
\mathcal{A_\xi}
=
\big(q_S^\xi)^2 \mathcal{A_{\xi \times \txn{Grav}}}
=
\big(q_S^\xi)^3 \mathcal{N}_\mathcal{B}
\spa\spa.
\end{align}

These anomalies are not independent constraints, so matching one automatically matches the other. From this we can determine the total number of massless composite fermions $\mathcal{N}_\mathcal{B}$ that must exist in the IR in order to match this anomaly. We report this in Table \tcolor{\ref{table_NB_massless_baryons_large_N_N2_rep_A}}, with the general expression given by

\begin{align}
\mathcal{N}_\mathcal{B}
= \frac{1}{2}
\frac{ 
\big( N_1 (N+4) + N_2(N-4) \big)
\big( N_1 (N+3) + N_2(N-3) \big)
}
{N_1 + N_2}
\spa\spa.
\label{NB_total_amount_baryons}
\end{align}

\renewcommand{\extable}{1ex}
\begin{center}
\begin{tabular}{||c | c  | c ||} 
\hline\hline
\rule{0pt}{3ex}  
\phantom{--}$R_i$\phantom{--}  & 
\phantom{--}$\mathcal{N}_\mathcal{B}$\phantom{--} &
Integer
\\ [0.5ex]
\hline\hline
\phantom{-} &   & \\ [-2ex]
$R_{3}$  & $N^2$ & \green{\cmark}  \\ [\extable]
$R_{4}$  & $\tfrac{1}{2}(3N-4)(N-1)$ & \green{\cmark} \\ [\extable]
$R_{5}$  & $(2N-3)(N-2)$ & \green{\cmark} \\ [\extable]
$R_{6}$  & $ \tfrac{1}{10}(5N-9)(5N-12)$ & \red{\xmark}\\ [\extable]
$R_{12}$ & $\tfrac{1}{2}(3N+4)(N+1)$ & \green{\cmark} \\ [\extable]
$R_{13}$ & $\tfrac{1}{10}(5N-3)(5N-4)$  & \red{\xmark} \\ [\extable]
$R_{17}$ & $(2N+3)(N+2)$ & \green{\cmark} \\ [\extable]
$R_{18}$ & $\tfrac{1}{10}(5N+4)(5N+3)$ & \red{\xmark} \\ [\extable]
$R_{23}$ & $\tfrac{1}{10}(5N+12)(5N+9)$& \red{\xmark}  \\ [\extable]
\hline
\end{tabular}
\end{center}
\captionof{table}{ \label{table_NB_massless_baryons_large_N_N2_rep_A}
Total number of massless composite fermions needed to match the U$_\xi^3$(1) anomaly.\\
}

From the matching of this anomaly alone we can already conclude that for theories $R_6$, $R_{13}$, $R_{18}$ and $R_{23}$, this pattern of symmetry breaking, as suggested by the large $N$ limit, must be incorrect since the predicted number of massless composites is not an integer. Note that these theories could develop a perturbative Banks-Zaks fixed point in the IR with a two-loop 't Hooft coupling of $\lambda_0 = 2/77 \approx 0.026$ (see Section \tcolor{\ref{BZ_FP_section}}).\\

We are left with theories $R_{4}$, $R_{5}$, $R_{12}$ and $R_{17}$. Up to this point, we know the total number of massless baryons that there should be in the IR, but we do not know the individual number of $\mathcal{B}_S$ and $\mathcal{B}_A$ baryons given by $\mathcal{N}_S$ and $\mathcal{N}_A$. This will depend on how the flavor symmetries SU$(N_i)$ break (or not). Consider that the symmetry group is broken as 

\begin{align}
G_{\txn{IR}} =
\txn{SU}(N_1 - n_1) \times \txn{SU}(N_2 - n_2) \times \txn{SU}(N_3 - n_3) \times \txn{U}_\xi(1)
\times 
\mathbbm{Z}^\chi_{2N_3/\me_\chi}
\spa\spa,
\end{align}

then the individual numbers of baryons are

\begin{align}
\mathcal{N}_S = 
(N_1 - n_1) \esp \txn{dimA}(N_3 - n_3)
\spa\spa,\spa\spa
\mathcal{N}_A =
(N_2 - n_2) \esp \txn{dimS}(N_3 - n_3)
\spa\spa,
\end{align}

where $\txn{dimS}(M)$ and $\txn{dimA}(M)$ are the dimensions for the symmetric and anti-symmetric representations of SU$(M)$. Using the result in (\tcolor{\ref{NB_total_amount_baryons}}) and the fact that $\mathcal{N}_{\mathcal{B}} = \mathcal{N}_S + \mathcal{N}_A$, we can find a solution for $n_3$ in terms of $n_1$ and $n_2$. This expression for $n_3$ is rather complicated and not very insightful so we do not show it here. We are only interested in the cases where $n_{1,2} = 0, ..., \esp N_{1,2}$. We can also consider $n_3 = 0$, but this does not yield solutions. The solutions are summarized in Table \tcolor{\ref{table_ni_solutions_large_N}}. We still have to check whether the solutions match the rest of the anomalies. From this analysis we can conclude that theories $R_4$ and $R_{12}$ do not allow for a symmetry breaking pattern as suggested by the large $N$ limit since it yields a negative value for $N_3 - n_3$.\\

\renewcommand{\extable}{0.3ex}
\begin{center}
\begin{tabular}{||c | c  | c | c | c | c | c ||} 
\hline\hline
\rule{0pt}{3ex}  
$R_i$  & 
$n_1$  & 
$n_2$  & 
$n_3$  & 
\phantom{-}$\mathcal{N}_S$\phantom{-} &
\phantom{-}$\mathcal{N}_A$\phantom{-} &
\phantom{-}$N_3 - n_3 > 0$\phantom{-}\\ [0.09ex]
\hline\hline
\phantom{-} &  &  &  &  & & \\ [-2ex]
$R_{4}$  & 0 & 0 & $4N-5$ & $N(N-1)/2$ & $(N-2)(N-1)$ & \red{\xmark} \\ [\extable] 
\arrayrulecolor{gray}
\hline
\arrayrulecolor{black}
\phantom{-} &  &  &  &  &  &\\ [-2ex]
\multirow{5}{*}{$R_5$}  & 
0 & 0 & $3(N-2)$ & $(N-2) (N-3)/2$ & $3(N-2)(N-1)/2$ & \green{\cmark} \\ [\extable]
 & 1 & 2 & $2(N-2)$ & 0 & $ (2N-3)(N-2)$ & \green{\cmark} \\ [\extable]
 & 1 & 2 & $6N-11$  & 0 & $(2N-3)(N-2) $ &  \red{\xmark} \\ [\extable]
 & 0 & 3 & $2N-5$ & $(2N-3)(N-2)$ & 0    & \green{\cmark} \\ [\extable]
 & 0 & 3 & $6(N-2)$ & $(2N-3)(N-2)$ & 0  & \red{\xmark} \\ [\extable]
\arrayrulecolor{gray}
\hline
\arrayrulecolor{black}
\phantom{-} &  &  &  &  & & \\ [-2ex]
$R_{12}$  & 0 & 0 & $5N-4$ & $(N+1)(N+2)$ & $N(N+1)/2$ & \red{\xmark} \\ [\extable]
\arrayrulecolor{gray}
\hline
\arrayrulecolor{black}
\phantom{-} &  &  &  &  & &\\ [-2ex]
\multirow{5}{*}{$R_{17}$}  &
0 & 0 & $3(N+2)$ & $3(N+1) (N+2)/2$ & $(N+2)(N+3)/2$ &  \green{\cmark} \\ [\extable]
  & 2 & 1 & $2(N+2)$ & $(2N+3)(N+2)$ & 0 & \green{\cmark}  \\ [\extable]
  & 2 & 1 & $6N+11$ & $(2N+3)(N+2)$ & 0 & \red{\xmark} \\ [\extable]
  & 3 & 0 & $2N+5$ & 0 & $(2N+3)(N+2) $  &  \green{\cmark} \\ [\extable]
  & 3 & 0 & $6(N+2)$ & 0 & $(2N+3)(N+2)$ &  \red{\xmark} \\ [\extable]
\hline
\end{tabular}
\end{center}
\captionof{table}{ \label{table_ni_solutions_large_N}
Possible patterns of symmetry breaking compatible with the matching of the U$_\xi^3$(1) anomaly.
}

The strategy presented here to match the  $\txn{U}_\xi(1)^3$ and $\txn{U}_\xi(1) \times \txn{Grav}^2$ anomalies is a general treatment; it accounts for an arbitrary symmetry breaking pattern for any of the $\txn{SU}(N_i)$ symmetries. To match the rest of the anomalies, we must specify how the SU$(N_i)$ branches under the unbroken subgroup, i.e. find a proper embedding. In what follows, we work with a specific ansatz in which the flavor symmetries break diagonally.\\

For $R_3$, the SU$(N_3)$ symmetry breaks such that the fundamental irrep decomposes as $\bm{2} \txn{\textbf{N}} \longrightarrow \txn{\textbf{N}}  \oplus \txn{\textbf{N}}$. The natural extension is to consider the $k$-fold diagonal $\txn{SU}(N_3) \longrightarrow \txn{SU}(K_3)$ with $N_3 = k K_3$, where the fundamental $N_3$ branches to $k$ copies of the fundamental $K_3$. The UV anomalies will contain multiplicities $\txn{dim}(\mathcal{R}_3) k_3$. Below we include factors of $k_i$ for generality.\\ 

The anomalies left to match are $\txn{U}_\xi(1) \times \txn{SU}(N_i)^2$ 
\setlength{\jot}{13pt}
\begin{align*}
\mathcal{A}_{\xi \times N_i}^{\txn{UV}}
&=
\frac{1}{2} q^\xi_i \esp \txn{dim}(\mathcal{R}_i) k_i
\spa\spa,
\numberthis
\\
\mathcal{A}_{\xi \times N_1}^{\txn{IR}}
&= \frac{1}{2} q^\xi_S \esp \txn{dimA}(N_3 - n_3) 
\spa\spa,
\numberthis
\\
\mathcal{A}_{\xi \times N_2}^{\txn{IR}}
&= \frac{1}{2} q^\xi_A \esp \txn{dimS}(N_3 - n_3) 
\spa\spa,
\numberthis
\\
\mathcal{A}_{\xi \times N_3}^{\txn{IR}}
&=
\frac{1}{2}\Big( (N_3 - n_3) - 2 \Big)
(N_1 - n_1) \esp  q^\xi_S 
+
\frac{1}{2}\Big( (N_3 - n_3) + 2 \Big)
(N_2 - n_2) \esp  q^\xi_A
\spa\spa,
\numberthis
\end{align*}
\setlength{\jot}{13pt}

and the non-abelian cubic anomalies $\txn{SU}(N_i)^3$
\setlength{\jot}{13pt}
\begin{align*}
\mathcal{A}_{N_i}^{\txn{UV}}
&= \txn{dim}(\mathcal{R}_i) k_i
\spa\spa,
\numberthis
\\
\mathcal{A}_{N_1}^{\txn{IR}} 
&= \txn{dimA}(N_3 - n_3)
\spa\spa,
\numberthis
\\
\mathcal{A}_{N_2}^{\txn{IR}} 
&= \txn{dimS}(N_3 - n_3)
\spa\spa,
\numberthis
\\
\mathcal{A}_{N_3}^{\txn{IR}} 
&=
(N_1 - n_1) \Big( (N_3 - n_3) - 4 \Big)
+
(N_2 - n_2) \Big( (N_3 - n_3) + 4 \Big)
\spa\spa.
\numberthis
\end{align*}
\setlength{\jot}{13pt}

$\bullet$ \textbf{The solution for} $R_5$\\

For $R_5$ there is $N_1 = 1$ Weyl fermion in the symmetric representation, and thus there are no anomalies involving SU$(N_1$). The solution that matches the complete set of 't Hooft anomalies is that with $n_1 = 1$, $n_2 = 2$ and $n_3 = 2(N-2)$, for which $\txn{SU}(N_2 = 3)$ is completely broken, and thus there are no anomalies involving it. There are $\mathcal{N}_A = (2N-3)(N-2)$ massless composite fermions given by 

\begin{align}
\mathcal{B}_A^{a,b} = \eta^a_{i} \lambda^{ij} \eta^b_{j}
\spa\spa,\spa\spa
a,b = 1, \esp... \esp , \esp 2(N-2)
\spa\spa,
\label{sol_R5_large_N}
\end{align}

with the IR symmetry group 

\begin{align}
G_{\txn{IR}} =
\txn{SU}(2N - 4) \times \txn{U}_\xi(1)
\times 
\mathbbm{Z}^\chi_{4(2N - 4)/\me_\chi}
\spa\spa.
\label{R5_solution_symm_group}
\end{align}

This is a valid candidate for the low-energy behavior of the $R_5$ theory. The $\psi$ fermions are confined in some set of baryons different from $\mathcal{B}_S$, and those happen to be heavy. Here $\me_\chi = 2,4,8$ for $N= \txn{odd}, \esp 4n+2, \esp 4n$ respectively with $n$ an integer.\\

$\bullet$ \textbf{The solution for} $R_{17}$\\

For $R_{17}$ there is $N_2 = 1$ Weyl fermion in the anti-symmetric representation, and thus there are no anomalies involving SU$(N_2$). The solution that matches the complete set of 't Hooft anomalies is that with $n_1 = 2$, $n_2 = 1$ and $n_3 = 2(N+2)$, for which $\txn{SU}(N_1 = 3)$ is completely broken, and thus there are no anomalies involving it. There are $\mathcal{N}_S = (2N+3)(N+2)$ massless composite fermions given by 
\begin{align}
\mathcal{B}_S^{a,b} = \eta^a_{i} \psi^{ij} \eta^b_{j}
\spa\spa,\spa\spa
a,b = 1, \esp..., \esp 2(N+2)
\spa\spa,
\label{sol_R17_large_N}
\end{align}

with the IR symmetry group 
\begin{align}
G_{\txn{IR}} =
\txn{SU}(2N + 4) \times \txn{U}_\xi(1)
\times 
\mathbbm{Z}^\chi_{4(2N + 4)/\me_\chi}
\spa\spa.
\label{R17_solution_symm_group}
\end{align}

This is a valid candidate for the low-energy behavior of the $R_{17}$ theory. The $\lambda$ fermions are confined in some set of baryons different from $\mathcal{B}_A$, and those happen to be heavy. Here $\me_\chi = 2,4,8$ for $N= \txn{odd}, \esp 4n+2, \esp 4n$ respectively with $n$ an integer.\\

For both $R_5$ and $R_{17}$, part of the $\mathbbm{Z}^\chi_{q^\chi_\mathcal{C}}$ is redundant in general, and furthermore it is completely redundant when considering the faithful symmetry restricted to the massless field content in the IR.\\

For $R_5$ the $\psi$ fermions are confined in massive composite states different from $\mathcal{B}_S = \eta \psi \eta $. Similarly, for $R_{17}$ the $\lambda$ fermions are confined in massive composite states different from $\mathcal{B}_A = \eta \lambda \eta $. As discussed around equation (\tcolor{\ref{large_N_yukawa_operator}}), it might be that both baryons $\mathcal{B}_S$ and $\mathcal{B}_A$ remain massless in the large $N$ limit. It seems like these arguments are in conflict, but this is not necessarily the case. It could be that $\mathcal{B}_S$ is not part of the spectrum (whether or not it gets a mass). The massive composite states formed by $\psi$ or $\lambda$ should be different from $\mathcal{B}_S$ or $\mathcal{B}_A$.\\

Finally, we remind the reader that, while the solutions presented here for both $R_5$ and $R_{17}$ were obtained by following the suggestions inferred from a large $N$ analysis, the validity of these solutions does not rely on the large $N$ limit itself. The solutions match the complete set of 't Hooft anomalies independently of the large $N$ limit, and thus are viable candidates for the low-energy description of these theories.

\subsubsection{$N_2 < 0$}
\label{large_N_negative_N2_and_tumbling}

The simplest of the models is the $R_7$ theory in Table \tcolor{\ref{table_N1_N2_N3_values}} with a single Weyl fermion in the $\overline{\textbf{A}}_2$ representation and $N_3 = 8$ in the anti-fundamental. The large $N$ limit of this model was studied in \tcolor{\cite{Eichten:Peccei:Preskill:Zeppenfeld:1986}}. Assuming that the gauge group confines, the authors of this study argue that the large $N$ limit yields a theory with an infinite number of non-interacting zero-width mesons and glueballs, with no baryons at all, similar to the large $N$ limit of QCD\hlfootnote{Arguing that baryons decouple in the large $N$ limit assumes that chiral symmetry breaking takes place. Then baryon masses scale as $N \Lambda$ with $\Lambda$ the confinement scale. An unbroken chiral symmetry can protect baryons, keeping them light.}. This result is blind to the number of flavors $N_i$.\\

For models with $N_3 > 0$ meaning the representation for these fermions is the anti-fundamental, we note that the following two features hold in general. As shown in Table \tcolor{\ref{appendix_table_MAC_2_S_A*_F*_formulas}}, in the large $N$ limit the colored condensates $\langle \psi \lambda \rangle$ in the adjoint and $\langle \psi \eta \rangle$ in the fundamental become as attractive as possible, signaling a possible breaking of the gauge group, at least as suggested by the MAC hypothesis \tcolor{\cite{Raby:Dimopoulos:Susskind:1980:tumbling}}\hlfootnote{Note however that this is also true for the colored condensates $\langle \psi \lambda \rangle$ and $\langle \psi \eta \rangle$ that can be formed in models where the $N_2$ Weyl fermions transform in $\textbf{A}_2$, see Table \tcolor{\ref{appendix_table_MAC_1_S_A_F*_formulas}}.}. The second feature is that, in the large $N$ limit there are no candidate bosonic operators that could condense and break the $\txn{U}(1)_{\xi}$ symmetry.\\

In what follows we will work in the $\widetilde{q}^{\alpha,\beta}_i$ basis. For $R_7$ there is no $\txn{U}(1)_\alpha^3$ anomaly, but there is a $\txn{U}(1)_\alpha \times \txn{Grav}^2$ anomaly given by $-6N$. Since there is no candidate order parameter to break the $\txn{U}(1)_{\alpha}$ symmetry, the $\txn{U}(1)_\alpha \times \txn{Grav}^2$ anomaly should be matched in the IR. However, one can check that all\hlfootnote{Recall that in the large $N$ limit we can neglect operators with $\eps$ tensor contractions.} gauge singlet states in the large $N$ limit carry zero $\txn{U}(1)_{\alpha}$ charge, so the $\txn{U}(1)_\alpha \times \txn{Grav}^2$ anomaly cannot be matched in the IR, and thus it should be broken. As discussed in \tcolor{\cite{Eichten:Peccei:Preskill:Zeppenfeld:1986}}, a possible solution to this contradiction is to abandon the assumption of color confinement, and furthermore, it is possible that the gauge symmetry could break as suggested by the MAC hypothesis. In fact, the authors of \tcolor{\cite{Goity:Peccei:Zeppenfeld:1985}} found a solution that matches the zero-form 't Hooft anomalies for a scenario where the gauge symmetry is broken in a tumbling fashion. See Section \tcolor{\ref{section_massless_general_commposite_operators}} for a brief review on the relevant composite operators used to construct this solution.\\  

For all other models both the $\txn{U}(1)_\alpha^3$ and $\txn{U}(1)_\alpha \times \txn{Grav}^2$ anomalies do not vanish. However, the $\txn{U}(1)_\alpha \times \txn{Grav}^2$ anomaly is not an independent constraint: it is a multiple of the $\txn{U}(1)_\alpha^3$ anomaly, and matching one automatically matches the other. In any case, since there is no candidate order parameter to break the $\txn{U}(1)_{\alpha}$ symmetry, the $\txn{U}(1)_\alpha \times \txn{Grav}^2$ anomaly should be matched in the IR. Consider a general composite operator 
\begin{align}
\mathcal{O} \sim 
\psi^{(\ell)} \lambda^{(r)} \eta^{(m)}
\spa\spa,
\end{align}

where the powers $\ell, r$ and $m$ denote UV fermion insertions. This operator is a gauge singlet only if $2\ell - 2r - m =0$. The charges for this operator are 

\begin{align}
\widetilde{q}^{\alpha,\beta}_\mathcal{O} = 
\ell q^{\alpha,\beta}_1 - r q^{\alpha,\beta}_2 + m q^{\alpha,\beta}_3
\spa\spa.
\end{align}

If we take $m = 2(\ell - r)$ enforcing the gauge singlet condition then 
\begin{align}
\widetilde{q}^{\alpha}_\mathcal{O} 
=
- N (\ell - r) (N_1 + N_2) 
\spa\spa,
\end{align}

which vanishes only for $r = \ell$ or the $R_7$ case in which $N_1 = 1$  and $N_2 = -1$. Gauge singlet states with $r = \ell$ have a total of $2\ell$ fermion insertions and thus are bosonic operators. We can say that gauge singlet bosonic operators with $r = \ell$ have charge $\widetilde{q}^{\alpha}_\mathcal{O} = 0$. However, in general baryons can have a non-zero $\widetilde{q}^{\alpha}_\mathcal{O}$ charge and thus the anomalies involving $\txn{U}(1)_\alpha$ could be matched by composite fermions in the IR. Equivalently, a composite is a fermionic operator only if $\ell + r + m = k$ for $k$ an odd number. For a gauge singlet we can write $3\ell -r = k$ and the $\txn{U}(1)_\alpha$ charge reads $\widetilde{q}^{\alpha}_\mathcal{O} \sim -2\ell + k$, which vanishes only for $2\ell = k$, which leads to a contradiction. Composite fermionic states can carry a non-zero $\widetilde{q}^{\alpha}_\mathcal{O}$ charge. By following the same line of reasoning we can see that gauge singlet operators also carry in general a non-zero  $\widetilde{q}^{\beta}_\mathcal{O}$ charge. There is no contradiction as in the $R_7$ case that needs to be solved, and abandoning the assumption of color confinement is not needed.\\ 

If a tumbling scenario such as the one discussed in \tcolor{\cite{Goity:Peccei:Zeppenfeld:1985}} could take place for the rest of this class of models with $N_3 > 0$, demanding that the final result of the tumbling process yields a global symmetry as large as possible, the gauge symmetry could be broken by MAC in the fundamental given by 
\begin{align}
\langle \psi^{ij, a} \eta_j^b \rangle
= \Lambda 
\delta^{ i \esp,\esp b - (N_3 - N)}
\spa\spa.
\end{align}

For large $N$ the condensate in the adjoint

\begin{align}
\langle \psi^{i k , a} \lambda_{kj}^b \rangle
= \Lambda ( \delta^i_j - N \delta^i_N \delta^N_j)
= \Lambda \esp \txn{diag}(1,...,1,1-N)
\spa\spa,
\label{MAC_adjoint_VEV_expresion_N2_negative}
\end{align}
is as attractive as the one in the fundamental, and thus it could also contribute in a non-trivial way to the tumbling processes. Note that the set of massless composites that match the anomalies for the $R_7$ theory derived in \cite{Goity:Peccei:Zeppenfeld:1985} makes use of complementarity \tcolor{\cite{Fradkin:Shenker:1979,Elitzur:1975,Banks:Rabinovici:1979,Osterwalder:Seiler:1978}}, where it is assumed that the composite that condenses is in the fundamental. If a composite in the adjoint breaks the gauge symmetry the resulting dynamics could be different.\\

If a tumbling scenario such as the one discussed in \tcolor{\cite{Goity:Peccei:Zeppenfeld:1985}} could take place for models where $N_3 < 0$ meaning that the $\eta$ fermions transform in the fundamental, as shown in Table \tcolor{\ref{appendix_table_MAC_3_S_A*_F_formulas}} and Figure \tcolor{\ref{figure_MAC_plot_3_2_fermion_op_S_Acc_F}}, a global symmetry as large as possible could be realized after the gauge group is broken by the MAC in the adjoint as in (\ref{MAC_adjoint_VEV_expresion_N2_negative}). For large $N$ the anti-fundamental 

\begin{align}
\langle \lambda_{ij}^a \eta^{j,b} \rangle
= \Lambda \delta_i^{b - (N_3 - N)}
\spa\spa,
\end{align}

is as attractive as the one in the adjoint, so it could also contribute in a non-trivial way to the tumbling processes. Note that for SU$(5)$ the MAC is $\langle \lambda_{ij}\lambda_{k\ell} \rangle$ in the $\overline{\textbf{A}}_4$ representation.\\

We conclude this section with the following observation for the two-species models where $N_3 = 0$ with $N$ fixed (see Appendix \tcolor{\ref{appendix_4d_two_species_models}}). As shown in Appendix \tcolor{\ref{MAC_appendix_two_species_models}}, the colored fermion bilinear with the MAC is $\langle \psi \lambda \rangle$ in the adjoint. The approach followed in \tcolor{\cite{Goity:Peccei:Zeppenfeld:1985}} relies on complementarity \tcolor{\cite{Fradkin:Shenker:1979,Elitzur:1975,Banks:Rabinovici:1979,Osterwalder:Seiler:1978}}, which works when the colored composite responsible for the breaking of the gauge group transforms in the fundamental. In this sense, it could be that the strategy needed to find a solution (if it exists) that matches the anomaly constraints for the two-species models is different from the one presented in \tcolor{\cite{Goity:Peccei:Zeppenfeld:1985}}.

%% file: 08_anomaly_matching.tex
\setlength{\jot}{13pt}

\section{Matching the anomalies of fractional fluxes}\label{section_anomaly_matching}

We now discuss the matching of mixed anomalies involving generalized symmetries. Here we follow the description outlined in Section \tcolor{\ref{section_twisted_gauge_field}}, where anomalies are computed by using twisted configurations of one-form gauge fields on a four-torus.\\

For $R_3$, as suggested by the large $N$ limit, the massless baryons in (\tcolor{\ref{R3_large_N_baryons}}) with symmetry group in (\tcolor{\ref{R3_large_N_symm_group}}) match the zero-form 't Hooft anomalies. Similarly, the baryons in (\tcolor{\ref{sol_R5_large_N}}) for $R_5$ and those in (\tcolor{\ref{sol_R17_large_N}}) for $R_{17}$ match the ordinary triangle-diagram zero-form anomalies. Therefore, it might be interesting to test whether they also match anomalies associated with fractional fluxes. In this section we show that along the unbroken directions,  matching these anomalies follows from zero-form matching together with a condition on the embedding of fluxes, and that the candidate IR descriptions of these theories match the anomalies.\\

Let us discuss equations (\tcolor{\ref{CFU_anomalies_twisted_fields}}) and (\tcolor{\ref{partition_function_shift_twisted_fluxes}}) more
closely. What these tell us is that $\mathcal{I}^{(s)}_\delta$ is a sum of topological charges with coefficients given by ordinary zero-form anomaly coefficients. The only difference between it and an ordinary zero-form anomaly is that the generalized anomalies require correlated fluxes and allow fractional values. If the individual ordinary zero-form anomalies match term by term in (\tcolor{\ref{CFU_anomalies_twisted_fields}}), then it is not relevant whether the fluxes are integral or fractional, and matching of the
generalized anomalies is almost guaranteed:\\

\begin{enumerate}[label=\textbf{\Roman*}]
    \item All that can go wrong is a failure of the fluxes to match.\\

    \item The fluxes match if every unbroken global symmetry of the UV appears in the IR with the same faithful quotient. Here the UV quotient is defined by the action on gauge invariant operators rather than on the fields, so that a global transformation that can be undone by an element of the gauged center already counts as redundant in the UV.\\

    \item \label{item_3} Constructing the massless IR fields from explicit interpolating operators made of the elementary UV fields guarantees that no UV flux is lost in the IR. This is not guaranteed for abstractly specified spectra. Also, it can happen that there are unbroken transformations $h$ that act trivially on every massless composite fermion but nontrivially in the UV theory. Then $h$ is a new contribution to the denominator of the IR faithful symmetry, and it can be used to couple to twisted fluxes. \\
\end{enumerate}

Let us discuss both prongs of this last point in more detail. Suppose we have a UV theory and a proposed IR scenario in which the massless DOF are all Goldstones or composite fermions. First, to see that the IR cannot have fewer fluxes associated with the unbroken symmetry group than the UV, note that an element of the UV denominator $\Gamma_{\txn{UV}}$ acts trivially on
every UV local operator, and therefore also on every IR operator built explicitly from them. Discrete quotients cannot be \say{un-gauged} on the local operators. Thus the linearly realized (i.e. unbroken) UV faithful global symmetry descends into the IR, and every flux associated with UV numerator factors that are either gauged and confining in the IR, or global and not spontaneously broken, can be realized by an IR flux simply by deleting the gauged-confining part. \\

Second, suppose $h$ is a nontrivial element of the UV symmetry that acts trivially on all massless DOF of the IR theory. (E.g. suppose $h$ transforms some gauge invariant operators that are not interpolating operators for the massless fields.) If $h$ is spontaneously broken, it does not appear in the cover of the unbroken faithful IR symmetry. We can couple the UV theory to the corresponding twisted fluxes and there is an anomaly matching constraint, but the sectors are defect sectors, outside the anomalies matched by the massless fermions in the IR. Anomaly matching becomes an exercise in assigning charges to defects. If $h$ is unbroken, however, then it is a redundancy of the entire massless theory and appears directly as a denominator factor of the IR faithful symmetry. In this case, the candidate IR theory has extra fluxes which do not embed into the center-locked fluxes of the UV theory and the UV theory has an extra unbroken faithful global symmetry. \\

However, this does not lead to a new condition either, which can be seen as follows. Let $\widetilde{G}_{\txn{IR}}$ be the cover of the unbroken symmetry, some product of $\txn{SU}(N_i)$ and $\txn{U}(1)$ factors. The zero-form 't Hooft anomalies of $\widetilde{G}_{\txn{IR}}$ are computed by the triangle diagrams which we assume  have already been matched between the UV  and the candidate IR theories. 
There is also a $\txn{U}(1)$ subgroup of $\widetilde{G}_{\txn{IR}}$ that can be parametrized as $g(\theta) = \eu{i\theta T}$, $0 \leq \theta \leq 2\pi$, such that $g(2\pi)=h$ acts trivially on the IR massless fermions but nontrivially on the UV theory. On a twisted background $s$ that is allowed in both the UV and IR theories, $g(\theta)$ induces a phase in the partition function

\begin{align}
\mathcal{Z} \longrightarrow \eu{i\theta \mathcal{A}_T}\,\mathcal{Z}
\label{path_anomaly}
\spa\spa.
\end{align}

When the zero-form anomalies match, $\mathcal{A}_T$ takes the same value in the UV and in the IR in every flux sector that is common to both. Taking $\theta \rightarrow 2\pi$, in the IR, $g(2\pi) = h$ acts as the identity on the massless fermions, so the phase in (\ref{path_anomaly}) is trivial. In the UV, $g(2\pi) = h$ acts nontrivially, but the zero-form anomaly matching guarantees that the UV partition function is also invariant. Thus matching the ordinary anomalies of $\widetilde{G}_{\txn{IR}}$ already implies that $h$ is anomaly-free in the UV and there is no obstruction to ``gauging it" (in the sense of having it appear in the denominator of the faithful symmetry group) in the IR.\\ 

It is also the case that the IR contains additional $h$-twisted sectors. However, these are not smooth backgrounds for the UV theory, so we do not obtain new anomaly constraints from them. Roughly speaking, the UV needs a defect surface in order to support these backgrounds, and anomaly matching can manifest again as inflow onto the surface.\\

If the massless states of the IR theory are specified more abstractly (not built from interpolating operators of the UV), then the only requirement we must impose is
\begin{align}
\Gamma^{\txn{inv}}_{\txn{UV}} \subseteq \Gamma_{\txn{IR}}
\spa\spa,
\label{flux_embedding_condition}
\end{align}

where $\Gamma^{\txn{inv}}_{\txn{UV}}$ is the set of elements of $\widetilde{G}_{\txn{IR}}$ that act trivially on the UV gauge invariants and $\Gamma_{\txn{IR}}$ the set that acts trivially on the massless IR fermions. This is a condition on the charges and $n$-alities of the massless fermions. Given (\ref{flux_embedding_condition}) and matching of the zero-form anomalies, the twisted-sector anomalies match, and no further computation is required.\hlfootnote{The argument assumes that $\widetilde{G}_{\txn{IR}}$ is connected, which holds for $N_3 \neq 0$, and that the Goldstone bosons are neutral under $h$, as in the cases below.}\\

To summarize, the new information (or lack thereof) that can be obtained from consideration of the fractional-flux mixed anomalies (beyond the information in ordinary zero-form anomalies) is as follows:\\

\begin{enumerate}[label=\textbf{\Alph*}]

    \item \label{item_A} On unbroken symmetries, the only condition beyond zero-form matching is~(\ref{flux_embedding_condition}). If  $\Gamma_{\rm IR}$ is strictly larger than $\Gamma^{\rm inv}_{\rm UV}$ there is no new condition.  How much of this condition is automatic depends on how the candidate IR theory is described. For abstractly specified spectra, with charges adjusted to match triangle diagrams but no operator realization, the condition is an additional arithmetic constraint on the charges of the massless fermions.\\

    \item \label{item_B} An unbroken discrete remnant of a broken factor, for example a $\mathbbm{Z}_k \subset \txn{U}(1)$ preserved by charge-$k$ condensates, appears in the IR as a separate discrete factor. These require additional ordinary discrete anomaly matching.\\

    \item \label{item_C} For backgrounds requiring flux along broken directions, the anomaly is not matched by the massless spectrum. It is deposited on defects by inflow, and it determines their quantum numbers.\\

    \item \label{item_D} Other directions of interest, outside our scope, are Witten-type global anomalies (see Section \tcolor{\ref{section_witten_anomalies}}), and anomalies of a genuine one-form symmetry (not present for the class of theories we consider unless $N_3=0$).
\end{enumerate}

\subsection{The case of $R_3$}
\label{higher_form_matching_R3}

Here we work in the $\widetilde{q}^{\alpha,\beta}_i$ basis. For $R_3$ the faithful symmetry group in the UV is
\begin{align}
G_{\txn{UV}}
=
\txn{SU}(N) \times 
\frac{
\txn{SU}(2N) \times 
\txn{U}_{\widetilde{\alpha}}(1)
\times
\txn{U}_{\widetilde{\beta}}(1)
}
{
\mathbbm{Z}_N \times \mathbbm{Z}_{2N/\me_\gamma}
\times \mathbbm{Z}_{\me_u}
}
\spa\spa.
\end{align}

We write $w_c \in \mathbbm{Z}_N$ for the color flux and $w_3$ for the flux of the $\txn{SU}(2N)$ flavor group. For integers $p$, $q$ and $r$, the
cocycle conditions (\tcolor{\ref{consistency_conditions_global_symm_section}}) read

\setlength{\jot}{5pt}
\begin{align}
\psi &\esp:\esp  
\frac{2 w_c}{N} + \phi_{\widetilde{\alpha}} 
+ (N-2) \phi_{\widetilde{\beta}}
\equiv p 
\spa\spa, 
\\
\lambda &\esp:\esp  
\frac{2 w_c}{N} + \phi_{\widetilde{\alpha}} 
- (N+2) \phi_{\widetilde{\beta}}
\equiv q
\spa\spa, 
\\
\eta &\esp:\esp  
- \frac{w_c}{N} + \frac{ w_3}{2N}
- \phi_{\widetilde{\alpha}}  
\equiv r
\spa\spa.
\end{align}
\setlength{\jot}{13pt}

In the IR there is a discrete putative symmetry given by a subgroup of the broken $\txn{U}_{\widetilde{\beta}}(1)$
\begin{align}
\mathbbm{Z}_{2N/\me_{\widetilde{\beta}}} 
\spa\spa,\spa\spa
\me_{\widetilde{\beta}}
= 
\txn{gcd}\big({2 N_1 T_1 , 2 N_2 T_2}\big)
\spa\spa.
\end{align}

Given that both baryons $\mathcal{B}_S$ and $\mathcal{B}_A$ have the same $\widetilde{\alpha}$ charge and the same flavor $n$-ality, $\mathbbm{Z}_{2N/\me_{\widetilde{\beta}}}$ can be trivially absorbed. The infrared faithful group is
\begin{align}
G_{\txn{IR}}
=
\frac{
\txn{SU}(N) \times  \txn{U}_{\widetilde{\alpha}}(1)
}
{ \mathbbm{Z}_N }
\spa\spa.
\label{IR_R3_group_faithful}
\end{align} 

To be an allowed bundle, a background of $G_{\txn{IR}}$ with $\txn{SU}(N)$ flux class $w_{\txn{IR}}$ must be
accompanied by the compensating fractional $\txn{U}_{\widetilde{\alpha}}(1)$ flux, as follows from the cocycle conditions
\begin{align}
\frac{1}{2\pi} \bigintsss_\Sigma F_{\widetilde{\alpha}}
=
\phi_{\widetilde{\alpha}} = \frac{ 2 w_{\txn{IR}} }{N}
\esp (\txn{mod } 1)
\spa\spa.
\label{fluxes_IR_R_3}
\end{align}

\subsubsection{Embedding IR fluxes in the UV}

We need to embed the IR symmetries and their nontrivial bundles into the UV description. Inside $\txn{SU}(2N)$, the generator of the diagonal $\txn{SU}(N)$ center is $z_N = e^{2\pi i /N } \mathbbm{1}_N$ acting in each of the two $N$-dimensional blocks, i.e. it is $e^{2\pi i /N } \mathbbm{1}_{2N}$ on the fundamental of $\txn{SU}(2N)$. This is
the same as $z^2_{2N}$. Therefore, a diagonal flavor flux of class $w_{\txn{IR}}$ embeds as the $\txn{SU}(2N)$ center class $w_3$
\begin{align}
w_3 = 2 w_{\txn{IR}} 
\spa (\txn{mod } 2N)
\spa\spa.
\label{fluxes_2_IR_R_3}
\end{align}

Here $w_3 \in  \mathbbm{Z}_{N_3}$ is constrained to be a multiple of $\me_\gamma$. To match the rest of the fluxes, set $\phi_{\widetilde{\beta}} = 0$ in the UV and identify $\phi_{\widetilde{\alpha}}^{\txn{UV}} = \phi_{\widetilde{\alpha}}^{\txn{IR}} = 2 w_{\txn{IR}}/ N$; accumulating everything
\setlength{\jot}{10pt}
\begin{align*}
w_c &= - w_{\txn{IR}}
\spa (\txn{mod } N)
\spa\spa, 
\\
w_3 &=2 w_{\txn{IR}} 
\spa (\txn{mod } 2N)
\spa\spa, 
\\
\phi_{\widetilde{\beta}} &= 0
\spa\spa, 
\\
\phi_{\widetilde{\alpha}}^{\txn{IR}} &= \frac{2 w_\txn{IR}}{N} \spa \txn{(mod 1)}
\spa\spa.
\numberthis
\label{R3_fluxes_embedding}
\end{align*}
\setlength{\jot}{13pt}

\subsubsection{Comparing the faithful quotients: Odd N}

Let us check (\ref{flux_embedding_condition}) explicitly at odd $N$. Fix the cover of the unbroken $\widetilde{G}_{\txn{IR}} = \txn{SU}(N) \times \txn{U}_{\widetilde{\alpha}}(1)$. An element $(z_N^k, e^{i\theta})$ acts on the UV fields by $e^{i\theta}$ on $\psi$ and $\lambda$ and by $e^{2\pi i k /N} e^{-i\theta}$ on $\eta$, so this trivially forces $\theta \in 2\pi \mathbbm{Z}$ and then $k \equiv 0$. The ultraviolet matter therefore realizes $\widetilde{G}_{\txn{IR}}$ faithfully, i.e. $\Gamma_{\txn{UV}} = \mathbbm{1}$. In particular, restricted to the unbroken directions the ultraviolet admits no fractional fluxes at all; setting $w_c = 0$ and $\phi_{\widetilde{{\beta}}} = 0$ forces $\phi_{\widetilde{\alpha}} \in \mathbbm{Z}$ and $w_3 \equiv 0$. The same element acts on the massless baryons by $e^{4\pi i k /N} e^{-i\theta}$, and doesn’t act on the Goldstones, so triviality implies $\theta \equiv 4\pi k / N$ and $\Gamma_{\txn{IR}} = \mathbbm{Z}_N$.\\

Let’s compare what this means for the twisted fluxes. First, $\widetilde{G}_{\txn{IR}}$ admits the fractional pairs in (\tcolor{\ref{fluxes_IR_R_3}}). The UV flavor symmetry acting on the matter fields can’t form these, but the gauged center supplies them. Checking if a $(z_N^k, e^{i\theta})$ element can be undone by a color center element $z_c^m$, triviality on $\psi$ and $\lambda$ fixes $\theta \equiv - 4\pi m/ N $, and triviality on $\eta$ then forces $m \equiv -k$. Therefore on gauge invariants the UV faithful quotient is $\Gamma_{\txn{UV}}^{\txn{inv}} = \mathbbm{Z}_N$. This is the same faithful quotient as in the IR. The result $m \equiv -k$ is equivalent to the flux relation $w_c = - w_{\txn{IR}}$, so all $N$ classes of (\tcolor{\ref{fluxes_IR_R_3}}) are reached.\\

Note that there is a twist the baryons do not detect. Since they have $n$-ality two, they are blind to $z^N_{2N} = - \mathbbm{1}_{2N}$, while the UV field $\eta$ flips sign under this transformation. Equivalently, the embedding (\tcolor{\ref{fluxes_2_IR_R_3}}) reaches only the even classes $w_3 = 2 w_{\txn{IR}}$. No color compensator exists, since $e^{2\pi i m/N} = -1$ has no solution at odd $N$. The theory handles it by spontaneously breaking the $- \mathbbm{1}_{2N}$, so these backgrounds fall under item \ref{item_C}. In the vacuum $z^N_{2N} = e^{i\pi \widetilde{\alpha} } e^{i\pi \widetilde{\beta} }$, i.e. an unbroken phase times
a broken one, and the broken factor shifts the $\widetilde{\beta}$-axion. This is why the twist doesn’t appear in the denominators of $G_{\txn{IR}}$ or of the full massless theory. Its twists do not stabilize the vacuum, the state contains topological defects, and the anomaly is pushed onto them by inflow. In summary, the U$_{\alpha}(1)$ symmetry has a $\mathbbm{Z}_{2}$ mixed anomaly with the minimal
flavor-center flux, so we learn something about the $\alpha$ charges of center vortices from this anomaly. \\

Third, we can ask whether the vacuum preserves a discrete remnant of the broken factors, the class of item \tcolor{\ref{item_B}}. Since the IR proposal in (\tcolor{\ref{IR_R3_group_faithful}}) specifies the unbroken symmetry completely, the answer is no.

\subsubsection{Explicit matching of the smooth sectors}

We now verify the matching explicitly in the smooth sectors. This is guaranteed on general grounds by the arguments above, but we can still check it. The matched quantity is the difference of $\txn{U}_{\widetilde{\alpha}}$(1) indices 

\begin{align}
\Delta_{\widetilde{\alpha}} 
=
\mathcal{I}_{\widetilde{\alpha}}^{\txn{IR}}(w_c)
-
\mathcal{I}_{\widetilde{\alpha}}^{\txn{UV}}(w_c)
\spa\spa.
\end{align}
There are no color fluxes in the IR, but they are needed in the UV in order to place the theory in the same consistent flavor bundle. There is no color contribution to the index since there is no ABJ anomaly, and thus we write 

\begin{align}
\mathcal{I}_{\widetilde{\alpha}}
=
\mathcal{A}_{{\widetilde{\alpha}} \times \txn{SU}(N)^2} 
\esp \mathcal{Q}_3 
+
\mathcal{A}_{{\widetilde{\alpha}}^3} 
\esp \phi^2_{\widetilde{\alpha}}
\spa\spa.
\label{HF_match_R3_index}
\end{align}

Here $\mathcal{Q}_3  \propto w_3^2$ is the flavor topological charge, with $\mathcal{A}_{{\widetilde{\alpha}} \times \txn{SU}(N)^2} $ and $\mathcal{A}_{{\widetilde{\alpha}}^3} $ the ordinary zero-form anomaly coefficients for $\txn{U}_{\widetilde{\alpha}}(1) \times \txn{SU}(N)^2$ and $\txn{U}_{\widetilde{\alpha}}(1)^3$. Equation (\tcolor{\ref{HF_match_R3_index}}) holds separately in the UV
and IR. Recall also that both zero-form anomaly coefficients are matched, so $\Delta_{\widetilde{\alpha}} = 0 $ and the fractional-flux mixed anomaly also matches.\\

The fractional quotient data enter only through which pairs of $\mathcal{Q}_3, \phi_{\widetilde{\alpha}}$ are allowed, and the previous subsection established that the UV and the IR admit exactly the same set:
the faithful quotients agree, and all $N$ classes are realized on both sides. The fractional-flux sectors associated with unbroken symmetries therefore impose no condition on the vacuum in (\tcolor{\ref{IR_R3_group_faithful}}) beyond ordinary anomaly matching.

\subsubsection{The $q^{\alpha,\beta}_i$ basis}

To match the anomalies in the $(\alpha,\beta)$ basis mainly used throughout this manuscript, consider the change of basis given in equation (\tcolor{\ref{U(1)_change_of_basis}}). The indices in this basis are 

\begin{align}
\mathcal{I}_{\alpha} 
= 
(N+2) \mathcal{I}_{\widetilde{\alpha}}
+ \mathcal{I}_{\widetilde{\beta}}
\spa\spa,\spa\spa
\mathcal{I}_{\beta} 
= 
(N-2) \mathcal{I}_{\widetilde{\alpha}}
- \mathcal{I}_{\widetilde{\beta}}
\spa\spa.
\end{align}

The baryon composites are taken ${\widetilde{\beta}}$-neutral (a field redefinition by the ${\widetilde{\beta}}$-axion removes their constituent ${\widetilde{\beta}}$-charges), so they inherit the $(\alpha,\beta)$ charges of $\eta$, namely $-(N-2)$ and $-(N+2)$.\\

With $\phi_{\widetilde{\beta}} = 0$ in the UV the ${\widetilde{\beta}}$ index is 
\begin{align}
\mathcal{I}_{\widetilde{\beta}}
= \mathcal{A}_{{\widetilde{\beta}} \times {\widetilde{\alpha}^2}} 
\esp \phi^2_{{\widetilde{\alpha}}}
\spa\spa,\spa\spa 
\mathcal{A}_{{\widetilde{\beta}} \times {\widetilde{\alpha}^2}} 
=
- N^2 
\spa\spa.
\end{align}

The baryons do not contribute to the infrared $\mathcal{I}_{\widetilde{\beta}}$, and thus $\Delta_{\widetilde{\beta}} = - N^2 \phi^2_{{\widetilde{\alpha}}} $. Then in the $(\alpha,\beta)$ basis 

\begin{align}
\Delta_\alpha = \Delta_{\widetilde{\beta}} 
= -  N^2 \phi^2_{{\widetilde{\alpha}}}
\spa\spa,\spa\spa
\Delta_\beta =-  \Delta_{\widetilde{\beta}} 
= N^2 \phi^2_{{\widetilde{\alpha}}}
\spa\spa.
\end{align}

The combination that the composite fermions match on their own is

\begin{align}
\Delta_\alpha + \Delta_\beta = 2N \Delta_{\widetilde{\alpha}} = 0
\spa\spa.
\end{align}

The orthogonal piece is $\Delta_{\widetilde{\beta}}$, the anomaly of the broken $\txn{U}_{\widetilde{\beta}}(1)$ which is saturated by the $\widetilde{\beta}$-axion rather than by the composites.

\subsubsection{Even $N$}

Consider now the case $N = 4n,4n+2$, for which $\me_\gamma = 2,4$ respectively. Formally, the embedding relations between fluxes are the same as those in (\ref{R3_fluxes_embedding}). However, the allowed classes that follow from $w_3 = 2 w_{\txn{IR}}$ are slightly different from those in the odd $N$ case.\\

The baryons are still blind to the twist $h = z^N_{2N}$ for all $N$. However, once the color center is accounted for, in this case the UV matter is also blind to $h$. The twist $h$ can be screened by $z_c^{N/2}$. There is no blind twist to match as in the odd $N$ case where $h$ is broken spontaneously. Now $-\mathbbm{1}_{2N}$ sits inside $\Gamma_{\txn{IR}}$ as $(z_N^{N/2},1)$, an ordinary redundancy present on both sides. No spontaneous breaking of a center element is required, no defect sectors enter, and the matching is satisfied by the massless spectrum alone. For the even $N$ classes fractional-flux anomaly matching provides no new information beyond that learned from ordinary zero-form 't Hooft anomaly matching.

\subsection{The case of $R_5$}

In what follows we work in the $\widetilde{q}^{\alpha,\beta}_i$ basis. For the theory $R_5$ where $N_3 = 4N-8$ the faithful symmetry group in the UV is

\begin{align}
G_{\txn{UV}}
=
\txn{SU}(N) \times 
\frac{
\txn{SU}(3) \times \txn{SU}(N_3) \times 
\txn{U}_{\widetilde{\alpha}}(1)
\times
\txn{U}_{\widetilde{\beta}}(1)
}
{
\mathbbm{Z}_N \times \mathbbm{Z}_3 \times \mathbbm{Z}_{N_3/\me_\gamma}
\times \mathbbm{Z}_{\me_u}
}
\spa\spa.
\end{align}

We write $w_c \in \mathbbm{Z}_N$ for the color flux and $w_i$ for the flux of the $\txn{SU}(N_i)$ flavor group. For integers $p$, $q$ and $r$, the cocycle conditions are

\setlength{\jot}{10pt}
\begin{align}
\psi &\esp:\esp  
\frac{2 w_c}{N} + (N-2)\phi_{\widetilde{\alpha}} 
+ \widetilde{q}^\beta_1 \phi_{\widetilde{\beta}}
\equiv p 
\spa\spa, 
\\
\lambda &\esp:\esp  
\frac{2 w_c}{N} + (N-2) \phi_{\widetilde{\alpha}} 
+ \widetilde{q}^\beta_2 \phi_{\widetilde{\beta}} + \frac{w_2}{3}
\equiv q
\spa\spa, 
\\
\eta &\esp:\esp  
- \frac{w_c}{N} -(N-1) \phi_{\widetilde{\alpha}}
+ \widetilde{q}^\beta_3 \phi_{\widetilde{\beta}}
+ \frac{w_3}{N_3}
\equiv r
\spa\spa.
\end{align}
\setlength{\jot}{13pt}

Here $w_3 \in \mathbbm{Z}_{N_3}$ is restricted to be a multiple of $\me_\gamma$. Since there is a single baryon operator $\mathcal{B}_A$ with charge $\widetilde{q}^\alpha_A = -N$, the discrete putative symmetry given by a subgroup of the broken $\txn{U}_{\widetilde{\beta}}(1)$ is redundant. Now we compare the faithful quotients. First fix the cover of the unbroken $\widetilde{G}_{\txn{IR}} = \txn{SU}(M) \times \txn{U}_{\widetilde{\alpha}}(1)$. An element $(z_M^k, e^{i\theta })$ acts on the UV fields by $e^{i\theta (N-2)}$ on $\psi$ and $\lambda$, and by $e^{2\pi i k /M} e^{-i\theta (N-1)}$ on $\eta$. Triviality forces $\theta (N-2) \in 2\pi \mathbbm{Z}$ and $k \equiv 2n \esp (\txn{mod } M)$, with $n \in \mathbbm{Z}_{M/2}$. This yields $\Gamma_{\txn{UV}} = \mathbbm{Z}_{M/2}$. The same element acts on the massless baryons by $e^{4\pi i k /M} e^{-i N \theta}$, and doesn’t act on the Goldstones. Triviality implies 

\begin{align}
G_{\txn{IR}}
=
\frac{
\txn{SU}(M) \times  \txn{U}_{\widetilde{\alpha}}(1)
}
{ \mathbbm{Z}_d \times \mathbbm{Z}_{N M / d } }
\spa\spa,
\label{IR_R5_group_faithful}
\end{align} 

where $M = N_3/2$ and $d = \txn{gcd}(N,M,2) = \txn{gcd}(N,2)$. Note that if we rescale the abelian charge as $\widetilde{q}^\alpha_A \to -1$, we would have to also rescale the UV fermion charges. The IR cocycle condition is
\begin{align}
N \phi_{\widetilde{\alpha}} = \frac{ 2 w_{\txn{IR}} }{M}
\esp (\txn{mod } 1)
\spa\spa.
\label{fluxes_IR_R_5}
\end{align}

Since SU$(N_3)$ breaks diagonally, the relation between generators $z_{N_3/2} = z^2_{N_3}$ yields the following embedding of fluxes

\begin{align}
w_3 = 2 w_{\txn{IR}} \esp (\txn{mod } N_3)
\spa\spa.
\end{align}

By solving the cocycle conditions we obtain the following set of allowed fluxes 
\begin{align*}
w_2 &= 0\esp (\txn{mod } 3)
\spa\spa,
\\
w_3 &= 2 w_{\txn{IR}} \esp (\txn{mod } N_3)
\spa\spa,
\\
w_c &= k - \frac{w_3}{4} \esp (\txn{mod } N_3/4)
\equiv k - \frac{w_\txn{IR}}{2}  \esp (\txn{mod } N_3/4)
\spa\spa,
\\
\phi_{\widetilde{\alpha}} &= 
\frac{k N - 2 w_c}{N(N-2)}
\spa\spa,
\\
\phi_{\widetilde{\beta}} &= 0 
\spa\spa.
\numberthis
\label{cocycle_solutions_R5_fluxes}
\end{align*}

with $k$ an arbitrary integer.\\

Checking if a $(z_M^k, e^{i\theta})$ element can be undone by a color center element $z_c^m$, triviality on $\psi$ and $\lambda$ fixes $\theta/2\pi \equiv k/M + m/N $, and on $\eta$ then forces $k/2 \in \mathbbm{Z} $. Therefore on gauge invariants the UV faithful quotient is $\Gamma_{\txn{UV}}^{\txn{inv}} = \mathbbm{Z}_N \times \mathbbm{Z}_{M / 2}$.  The quotients agree up to a $\mathbbm{Z}_2$ factor present for both odd and even $N$. Since $\Gamma^{\rm inv}_{\rm UV}$ is an index-two subgroup of $\Gamma_{\rm IR}$, condition~(\ref{flux_embedding_condition}) holds. The condition forcing $k$ to be even is independent of $m$, and thus there is no color twist that can eliminate this discrepancy. This $\mathbbm{Z}_2$ excess is generated by a transformation $h$ which acts on the elementary fermions exactly like $z_c^{1/2}$ for all $N$, the square root of the color center transformation.\\

The $h$ transformation is a redundancy of the IR massless spectrum that is not a redundancy of the UV theory for any $N$. This means that in the IR it can be used to create twisted fluxes or be gauged by background 2-form gauge fields, but in the UV it cannot. By the argument given at the beginning of this section, $h$ is anomaly-free in the UV as a consequence of zero-form matching. As a check, we compute its phase directly. The path integral phase induced by $h$ is like the one for half the color center anomaly. The induced phase under this transformation is
\begin{align}
\mathcal{Z}
\to 
e^{2\pi i \mathcal{A}_h}
\mathcal{Z}
\spa\spa,\spa\spa
\mathcal{A}_h = 
\frac{1}{2N} \sum_i n^c_i \mathcal{I}_i
\spa\spa.
\end{align}

Note that the $n^c_i$ factor prevents the contribution proportional to the topological charge $Q_c$ in $\mathcal{I}_i$ from vanishing. The anomaly reads
\begin{align}
\mathcal{A}_h
=
2 Q_c - \frac{1}{2} Q_3 - N(N-2) Q_{\widetilde{\alpha}}
\spa\spa.
\end{align}

Imposing the set of fluxes in (\ref{cocycle_solutions_R5_fluxes}) consistent with the cocycle conditions, as expected on general grounds this contribution is always an integer, and thus $h$ is anomaly-free
\begin{align}
\mathcal{A}_h \in \mathbbm{Z}
\spa\spa.
\end{align}

Therefore, the proposed IR scenario for $R_5$ matches the 't Hooft anomalies induced by fractional fluxes. These results also hold for $R_{17}$ with the IR proposal presented in (\ref{sol_R17_large_N}) and (\ref{R17_solution_symm_group}), which is related to $R_5$ by the replacement $N_1 \leftrightarrow N_2$.

%% file: 09_IR_dynamics.tex
\setlength{\jot}{13pt}

\section{The fate of the IR dynamics} \label{fate_IR_dyna}

The 't Hooft anomalies computed above do not vanish, so the infrared cannot be trivially gapped. Since the models contain no scalars, the gauge group can only break through fermion bilinear condensates, conjecturally in the Most Attractive Channel (MAC) \cite{Raby:Dimopoulos:Susskind:1980:tumbling}, initiating a sequence of tumbling steps\hlfootnote{In \cite{Eichten:Feinberg:1982} it is argued that, for theories satisfying anomaly cancellation and asymptotic freedom, tumbling as realized by the MAC hypothesis is disfavored as the mechanism responsible for the breakdown of the gauge group.}. The MAC ordering for all 30 theories is computed in Appendix \ref{appendix_clases_MAC} and invoked below where relevant. Assuming the gauge group remains unbroken in the IR instead, we briefly review how the global symmetry might be realized, linearly by massless composite fermions, or spontaneously broken by condensates. In Section \ref{subsection_tumbling_operators_R7} we review a tumbling spectrum proposed in \cite{Goity:Peccei:Zeppenfeld:1985} for $R_7$. We also invoke the $a$-theorem for a scenario where the continuous global symmetries are completely broken.\\

The main new result is presented in Section \ref{BZ_FP_section}: a three-loop analysis identifies 12 of the 30 theories as admitting a Banks--Zaks fixed point under perturbative control (Table \ref{table_Banks_Zaks}), for which the UV fermions supply the infrared degrees of freedom and all anomalies are matched automatically.

\subsection{Massless composite fermionic operators} \label{section_massless_general_commposite_operators}

Taking the presence of a genuine one-form electric center symmetry as a criterion for confinement, given that the theories studied here do not enjoy such a symmetry unless the full global symmetry group is gauged, there is no candidate order parameter to distinguish confinement from deconfinement. In what follows we assume that the UV fermions form massless composite fermionic operators while preserving the global symmetry group in a strongly coupled regime.\\

For 11 out of the 30 theories in Table \ref{table_N1_N2_N3_values}, the $\lambda$ operators transform under the two-index anti-symmetric representation $\txn{\textbf{A}}_{2}$, no complex-conjugation required. By contrast, in 19 of these models (for which $N_3$ is still non-zero in general) these fermions transform in the complex-conjugated representation\hlfootnote{For these cases $N_2$ is negative in Table \ref{table_N1_N2_N3_values}.} $\overline{\txn{\textbf{A}}}_{2}$. Note also that for $R_9$, $R_8$, $R_{10}$, $R_{11}$, $R_{15}$, $R_{16}$, $R_{21}$, $R_{22}$ and $R_{26}$, $N_3 < 0$ for some values of $N$; in these cases the $\eta^i$ Weyl fermions transform in the fundamental representation. In the discussion below we do not consider the composite operators that can be built in these scenarios.\\

Here we follow the convention that composite operators are formed by UV fermion operators with \textit{upper} color indices only. Insertions of the SU($N$) Levi-Civita symbol will have lower indices as $\eps_{a_1,...,a_N}$. To do this, we use the following dualized fields 
\begin{align*}
\widetilde{\lambda}^{a_1,...,a_{N-2}}
&=
\epss^{ij \esp a_1,...,a_{N-2}} \lambda_{ij}
\spa\spa \txn{ in } \spa\spa \dualccasimtworep
\spa\spa,
\\
\widetilde{\eta}^{a_1,...,a_{N-1}}
&=
\epss^{i \esp a_1,...,a_{N-1}} \eta_{i}
\spa\spa \txn{ in } \spa\spa \txn{\textbf{A}}_{N-1}
\spa\spa.
\numberthis \label{dual_fermion_operators}
\end{align*}
Crucially, the dimension of the dual representations is the same as for the regular representations, and thus anomalies sourced by the regular fermion operators are not altered. We will use the $\widetilde{\lambda}$ fermions when needed, that is, when the $\lambda$ operators transform under $\ccasimtworep$. Since for most cases the $\eta$ fermions transform under $\ccfundrep$, then we use the dual $\widetilde{\eta}$ operators in $\txn{\textbf{A}}_{N-1} \cong \overline{\txn{\textbf{F}}}$ to construct composite massless fermions.\\

As first done in \tcolor{\cite{Anber:Chan:2023}}, we can start by studying the possible composites that can be built for the 2-index theories, for which $N_3 = 0$ with $N$ fixed (see Appendix \ref{appendix_4d_two_species_models}). For these theories, the $\lambda$ fermions \textit{always} transform in $\overline{\txn{\textbf{A}}}_2$ and thus we use the $\widetilde{\lambda}$ operators. Consider the following operator

\begin{align}
\Theta \sim 
\varepsilon^{(p_j)}  \esp
\psi^{(\ell_j)}  \esp
\widetilde{\lambda}^{(r_j)} \esp 
F^{(n_j)} \esp
\label{2_index_composites_lambda_dual_p}
\spa\spa,
\end{align}

here $\ell_j$ and $r_j$ are powers of UV fermion operator insertions; similarly $p_j$ denotes powers of the antisymmetric Levi-Civita symbol and $n_j$ powers of the SU($N$) field strength. All color indices must be contracted. Thus, requiring the composites to be gauge-invariant (i.e. color singlets) and the operators to be fermionic yields the following conditions
\begin{align}
2 \ell_j + (N-2) r_j - p_j N = 0  
\spa\spa\spa,\spa\spa\spa
\ell_j + r_j  = 2k + 1
\spa\spa,\spa\spa k \in \mathbbm{Z}^+
\spa\spa.
\label{composites_constraint_2_index_th_p}
\end{align}

Recall that the color field strength transforms in the adjoint representation which is built from a fundamental and anti-fundamental. Thus $F$ carries one upper and one lower index and it does not contribute to the above conditions. However, insertions of this operator might be necessary so that $\Theta$ does not vanish for symmetry reasons among color indices. Combining both conditions in (\ref{composites_constraint_2_index_th_p}) we get
\begin{align}
(N-4) r_j - p_j N + 4k + 2 = 0
\spa\spa,
\end{align}

if we set $N = 4z$ for $z$ a positive integer, then 
\begin{align}
1 = 2 \big( z p_j + (1-z)r_j - k \big) 
\spa\spa,
\label{SU(4z)_no_composites_p}
\end{align}

thus the SU($4z$) 2-index theories do not admit composite \textit{fermionic} operators of the type $\Theta$.\\

Before considering operators such as those in (\tcolor{\ref{2_index_composites_lambda_dual_p}}) for the class of theories studied here (including anti-fundamentals $\eta_i$), note that there are simpler fermionic composites that do not require explicit insertions of Levi-Civita symbols. Such operators can be written as
\begin{align}
\mathcal{O}_1^{ABC}
=
\psi^{A , ij} \eta_i^B \eta_j^C
\spa\spa,\spa\spa
\mathcal{O}_2^{ABC}
=
\lambda^{A , ij} \eta_i^B \eta_j^C
\spa\spa,
\end{align}

where $A, B$ and $C$ are flavor indices. In fact, as is well known, the $\mathcal{O}_1^{ABC}$ operators match the UV 't Hooft anomalies for the simplest of the theories $R_1$ where $N_1 = 1$, $N_2 = 0$ and $N_3 = N+4$. Remarkably, these composites saturate the anomalies for all $N$ without chiral symmetry breaking \cite{Bars:Yankielowicz:1981}. These operators are written in terms of the original (not dual) anti-fundamentals $\eta_i$. Equivalently, we can use the dual fields with two insertions of Levi-Civita symbols with lower indices
\begin{align}
\eps \eps
\psi \widetilde{\eta} \widetilde{\eta}
\spa\spa,\spa\spa
\eps \eps
\lambda \widetilde{\eta} \widetilde{\eta}
\spa\spa.
\end{align}

Operators such as $\mathcal{O}_{1}$ and $\mathcal{O}_{2}$ with odd insertions of the fermions $\psi^{ij}$, $\lambda^{ij}$ or a mix of these, are always fermionic, since we need an even number of anti-fundamentals to contract the indices. Operators with even insertions of  $\psi^{ij}$ or $\lambda^{ij}$ are bosonic. In general we can write 
\setlength{\jot}{12pt}
\begin{align*}
&\mathcal{O}_\psi
=
\psi^{i_1 i_2} 
\psi^{i_3 i_4} 
\esp ... \esp 
\psi^{i_{s-1} i_s}
\eta_1 \eta_2
\esp ... \esp 
\eta_{i_n}
\spa\spa,\spa\spa
\mathcal{O}_\lambda
=
\lambda^{i_1 i_2} 
\lambda^{i_3 i_4} 
\esp ... \esp 
\lambda^{i_{s-1} i_s}
\eta_1 \eta_2
\esp ... \esp 
\eta_{i_n}
\spa\spa,
\\
&\mathcal{O}_{\psi\lambda}
=
\psi^{i_1 i_2} 
\psi^{i_3 i_4} 
\esp ... \esp 
\lambda^{i_a i_b} 
\lambda^{i_c i_d} 
\esp ... \esp 
\lambda^{i_{s-1} i_s}
\eta_1 \eta_2
\esp ... \esp 
\eta_{i_n}
\spa\spa,\spa\spa
s = n \txn{ : even}
\spa\spa.
\numberthis
\end{align*}
\setlength{\jot}{13pt}

The total number of fermion operators is $3s/2$, and its parity decides the statistics.

In the case where the $\lambda$ fermions transform under $\overline{\txn{\textbf{A}}}_{2}$, we still can write $\mathcal{O}_\psi$ composites, but we cannot write $\mathcal{O}_\lambda$ or $\mathcal{O}_{\psi\lambda}$ operators. Using the duals $\widetilde{\lambda}^{i_1 ,..., i_{N-2}}$ and regular $\eta_i$ operators we can write
\begin{align}
\widetilde{\mathcal{O}}^{(j)}_{\psi\lambda}
=
\psi^{(\ell_j)}
\widetilde{\lambda}^{(r_j)}
\eta^{(m_j)}
\spa\spa.
\end{align}

The constraints are 
\begin{align}
2\ell_j + (N-2) r_j - m_j = 0
\spa\spa,\spa\spa
\ell_j + r_j + m_j = 2k+1
\spa\spa,\spa\spa k \in \mathbbm{Z}^+
\spa\spa.
\end{align}

If we set $\ell_j = 0$, $r_j = 1$ and $m_j=N-2$ then 

\begin{align}
\widetilde{\mathcal{O}}_{\lambda}^{\esp A_0,...,A_{N-2}}
=
\widetilde{\lambda}^{ A_0 , i_1 ,..., i_{N-2}}
\eta^{A_1}_{i_1}
\eta^{A_2}_{i_2}
\esp ... \esp 
\eta^{A_{N-2}}_{i_{N-2}}
\spa\spa,
\end{align}

this operator is fermionic if $1 + (N-2) = 2k+1$, which holds only for $ N = 2(k+1)$, an even number. Thus SU($2k+1$), where the $\lambda$ fermions transform in $\overline{\txn{\textbf{A}}}_{2}$, does not admit \textit{fermionic} composite operators of the type $\widetilde{\mathcal{O}}_{\lambda}^{\esp A_0,...,A_{N-2}}$.\\

As mentioned above, when including $N_3$ anti-fundamentals the $\lambda$ operators can transform under $\ccasimtworep$ or $\asimtworep$. If the representation is $\asimtworep$ such that color indices are up, a general possible set of composite massless fermions can be written as  

\begin{align}
\mathcal{B}_j \sim 
\varepsilon^{(p_j)} 
\psi^{(\ell_j)} \lambda^{(r_j)} \widetilde{\eta}^{(m_j)}
\esp 
F^{(q_j)}
\spa\spa,
\end{align}

the conditions for the operator to be a fermionic color singlet are 

\begin{align}
2 \ell_j + 2 r_j + (N-1) m_j -  N p_j = 0
\spa\spa\spa,\spa\spa\spa
\ell_j + r_j + m_j = 2k + 1
\spa\spa,\spa\spa k \in \mathbbm{Z}^+
\spa.
\label{IR_baryons_gauge_invariace_condition}
\end{align}

If the representation for the $\lambda$ fermions is $\ccasimtworep$, a general possible set of composite massless fermions can be written as  
\begin{align}
\widetilde{\mathcal{B}}_j \sim 
\varepsilon^{(p_j)} 
\psi^{(\ell_j)} \widetilde{\lambda}^{(r_j)} \widetilde{\eta}^{(m_j)}
\esp 
F^{(q_j)}
\spa\spa,
\end{align}

the conditions for the operator to be a fermionic color singlet are 
\begin{align}
2 \ell_j + (N-2) r_j + (N-1) m_j - N p_j = 0
\spa\spa\spa,\spa\spa\spa
\ell_j + r_j + m_j = 2k + 1
\spa\spa,\spa\spa k \in \mathbbm{Z}^+
\spa.
\label{IR_baryons_gauge_invariace_condition}
\end{align}

One should keep in mind that, despite the hope that a set of composites can saturate all the anomalies for arbitrary $N$ as in the $R_1$ case, in general, different $N$ allows for different spectra of composite operators.\\

Regardless of the particular solution $(\ell_j,r_j,m_j)$, given the set of UV charges in  (\tcolor{\ref{charges_U1_3_fermions_alpha}}) and (\tcolor{\ref{charges_U1_3_fermions_beta}}), the charges for the composites under the U$_\alpha(1)$ and U$_\beta(1)$ symmetries are given by
\begin{align}
\mathrm{Q}_j^\alpha 
= 
\big( 2N_3 T_3 \esp \ell_j - 2 N_1 T_1 \esp m_j \big)
\frac{1}{\me_\alpha}
\spa\spa,\spa\spa
\mathrm{Q}_j^\beta
= 
\big( 2N_3 T_3 \esp r_j - 2N_2 T_2 \esp m_j \big)
\frac{1}{\me_\beta}
\spa\spa.
\end{align}

Finding a set of composites that match all the 't Hooft anomalies is not a trivial task. One can attempt to match the U$_{\alpha,\beta}(1)^3$ and U$_{\alpha,\beta}(1) \times \txn{Grav}^2$ anomalies first, and then check whether the solution matches the rest of the anomalies. If there are $\mathcal{N}_i$ copies of the $\mathcal{B}_i$ composites, then the above mentioned matching conditions are 

\setlength{\jot}{9pt}
\begin{align*}
\sum_i \mathcal{N}_i \big(\mathrm{Q}_i^\alpha\big)^3 
=
\sum_i \big(q_i^\alpha\big)^3  N_i \esp \txn{dim}(\mathcal{R}_i) 
\spa\spa &,\spa\spa
\sum_i \mathcal{N}_i \big(\mathrm{Q}_i^\beta\big)^3 
=
\sum_i \big(q_i^\beta\big)^3 N_i \esp \txn{dim}(\mathcal{R}_i) 
\spa\spa,
\numberthis 
\\
\sum_i \mathcal{N}_i  \mathrm{Q}_i^\alpha
=
\sum_i q_i^\alpha  N_i \esp  \txn{dim}(\mathcal{R}_i)
\spa\spa &,\spa\spa
\sum_i \mathcal{N}_i \mathrm{Q}_i^\beta 
=
\sum_i  q_i^\beta  N_i \esp \txn{dim}(\mathcal{R}_i)
\spa\spa.
\numberthis
\end{align*}
\setlength{\jot}{13pt}

\subsubsection{The special case of $R_7$}
\label{subsection_tumbling_operators_R7}

Here we briefly discuss a particular set of composites relevant for the $R_7$ theory for which $N_1 = 1$, $N_2 = -1$ and $N_3 = 8$. We abandon the assumption of color confinement (and motivate why to do this in what follows). As discussed in Section \tcolor{\ref{section_large_N}}, in \cite{Goity:Peccei:Zeppenfeld:1985} a set of massless composites that match the  't Hooft anomalies which follow from tumbling and complementarity are presented for $R_7$ (see also \cite{Eichten:Peccei:Preskill:Zeppenfeld:1986}). As discussed in \cite{Eichten:Peccei:Preskill:Zeppenfeld:1986}\hlfootnote{See also \cite{Preskill:1982}.} one can construct the following sequence of composite operators

\begin{align}
{\eta^{i}_a}^{\dagger}  \lambda_{ij} {\eta^{j}_b}^{\dagger} 
\esp \longrightarrow \esp
{\eta^{i}_a}^{\dagger}  \lambda_{ij}
\Big( \psi^{jk} \lambda_{k\ell} \Big)
{\eta^{\ell}_b}^{\dagger} 
\esp \longrightarrow \esp
{\eta^{i_1}_a}^{\dagger}  \lambda_{i_1 i_2}
\Big( 
\psi^{i_2 i_3} \lambda_{ i_3 i_4} 
\esp ... \esp 
\psi^{i_{n-2} i_{n-1}} \lambda_{ i_{n-1} i_{n}} 
\Big)
{\eta^{i_n}_b}^{\dagger} 
\spa,
\label{baryon_sequece_s_R7_2}
\end{align}
or equivalently
\begin{align}
\mathcal{O}^{ab}_s
\equiv
{\eta_{i}^a}  \psi^{ij} {\eta_{j}^b}
\esp \longrightarrow \esp
{\eta_{i}^a}  \psi^{ij}
\Big( \lambda_{jk} \psi^{k\ell} \Big)
{\eta_{\ell}^b} 
\esp \longrightarrow \esp
{\eta_{i_1}^a}  \psi^{i_1 i_2}
\Big( 
\lambda_{i_2 i_3} \psi^{ i_3 i_4} 
\esp ... \esp 
\lambda_{i_{n-2} i_{n-1}} \psi^{ i_{n-1} i_{n}} 
\Big)
{\eta_{i_n}^b}
\spa.
\label{baryon_sequece_s_R7}
\end{align}

The sequence is constructed by inserting the following \say{blocks} in the adjoint of SU($N$)  

\begin{align}
\Delta^i_j = \psi^{ik}\lambda_{kj}
\spa\spa.
\label{eq:adjoint-block}
\end{align}

Here $s$ denotes the number of $\Delta^i_j$ insertions. Because the block $\Delta^i_j$ is an $\mathrm{SU}(8)$ singlet but carries a non-zero $\mathrm{U}_\beta(1)$ charge, each insertion shifts both abelian charges by the same constant amount. The tower of operators $\mathcal{O}_s$ is thus an arithmetic progression of charges. In the normalization of \cite{Goity:Peccei:Zeppenfeld:1985} these charges are 

\begin{equation}
  Q_s \;=\; N-2-4s\,, \qquad s=0,1,2,\dots
  \label{O_s_charges_GPZ_basis}
  \spa\spa.
\end{equation}

It is only through this $\mathrm{U}(1)$ label that the different composites are distinguished, so they are indistinguishable under the non-abelian flavor symmetry. The block $\Delta^i_j$ is neutral only under the diagonal combination $\mathrm{U}_\alpha(1)-\mathrm{U}_\beta(1)$ and carries the same charge $2N_3T_3$ under both $\mathrm{U}_\alpha(1)$ and $\mathrm{U}_\beta(1)$. Insertions of $\Delta^i_j$ leave the non-abelian flavor quantum numbers untouched. The charge for these operators is
\begin{align}
 q^{\beta}(\mathcal{O}_s) 
 = 
 2 q^{\beta}_{\eta} + s q^{\beta}_{\Delta}
  =
  -4N_2 T_2 + 2N_3 T_3 s \spa\spa,\spa\spa
s=0,1,2,\dots
  \label{eq:tower-charges}
\spa\spa.
\end{align}

The fact that $q^{\alpha}_{\Delta}=q^{\beta}_{\Delta}$ is what allows the sequence to be labeled by a single abelian charge.\\

Kinematically, the sequence $\mathcal{O}_s$ contains only $\mathcal{O}(N)$ independent gauge invariants of this type. Repeated adjoint insertions of \eqref{eq:adjoint-block} are related by the Cayley--Hamilton identity. The insertion $\Delta^i_{\;j}$ is a $N\times N$ matrix. The Cayley--Hamilton identity states that $\Delta$ satisfies its own characteristic equation of degree $N$, so that $\Delta^N$, and by iteration every higher power, can be written as a linear combination of  $\{\mathbb{1},\Delta,\dots,\Delta^{N-1}\}$ with coefficients built from the gauge-singlet traces $\txn{Tr}(\Delta^p)$. At most $N$ of the $\mathcal{O}_s$ are thus independent: longer chains reduce to shorter ones dressed by singlet factors, and produce no new multiplets. In the tumbling scenario, writing $N = 8n+k$ with $k = 0,\dots,7$, the anomaly conditions retain only the first $2n+2$ operators, $s = 0,1,\dots,2n+1$, which renders the number of massless multiplets finite. On general grounds the set of operators retained by tumbling is therefore smaller than the maximal set of independent operators allowed by the Cayley--Hamilton identity.\\

The importance of these composites is that the truncated tower, with the charges \eqref{eq:tower-charges}, constitutes the massless spectrum that saturates the 't Hooft anomalies of the final global symmetry group that follows from a series of tumbling steps \cite{Goity:Peccei:Zeppenfeld:1985}. In this scenario the final global symmetry group is as large as possible. The sums $\sum_s \bigl(2q^{\beta}_{3}+2N_3T_3\,s\bigr)^{p}$ that enter the cubic and gravitational anomaly conditions, where $p$ denotes the number of abelian U(1) generators appearing in the anomaly being evaluated, are precisely sums over the $\mathcal{O}_s$. Physically, the abelian charge acts as a quark \say{counter}. It produces repeated copies of the same $\mathbf{28}$ of $\mathrm{SU}(8)$, distinguished only by their $\mathrm{U}_\alpha(1)$ label. This is the mechanism by which $R_7$ allows for a repeated family structure at the fermionic composite level.\\

The charges \eqref{eq:tower-charges} are written in the $(\alpha,\beta)$ basis. They are related to the basis (\tcolor{\ref{O_s_charges_GPZ_basis}}) used in \cite{Goity:Peccei:Zeppenfeld:1985} as follows. For $R_7$, the charges in \eqref{eq:tower-charges} become an arithmetic progression with base $q^{\alpha}_{1}  + 2q^{\alpha}_{3}= 2N_3T_3 - 4N_1 T_1$ and step $q^{\alpha}_{\Delta}=2N_3T_3=8$. In \cite{Goity:Peccei:Zeppenfeld:1985} the same progression is written $N-2-4s$, i.e. their $\mathrm{U}_2(1)$ generator is identified with our $\mathrm{U}_\alpha(1)$ up to an overall rescaling by $-1/2$. Their second abelian generator $\mathrm{U}_1(1)$, which counts the number of $\varepsilon$-tensors in a singlet state, is a linear combination of our $\mathrm{U}_\alpha(1)$ and $\mathrm{U}_\beta(1)$ and does not vary along the tower.\\

Finally, the same $\Delta^i_j$ that generates the tower is the adjoint channel $\langle \psi^{ij}\lambda_{jk}\rangle$, which after condensation, at large $N$ becomes as attractive as the fundamental channel $\langle \psi^{ij} \eta_{j}\rangle$ that drives the tumbling processes in the scenario discussed in \cite{Goity:Peccei:Zeppenfeld:1985} (see Appendix \tcolor{\ref{appendix_clases_MAC}}). Above some critical $N_*$ the theory may therefore tumble along the adjoint direction into a Higgs or Coulomb phase and fail to confine altogether \cite{Eichten:Peccei:Preskill:Zeppenfeld:1986}. Consistently, the bound states realizing \eqref{eq:tower-charges} in the confining phase are the color singlets $\mathcal{O}_s$, which contain up to $\mathcal{O}(N)$ quarks and whose self-interactions grow with $N$. In the strict planar limit these composites are suppressed: they require a $\lambda_{ij}$ loop, costing a factor $1/N$, and they are neutral under the U$(1)$ symmetry whose charge counts the number of $\txn{SU}(N)$ color indices left uncontracted by pairing the generator $Q_1$ in \cite{Goity:Peccei:Zeppenfeld:1985}. This U$(1)$ has a non-zero mixed gravitational anomaly, $\txn{Tr}\,Q_1\neq 0$, and matching it would require massless bound states charged under U$(1)$. The tower provides none, and no gauge-singlet condensate charged under U$(1)$ is available to break it either. This is the contradiction that suggests the loss of confinement at large $N$ as discussed in Section \tcolor{\ref{section_large_N}}. The generation of the $\mathcal{O}_s$ tower might be a finite-$N$-only phenomenon.\\

The sequence of operators $\mathcal{O}_s$ is a special case that we do not expect to be allowed for the remaining theories in Table \ref{table_N1_N2_N3_values}. Insertions of $\Delta^i_j$ do not alter the flavor content, and every operator in the sequence transforms in the same antisymmetric $\mathbf{28}$ representation of $\mathrm{SU}(8)$ carried by the two end $\eta$'s (once the $\eta$ spins are symmetrized, as required for a $(\tfrac12,0)$ composite fermion). For $|N_1|>1$ or $|N_2|>1$, the fermions $\psi$ and $\lambda$ carry their own $\mathrm{SU}(N_1)$ and $\mathrm{SU}(N_2)$ indices, and thus operators of different length fall into different flavor representations.

\subsection{Bosonic operators}

In this scenario the global symmetry group is spontaneously broken by condensates. Taking the presence of a genuine one-form electric center symmetry as a criterion for confinement, given that the theories studied here do not enjoy such a symmetry unless the full global symmetry group is gauged, there is no candidate order parameter to distinguish confinement from deconfinement. Since in this case we are interested in bosonic and not fermionic operators, in what follows we do not employ the duals $\widetilde{\lambda}$ and $\widetilde{\eta}$.\\

Consider the general composite 

\begin{align}
\mathcal{C}_j \sim 
\psi^{(a_j)} \lambda^{(b_j)} \eta^{(c_j)}
\spa\spa,
\end{align}

some gauge invariant combinations of the fermion operators can yield a vanishing condensate because of possible symmetries between their spinor indices $s_i$. This can be avoided by inserting gluonic operators of the form $(F_{\mu\nu})^{s_i}_{s_j} \sigma^{\mu\nu}$.\\

If the $\lambda$ fermions transform in $\asimtworep$ then the constraints are
\begin{align}
2 a_j + 2 b_j - c_j = 0
\spa\spa,\spa\spa
a_j + b_j + c_j = 2k 
\spa\spa,\spa\spa
k \in \mathbbm{Z}^+
\spa\spa.
\end{align}

The simplest condensate that can be formed in this case is

\begin{align}
\mathcal{C}^{ABCDEF}
=
\psi^{A, i_1 i_2 }
\lambda^{B, i_3 i_4 }
\eta^C_{i_1}
\eta^D_{i_2}
\eta^E_{i_3}
\eta^F_{i_4}
\spa\spa.
\end{align}

More generally, for even insertions of $\psi^{ij}$ or $\lambda^{ij}$ operators, we need twice as many anti-fundamentals $\eta_i$ to contract all the indices. There is always an even number of fermions, so this type of composite is always bosonic.\\

If the $\lambda$ fermions transform in $\ccasimtworep$ then the constraints are
\begin{align}
2 a_j - 2 b_j - c_j = 0
\spa\spa,\spa\spa
a_j + b_j + c_j = 2k 
\spa\spa,\spa\spa
k \in \mathbbm{Z}^+
\spa\spa.
\end{align}

The simplest condensates that can be formed in this case are those with $c_j = 0$
\begin{align}
\mathcal{C}_j^{n_1,n_2}
=
\Big( \psi^{ n_1 , ij }  \Big)^{\ell_j}
\Big( \lambda^{n_2}_{ij} \Big)^{r_j}
\spa\spa,
\end{align}

gauge invariance dictates $\ell_j = r_j$. Since $\ell_j + r_j$ is always even, this operator is bosonic.\\

 We do not further explore the possible IR condensates and their implications for symmetry breaking and anomaly matching. Determining how many condensates form in the IR in general remains an open problem, with only a few examples where the global symmetry breaking pattern is known (see \cite{Li:1974,Swift:Eliezer:Elias:1975,Ruegg:1980,Wu:1983,Wyler:Gerard:Jetzer:1984} and references therein).\\

As was noted in \cite{Anber:Chan:2023}, for the class of theories where $N$ is fixed such that there is no matter in the anti-fundamental representation, in some cases the condensates break the center flavor symmetries, which render the mixed U$_{\alpha,\beta}(1) \times$ CFU anomaly irrelevant.

\subsection{The $a$-theorem}

The \textit{a}-theorem reflects the fact that the number of degrees of freedom must decrease under the renormalization group flow \cite{Cardy:1988}. As proved in \cite{Komargodski:Schwimmer:2011}, this imposes the following bound between the UV and IR degrees of freedom 

\begin{align}
a_{UV} > a_{IR}
\spa\spa.
\label{a_theorem}
\end{align}

In the non-interacting limit, the $a$-function is

\begin{align}
a = N_s + 11 N_f + 62 N_v
\spa\spa,
\end{align}
here $N_s$, $N_f$ and $N_v$ stand for the number of scalars, Dirac fermions and vector bosons respectively\hlfootnote{Here we use units where $a = 1$ for a single real massless scalar field.}. For the theories considered in this work the $a$-function in the UV is  

\begin{align}
a_{UV}
=
62  (N^2 - 1) 
+ 
\frac{11}{2}
\Big(
\txn{dim}(\txn{\textbf{S}}) |N_1| + \txn{dim}(\txn{\textbf{A}}) |N_2| + \txn{dim}(\txn{\textbf{F}}) |N_3|
\Big)
\spa\spa.
\end{align}

If we consider the case where the global symmetry group is fully broken in the IR, the $a$-function would be 
\begin{align}
a_{IR}
=
\sum_i^3 \Big( N_i^2 - 1 \Big) + 2
\spa\spa.
\end{align}

For the cases where some $N_j = 0$, the sum is restricted to $i\neq j$, with the abelian contribution being 1 instead of 2. This happens only for $R_1$, $R_2$ and the 2-index theories with $N_3 = 0$ (see Table \ref{two_species_table_appendix}). These values for $a_{IR}$ satisfy the bound in (\tcolor{\ref{a_theorem}}) for all theories in Table \ref{table_N1_N2_N3_values} for all $N$. Given this result, we do not learn anything new besides the fact that a completely broken chiral symmetry in the IR is compatible with the $a$-theorem for all cases studied here.

\subsection{Banks-Zaks fixed points} \label{BZ_FP_section}

It may also be possible for these theories to develop a fixed point flowing to a CFT in the IR where the global symmetry group is preserved. In the scenario where the CFT is weakly coupled, the UV chiral fermions can be regarded as the IR gapless excitations, and thus the set of anomalies is matched automatically. A weakly coupled CFT can be realized by the existence of a Banks-Zaks fixed point \tcolor{\cite{Banks:Zaks:1982}}.\\

The three-loop beta function is 
\begin{align}
\beta(g) 
&=
 \beta_0 \esp \frac{g^3}{(4\pi)^2}
+ \beta_1 \esp \frac{g^5}{(4\pi)^4}
+ \beta_2 \esp \frac{g^7}{(4\pi)^6}
\spa\spa,
\end{align}

at two-loop order, there is a possible fixed point given by\hlfootnote{In the convention followed here, a fixed point requires $\beta_0 < 0$ and $\beta_1 > 0$ (see Appendix \ref{appendix_beta_func}).}
\begin{align}
\frac{g_0^2}{16\pi^2} = - \frac{\beta_0}{\beta_1}
\spa\spa.
\end{align}

Nevertheless, what really controls the perturbative expansion at the fixed point is the 't Hooft coupling $\lambda = g^2 N$. We define the value of the coupling at the fixed point as

\begin{align}
\lambda_0
=
\lim_{N \to \infty}  \frac{N g_0^2 }{16\pi^2}
=
\frac{2 ( 11 - ( |N_1| + |N_2| ) - |N_1 + N_2| ) } 
{16( |N_1| + |N_2| ) + 13 |N_1 + N_2| - 68}
\spa\spa.
\label{Hooft_coupling_large_N}
\end{align}

If the coupling is small, then we can assess its stability by comparing with the 3-loop beta function, i.e. with the $\beta_2$ coefficient. This quantity which we define as $\eps$ has to be small in order for the fixed point to be under control
\begin{align}
\eps = 
\lim_{N \to \infty}
\left| 
\dfrac{ \left( g_0^2/16\pi^2 \right)^2  }{\beta_0/\beta_2}
\right|
\spa\spa.
\end{align} 

For the 30 possible theories presented in Table \tcolor{\ref{table_N1_N2_N3_values}}, we found that only 12 possibly develop a well-controlled Banks-Zaks fixed point. The value for the coupling at the fixed point for these cases is reported in Table \tcolor{\ref{table_Banks_Zaks}}. This might not be shocking given the large matter content and in particular that the number of fermions in the anti-fundamental representation of the gauge group scales linearly with $N$. However, there are some cases like that of $R_6$ where the matter content is not too large but still a well-controlled fixed point could be developed. For those theories where $\lambda$ is $\mathcal{O}(1)$, we cannot reliably establish the presence of a weakly coupled Banks-Zaks fixed point. Recall also that this class of theories admits a large $N$ limit.\\

For $R_{6}$, $R_{13}$, $R_{18}$ and $R_{23}$, a symmetry breaking pattern as suggested by the large $N$ limit cannot take place (see Section \tcolor{\ref{section_large_N}}). In this scenario, a condensate breaks the $q_\alpha - q_\beta$ direction in the $\txn{U}_\alpha(1) \times \txn{U}_\beta(1)$ combination leaving the $q_\xi = q_\alpha + q_\beta$ direction intact. For these theories, the total number of massless baryons needed to match the $\txn{U}_\xi^3(1)$ anomaly is fractional. This failure to match the anomalies further supports the scenario where a weakly coupled CFT is realized by a perturbative Banks-Zaks fixed point.\\

\renewcommand{\extable}{1ex}
\begin{center}
\begin{tabular}{||c | c | c||} 
\hline\hline
\rule{0pt}{3ex}  
\phantom{--}$R_i$\phantom{--}  & 
\phantom{--}$\lambda_0$\phantom{--}  & 
\phantom{--}$\eps$\phantom{--}  \\ [1ex]
\hline\hline
\phantom{-} &  &  \\ [-2ex]
$R_{6}$  &  $2/77 \approx 0.026$            &  $898/7623 \approx 0.12$     \\ [\extable]
$R_{11}$ &  $1/40 = 0.025$  &  $1597/14400 \approx 0.11$  \phantom{0} \\ [\extable]
$R_{13}$ &  $2/77 \approx 0.026$            &  $898/7623 \approx 0.12$     \\ [\extable]
$R_{16}$ &  $2/83 \approx 0.024$            &  $6490/62001 \approx 0.1$  \phantom{0.} \\ [\extable]
$R_{18}$ &  $2/77 \approx 0.026$            &  $898/7623 \approx 0.12$     \\ [\extable]
$R_{22}$ &  $1/43 \approx 0.023$            &  $1648/16641 \approx 0.1$ \phantom{0.} \\ [\extable]
$R_{23}$ &  $2/77 \approx 0.026$            &  $898/7623 \approx 0.12$     \\ [\extable]
$R_{26}$ &  $2/89 \approx 0.022$            &  $6694/71289 \approx 0.1$ \phantom{0.} \\ [\extable]
$R_{27}$ &  $1/40 = 0.025$  &  $1597/14400 \approx 0.11$  \phantom{.}   \\ [\extable]
$R_{28}$ &  $2/83 \approx 0.024$            &  $6490/62001 \approx 0.1$ \phantom{0.}   \\ [\extable]
$R_{29}$ &  $1/43 \approx 0.023$            &  $1648/16641 \approx 0.1$ \phantom{0.} \\ [\extable]
$R_{30}$ &  $2/89 \approx 0.022$            &  $6694/71289 \approx 0.1$ \phantom{0.} \\ [\extable]
\hline
\end{tabular}
\end{center}
\captionof{table}{ \label{table_Banks_Zaks}
Value for the 't Hooft coupling $\lambda_0$ at the Banks-Zaks fixed point and the stability test value $\eps$.
}

%% file: 10_concluding_remarks.tex
\section{Concluding remarks}
\label{concluding_remarks}

Let us briefly summarize our findings and comment on some directions for future work.\\

We have determined the global structure of the symmetries in a large catalog of 4$d$ SU($N$) chiral gauge theories.  Using the descent procedure we obtained the ordinary zero-form anomalies and the anomalies sensitive to the global structure of the faithful zero-form symmetry for all theories in the catalog. The same results (up to predictable linear combinations) were obtained with the method of fractional 't Hooft fluxes on $\mathbbm{T}^4$.\\

A convenient feature of the catalog is that these theories remain asymptotically free in the large $N$ limit. In this limit certain diagrammatic arguments can be used to posit consistent IR behaviors. We have conjectured a new IR phase for theories $R_5$ and $R_{17}$, consistent with large $N$ reasoning and the ordinary zero-form 't Hooft anomaly constraints. This scenario is structurally similar to the old proposal \cite{Eichten:Peccei:Preskill:Zeppenfeld:1986} for $R_3$; however, it does not work when extended to the rest of the theories. \\

For the unbroken sectors with the same smooth fluxes available in the UV and IR, matching of the fractional-flux mixed anomalies was found to be a corollary of the ordinary zero-form anomaly matching. This holds provided that the smooth flux backgrounds associated with unbroken symmetries of the UV theory remain valid backgrounds for the massless IR fermions. This is automatic for explicitly composite spectra. The IR faithful symmetry group can have more redundancies than its UV counterpart, as we observed in $R_5$ and $R_{17}$, but the corresponding UV symmetries are anomaly-free as a consequence of ordinary zero-form matching. The remaining content of these anomalies concerns fluxes along spontaneously broken directions, which fix the quantum numbers of defects, and unbroken discrete remnants of broken symmetries.\\

Of the theories in the catalog, 12 could flow to a Banks-Zaks weakly coupled CFT in the IR. In particular, theories $R_{6}$, $R_{13}$, $R_{18}$ and $R_{23}$ could develop a fixed point in the IR with a two-loop 't Hooft coupling of $\lambda_0 = 2/77 \approx 0.026$. In addition to this, for these theories the pattern of symmetry breaking suggested by the large $N$ limit is in any case excluded. Of course, some other theories in the catalog might flow to a strongly coupled CFT, but our work does not give particular further insight into this possibility.\\

We conclude by discussing a few directions for future work. First, there are additional  anomalies we did not investigate; for example, mixed  Witten–CFU anomalies. Furthermore, our discussion of the construction of massless composite fermion operators was brief; for instance, we did not discuss cases where $N_3 < 0$ (the $|N_3|$ Weyl fermions transform in the fundamental representation). Finally, it would  be of interest to undertake a systematic investigation of tumbling scenarios across the catalog.

%% file: Appendix_0.tex
\section{Dirac indices}
\label{appendix_fluxes}

The anomalies induced in each twisted sector $s$ are linear combinations of those sourced by $B^{(2)}_i$ field fractional fluxes. The coefficients of this expansion are given by the ratio between fluxes of twisted sectors and those that follow from the descent prescription. We write these coefficients as $\mathcal{K}^{(s)}_{j}$ where $j \equiv c,1,2,3,u$ stands for each of the center symmetries participating in the anomaly

\begin{align}
\mathcal{I}^{(s)}_{\delta}
=
\sum_j \mathcal{K}^{(s)}_{j}
\Phi^{j}_{\delta} 
\spa\spa,\spa\spa
\mathcal{K}^{(s)}_{j}
=
\frac{ \mathcal{Q}^{(s)}_j }{Q_j}
\spa\spa.
\end{align}

Here $\mathcal{I}^{(s)}_{\delta}$ is the total Dirac index for each twisted sector $s$. The $\Phi^{j}_{\delta}$ are partial indices sourced by fractional fluxes computed from the descent prescription.

\subsection*{$\bullet$ Example: $R_{12}$ at $N = 6$}

Consider the $R_{12}$ theory at $N = 6$, for which $\me_\gamma = 2$ and $\me_\delta = 4$. The U$(1)_\alpha$ total index in the $s = 1$ twisted sector computed by direct evaluation of (\ref{total_index_twisted}) is $\mathcal{I}_\alpha^{(1)} = - 2841$ (see Table \ref{I_alpha_Z_N1_anomaly_twisted_fields}). We can also obtain this by evaluating the linear combination

\begin{align*}
\mathcal{I}_\alpha^{(1)} 
&= 
\mathcal{K}^{(1)}_{1} \Phi^{1}_{\alpha} 
+ \mathcal{K}^{(1)}_{3} \Phi^{3}_{\alpha}  
+ \mathcal{K}^{(1)}_{\alpha} \Phi^{\alpha}_{\alpha} 
\\
&=
- \Phi^{1}_{\alpha} - 336 \esp \Phi^{3}_{\alpha}  
+ \frac{1}{4} \esp \Phi^{\alpha}_{\alpha} 
\spa\spa.
\numberthis
\end{align*}

Taking the partial indices $\Phi^i_\alpha$ by direct evaluation of (\ref{partial_index_B_flux}) and (\ref{partial_index_gu_B_flux}), we reproduce $\mathcal{I}_\alpha^{(1)}$ 

\begin{align*}
\mathcal{I}_\alpha^{(1)} 
&= 
(-1)\bigg(-\frac{231}{2}\bigg)
- 336 \esp \frac{96}{11}  
+ \frac{1}{4} \bigg(- \frac{1062}{11} \bigg)
= - 2841
\spa\spa.
\numberthis
\end{align*}

\subsection*{$\bullet$ Example: $R_{18}$ at $N = 7$}

Consider the $R_{18}$ theory at $N =7$, for which $\me_\gamma = 1$ and $\me_\delta = 1$. The U$(1)_\beta$ total index in the $s = 2$ twisted sector computed by direct evaluation of (\ref{total_index_twisted}) is $\mathcal{I}_\beta^{(2)} = - 931$ (see Table \ref{I_beta_Z_N2_anomaly_twisted_fields}). We can also obtain this by evaluating the linear combination 

\begin{align*}
\mathcal{I}_\beta^{(2)} 
&= 
\mathcal{K}^{(2)}_{2} \Phi^{2}_{\beta} 
+ \mathcal{K}^{(2)}_{3} \Phi^{3}_{\beta}  
+ \mathcal{K}^{(2)}_{\beta} \Phi^{\beta}_{\beta}
\\
&=
(- 1) \esp  \Phi^{2}_{\beta} 
- 950 \esp \Phi^{3}_{\beta}  
+ \frac{1}{2916} \esp \Phi^{\beta}_{\beta} 
\spa\spa.
\numberthis
\end{align*}

Taking the partial indices $\Phi^i_\beta$ by direct evaluation of (\ref{partial_index_B_flux}) and (\ref{partial_index_gu_B_flux}) we reproduce  $\mathcal{I}_\beta^{(2)}$ 

\begin{align*}
\mathcal{I}_\beta^{(2)}
&= 
(- 1) \bigg( - \frac{819}{2}\bigg)
- 950 \esp \frac{70}{39}
+ \frac{1}{2916} \esp \frac{13822326}{13}
=
-931
\spa\spa.
\numberthis
\end{align*}

\subsection*{$\bullet$ Example: $R_{7}$ at $N = 8$}

Consider the $R_{7}$ theory at $N = 8$, for which $\me_\gamma = \me_\delta = 2$. The U$(1)_\beta$ total index in the $s = 3$ twisted sector computed by direct evaluation of (\ref{total_index_twisted}) is $\mathcal{I}_\beta^{(3)} = 2960$ (see Table \ref{I_beta_Z_N3_anomaly_twisted_fields}). We can also obtain this by evaluating the linear combination

\begin{align*}
\mathcal{I}_\beta^{(3)} 
&= 
\mathcal{K}^{(3)}_{3} \Phi^{3}_{\beta} 
+ \mathcal{K}^{(3)}_{\alpha} \Phi^{\alpha}_{\beta}  
+ \mathcal{K}^{(3)}_{\beta} \Phi^{\beta}_{\beta} 
+ \mathcal{K}^{(3)}_{\alpha\beta} \Phi^{\alpha\beta}_{\beta} 
\\
&=
(-252) \Phi^{3}_{\beta} 
+ \frac{1}{9} \esp \Phi^{\alpha}_{\beta}  
+ \frac{1}{25} \esp \Phi^{\beta}_{\beta} 
+ \frac{1}{15} \esp \Phi^{\alpha\beta}_{\beta} 
\spa\spa.
\numberthis
\end{align*}

Taking the partial indices $\Phi^i_\beta$ by direct evaluation of   (\ref{partial_index_gu_B_flux}) we reproduce the value for $\mathcal{I}_\beta^{(3)}$ 

\begin{align*}
\mathcal{I}_\beta^{(3)} 
&= 
(-252)(-12) 
+ \frac{1}{9} (2700) 
+ \frac{1}{25} (-100)
+ \frac{1}{15} (-5400)
= 2960
\spa\spa.
\numberthis
\end{align*}

\subsection*{$\bullet$ Example: $R_{22}$ at $N = 8$}

Consider the $R_{22}$ theory at $N =8$, for which $\me_\gamma = 2$, $ \me_\delta = 30$ and $\me_u = 8$. The U$(1)_\beta$ total index in the $s = 4$ twisted sector computed by direct evaluation of (\ref{total_index_twisted}) is $\mathcal{I}_\beta^{(4)} =  380$ (see Table \ref{I_beta_ZN_anomaly_twisted_fields}). We can also obtain this by evaluating the linear combination 

\begin{align*}
\mathcal{I}_\beta^{(4)} 
&= 
\mathcal{K}^{(4)}_{c} \Phi^{c}_{\beta} 
+ \mathcal{K}^{(4)}_{3} \Phi^{3}_{\beta}  
+ \mathcal{K}^{(4)}_{\alpha} \Phi^{\alpha}_{\beta}
+ \mathcal{K}^{(4)}_{\beta} \Phi^{\beta}_{\beta}
+ \mathcal{K}^{(4)}_{\alpha\beta} \Phi^{\alpha\beta}_{\beta}
\\
&=
(-7) \esp \Phi^{c}_{\beta} 
- 15 \esp \Phi^{3}_{\beta}  
+ \frac{1}{16} \esp \Phi^{\alpha}_{\beta}
+ \frac{1}{16} \esp \Phi^{\beta}_{\beta}
+ \frac{1}{16} \esp \Phi^{\alpha\beta}_{\beta}
\spa\spa.
\numberthis
\end{align*}

Taking the partial indices $\Phi^i_\beta$ by direct evaluation of (\ref{partial_index_B_flux}) and (\ref{partial_index_gu_B_flux}) we reproduce  $\mathcal{I}_\beta^{(4)}$ 

\begin{align*}
\mathcal{I}_\beta^{(4)} 
&= 
(-7)(0)
- 15 \esp (-30)
+ \frac{1}{16} \esp (6750)
+ \frac{1}{16} \esp (5630)
+ \frac{1}{16} \esp (-13500)
=  380
\spa\spa.
\numberthis
\end{align*}

\newpage

\subsection{Two-form gauge fields}
\label{appendix_fluxes_B_fields}

The partial indices $\Phi^i_{\alpha,\beta}$ shown here are obtained by direct evaluation of (\ref{partial_index_B_flux}). For all the tables in this section, the dash indicates that $N_3 = 0$ for that $N$ (see Table \tcolor{\ref{two_species_table_appendix}}). Theories that are not asymptotically free are denoted by the $\diamond$ symbol (see Tables \ref{table_rep_multiplictly_1} and \ref{table_rep_multiplictly_2}). The cases with $\me_u = 1$ are marked by the * symbol; in these cases the axial center is trivial and thus there is no $s = 5$ twisted sector (see Table \ref{g_u_values}).\\

$\bullet$ \textbf{Flavor center} $\mathbbm{Z}_{N_1}$\\

For all theories, the partial index $\Phi^1_\beta = 0$ for all $N$.\\

\renewcommand{\extable}{0.85ex}
\begin{center}
\begin{tabular}{||c | c c c c c c ||} 
\hline\hline
\multicolumn{7}{||c||}{
$\Phi^1_\alpha$ 
} \\
\hline
\rule{0pt}{2ex}  
\diagbox[width=3em,height=2em,font=\scriptsize\itshape]{$R_i$}{$N$}& 
  $5$  &  $6$  & $7$  & $8$ & $9$ & $10$        \\ [0.0ex]
\hline\hline
\phantom{-} &  &    &    &     &   &    \\ [-2.5ex]
$R_{12}$   &     -285/2 &     -231/2 &   -350 & -126 & -1395/2 &  -935/2 \\ [\extable]
$R_{13}$   &      -45/2 &     -273/2 &   -434 & -162 & -1845/2 & -1265/2 \\ [\extable]
$R_{14}$   &     -255/2 &     -189/2 &   -266 &  -18 &  -945/2 &  -605/2 \\ [\extable]
$R_{15}$   &     -225/2 &     -147/2 &   -182 &  -54 &   -45/2 &  -275/2 \\ [\extable]
$R_{16}$   &     -195/2 &     -105/2 &    -98 &  -18 &   -45/2 &    55/2 \\
\hline
\end{tabular}
\end{center}
\captionof{table}{ \label{I_alpha_Z_N1_anomaly_B_fields}
Partial index from the $\mathbbm{Z}_{N_1}$ flux.
}

\newpage

$\bullet$ \textbf{Flavor center} $\mathbbm{Z}_{N_2}$\\

For all theories, the partial index $\Phi^2_\alpha = 0$ for all $N$.\\

\renewcommand{\extable}{2ex}
\begin{center}
\begin{tabular}{||c | c c c c c c  ||} 
\hline\hline
\multicolumn{7}{||c||}{
$\Phi^2_\beta$ 
} \\
\hline
\diagbox[width=3em,height=2em,font=\scriptsize\itshape]{$R_i$}{$N$}& 
  $5$  &  $6$  & $7$  & $8$ & $9$ & $10$        \\ [0.0ex]
\hline\hline
\phantom{-} &  &    &    &     &   &    \\ [-2.5ex]
$R_{4}$    &        -55 &     -105/2 & -357/2 &    -70 &   -414 &  -585/2 \\ [\extable]
$R_{5}$    &      -40/3 &        -20 &    -28 & -112/3 &    -48 &     -60 \\ [\extable]
$R_{6}$    &      -65/2 &     -135/4 & -483/4 &    -49 &   -297 &  -855/4 \\ [\extable]
$R_{8}$    &         35 &       45/2 &   21/2 &     14 &     54 &    45/2 \\ [\extable]
$R_{9}$    &       20/3 &          5 &     14 &      - &    -24 &     -15 \\
\hline
\end{tabular}
\end{center}
\captionof{table}{ \label{I_beta_Z_N2_anomaly_B_fields}
Partial index from the $\mathbbm{Z}_{N_2}$ flux.
}

\newpage

$\bullet$ \textbf{Flavor center} $\mathbbm{Z}_{N_3 / \textsl{g}_\gamma }$\\

\renewcommand{\extable}{0.6ex}
\begin{center}
\begin{tabular}{||c | c c c c c c  ||} 
\hline\hline
\multicolumn{7}{||c||}{
$\Phi^3_\alpha$ 
} \\
\hline
\rule{0pt}{2ex}  
\diagbox[width=3em,height=2em,font=\scriptsize\itshape]{$R_i$}{$N$}& 
  $5$  &  $6$  & $7$  & $8$ & $9$ & $10$        \\ [0.0ex]
\hline\hline
\phantom{-} &  &    &    &     &   &    \\ [-2.5ex]
$R_{3}$    &        7/2 &         16 &    9/2 &     10 &   11/2 &     24 \\ [\extable]
$R_{4}$    &      35/11 &       48/7 &  63/17 &    8/5 &  99/23 & 120/13 \\ [\extable]
$R_{5}$    &      35/12 &          6 &  63/20 &   20/3 &  99/28 &     15 \\ [\extable]
$R_{6}$    &      35/13 &       16/3 &  63/23 &   40/7 &   3/11 & 120/19 \\ [\extable]
$R_{7}$    &       35/8 &         12 &   63/8 &     20 &   99/8 &     60 \\
\hline
\end{tabular}
\end{center}
\captionof{table}{ \label{I_alpha_Z_N3_anomaly_B_fields}
Partial index from the $\mathbbm{Z}_{N_3/ \textsl{g}_\gamma}$ flux. 
}

\renewcommand{\extable}{0.8ex}
\begin{center}
\begin{tabular}{||c | c c c c c c  ||} 
\hline\hline
\multicolumn{7}{||c||}{
$\Phi^3_\beta$
} \\
\hline
\rule{0pt}{2ex}  
\diagbox[width=3em,height=2em,font=\scriptsize\itshape]{$R_i$}{$N$}& 
  $5$  &  $6$  & $7$  & $8$ & $9$ & $10$        \\ [0.0ex]
\hline\hline
\phantom{-} &  &    &    &     &   &    \\ [-2.5ex]
$R_{3}$    &        3/2 &          8 &     5/2 &     6 &     7/2 &     16 \\ [\extable]
$R_{4}$    &      30/11 &       48/7 &   70/17 &  24/5 &  126/23 & 160/13 \\ [\extable]
$R_{5}$    &        5/4 &         18 &   21/20 &     4 &   27/28 &     15 \\ [\extable]
$R_{6}$    &      60/13 &       32/3 &  140/23 &  48/7 &   84/11 & 320/19 \\ [\extable]
$R_{7}$    &      -15/8 &        -12 &   -35/8 &   -12 &   -63/8 &    -20 \\
\hline
\end{tabular}
\end{center}
\captionof{table}{ \label{I_beta_Z_N3_anomaly_B_fields}
Partial index from the $\mathbbm{Z}_{N_3}  / \textsl{g}_\gamma$ flux.  
}

\newpage

$\bullet$ \textbf{Axial center} $\mathbbm{Z}_{\me_u}$\\

\begingroup 
\normalsize 
{\scalefont{0.95} 

\renewcommand{\extable}{1.25ex}
\begin{center}
\begin{tabular}{||c | c c c c c c ||} 
\hline\hline
\multicolumn{7}{||c||}{
$\Phi^{\alpha}_\alpha$ 
} \\
\hline
\rule{0pt}{2ex}  
\diagbox[width=3em,height=2em,font=\scriptsize\itshape]{$R_i$}{$N$}& 
  $5$  &  $6$  & $7$  & $8$ & $9$ & $10$        \\ [0.0ex]
\hline\hline
\phantom{-} &  &    &    &     &   &    \\ [-2.5ex]
$R_{3}$    &  -387/2  &  -1  &  1375/2  &  342  &  14161/2  &  236   \\ [\extable]
$R_{4}$    &  3600/11  &  261/7  &  298900/17  &  *  &  2317896/23  &  79600/13   \\ [\extable]
$R_{5}$    &  12015/4  &  162  &  30485/4  &  21492  &  1467963/4  &  1220   \\ [\extable]
$R_{6}$    &  118080/13  &  1244/3  &  3883600/23  &  364032/7  &  79968  &  994240/19   \\ [\extable]
$R_{7}$    &  -6795/8  &  *  &  -82775/8  &  -3204  &  -445851/8  &  *   \\
\hline
\end{tabular}
\end{center}
\captionof{table}{ \label{I_alpha_Zu_anomaly_B_fields}
Partial index from the $\mathbbm{Z}_{\mathrm{g}_u}$ flux.
}

}\endgroup

\begingroup 
\normalsize 
{\scalefont{0.95}

\renewcommand{\extable}{1.25ex}
\begin{center}
\begin{tabular}{||c | c c c c c c ||} 
\hline\hline
\multicolumn{7}{||c||}{
$\Phi^{\alpha}_\beta$ 
} \\
\hline
\rule{0pt}{2ex}  
\diagbox[width=3em,height=2em,font=\scriptsize\itshape]{$R_i$}{$N$}& 
  $5$  &  $6$  & $7$  & $8$ & $9$ & $10$        \\ [0.0ex]
\hline\hline
\phantom{-} &  &    &    &     &   &    \\ [-2.5ex]
$R_{3}$    &  -1323/2  &  -32  &  -10125/2  &  -1350  &  -41503/2  &  -576   \\ [\extable]
$R_{4}$    &  -52920/11  &  -768/7  &  -567000/17  &  *  &  -2988216/23  &  -92160/13   \\ [\extable]
$R_{5}$    &  -19845/4  &  -648  &  -8505/4  &  -8100  &  -205821/4  &  -540   \\ [\extable]
$R_{6}$    &  -423360/13  &  -2048/3  &  -4536000/23  &  -172800/7  &  -724416  &  -737280/19   \\ [\extable]
$R_{7}$    &  6615/8  &  *  &  70875/8  &  2700  &  373527/8  &  *   \\
\hline
\end{tabular}
\end{center}
\captionof{table}{ \label{I_alpha_beta_Zu_anomaly_B_fields}
Partial index from the $\mathbbm{Z}_{\mathrm{g}_u}$ flux.
}
}\endgroup

\begingroup 
\normalsize 
{\scalefont{0.95} 

\renewcommand{\extable}{1.25ex}
\begin{center}
\begin{tabular}{||c | c c c c c c ||} 
\hline\hline
\multicolumn{7}{||c||}{
$\Phi^{\beta}_\alpha$ 
} \\
\hline
\rule{0pt}{2ex}  
\diagbox[width=3em,height=2em,font=\scriptsize\itshape]{$R_i$}{$N$}& 
  $5$  &  $6$  & $7$  & $8$ & $9$ & $10$        \\ [0.0ex]
\hline\hline
\phantom{-} &  &    &    &     &   &    \\ [-2.5ex]
$R_{3}$    &  -3087/2  &  -64  &  -18225/2  &  -2250  &  -65219/2  &  -864   \\ [\extable]
$R_{4}$    &  -61740/11  &  -768/7  &  -510300/17  &  *  &  -2347884/23  &  -69120/13   \\ [\extable]
$R_{5}$    &  -46305/4  &  -216  &  -25515/4  &  -13500  &  -754677/4  &  -540   \\ [\extable]
$R_{6}$    &  -246960/13  &  -1024/3  &  -2041200/23  &  -144000/7  &  -25872  &  -276480/19   \\ [\extable]
$R_{7}$    &  -15435/8  &  *  &  -127575/8  &  -4500  &  -586971/8  &  *   \\
\hline
\end{tabular}
\end{center}
\captionof{table}{ \label{I_beta_alpha_Zu_anomaly_B_fields}
Partial index from the $\mathbbm{Z}_{\mathrm{g}_u}$ flux.
}
}\endgroup

\begingroup 
\normalsize 
{\scalefont{0.95}

\renewcommand{\extable}{1.25ex}
\begin{center}
\begin{tabular}{||c | c c c c c  c   ||} 
\hline\hline
\multicolumn{7}{||c||}{
$\Phi^\beta_\beta$ 
} \\
\hline
\rule{0pt}{2ex}  
\diagbox[width=3em,height=2em,font=\scriptsize\itshape]{$R_i$}{$N$}& 
  $5$  &  $6$  & $7$  & $8$ & $9$ & $10$        \\ [0.0ex]
\hline\hline
\phantom{-} &  &    &    &     &   &    \\ [-2.5ex]
$R_{3}$    &  8477/2  &  148  &  37503/2  &  4250  &  115313/2  &  1449   \\ [\extable]
$R_{4}$    &  65660/11  &  702/7  &  416178/17  &  *  &  1620432/23  &  44730/13   \\ [\extable]
$R_{5}$    &  3675/4  &  72  &  567/4  &  300  &  3267/4  &  0   \\ [\extable]
$R_{6}$    &  -92120/13  &  -428/3  &  -936684/23  &  -35600/7  &  -149424  &  -152460/19   \\ [\extable]
$R_{7}$    &  -24745/8  &  *  &  -37989/8  &  -100  &  94743/8  &  *   \\
\hline
\end{tabular}
\end{center}
\captionof{table}{ \label{I_beta_Zu_anomaly_B_fields}
Partial index from the $\mathbbm{Z}_{\mathrm{g}_u}$ flux.
}

}\endgroup

\begingroup 
\normalsize 
{\scalefont{0.9}

\renewcommand{\extable}{1ex}
\begin{center}
\begin{tabular}{||c | c c c c    ||} 
\hline\hline
\multicolumn{5}{||c||}{
$\Phi^{\alpha\beta}_\alpha$ 
} \\
\hline
\rule{0pt}{2ex}  
\diagbox[width=3em,height=2em,font=\scriptsize\itshape]{$R_i$}{$N$}& 
  $5$  &  $6$  & $7$  & $8$         \\ [0.0ex]
\hline\hline
\phantom{-} &  &    &    &       \\ [-2.5ex]
$R_{3}$    &       3087 &        128 &      18225 &      4500  \\ [\extable]
$R_{4}$    &  123480/11 &     1536/7 & 1020600/17 &         *  \\ [\extable]
$R_{5}$    &    46305/2 &        432 &    25515/2 &     27000  \\ [\extable]
$R_{6}$    &  493920/13 &     2048/3 & 4082400/23 &  288000/7  \\ [\extable]
$R_{7}$    &    15435/4 &          * &   127575/4 &      9000  \\
\hline
\end{tabular}
\end{center}
\captionof{table}{ \label{I_beta_Zu_anomaly_B_fields}
Partial index from the $\mathbbm{Z}_{\mathrm{g}_u}$ flux.
}

}\endgroup

\begingroup 
\normalsize 
{\scalefont{0.87}

\renewcommand{\extable}{1ex}
\begin{center}
\begin{tabular}{||c | c c c c    ||} 
\hline\hline
\multicolumn{5}{||c||}{
$\Phi^{\alpha\beta}_\beta$ 
} \\
\hline
\rule{0pt}{2ex}  
\diagbox[width=3em,height=2em,font=\scriptsize\itshape]{$R_i$}{$N$}& 
  $5$  &  $6$  & $7$  & $8$         \\ [0.0ex]
\hline\hline
\phantom{-} &  &    &    &       \\ [-2.5ex]
$R_{3}$    &        1323 &         64 &       10125 &     2700  \\ [\extable]
$R_{4}$    &   105840/11 &     1536/7 &  1134000/17 &        *  \\ [\extable]
$R_{5}$    &     19845/2 &       1296 &      8505/2 &    16200  \\ [\extable]
$R_{6}$    &   846720/13 &     4096/3 &  9072000/23 & 345600/7  \\ [\extable]
$R_{7}$    &     -6615/4 &          * &    -70875/4 &    -5400  \\
\hline
\end{tabular}
\end{center}
\captionof{table}{ \label{I_beta_Zu_anomaly_B_fields}
Partial index from the $\mathbbm{Z}_{\mathrm{g}_u}$ flux.
}

}\endgroup

\newpage

\subsection{Twisted one-form gauge fields}
\label{appendix_fluxes_twisted_fields}

The total indices $\mathcal{I}^{(s)}_{\alpha,\beta}$ shown here are obtained by direct evaluation of (\ref{total_index_twisted}). For all the tables in this section, the dash indicates that $N_3 = 0$ for that $N$ (see Table \tcolor{\ref{two_species_table_appendix}}). Theories that are not asymptotically free are denoted by the $\diamond$ symbol (see Tables \ref{table_rep_multiplictly_1} and \ref{table_rep_multiplictly_2}). The cases with $\me_u = 1$ are marked by the * symbol; in these cases the axial center is trivial and thus there is no $s = 5$ twisted sector (see Table \ref{g_u_values}).\\

$\bullet$ \textbf{Flavor center} $\mathbbm{Z}_{N_1}$\textbf{: \textit{s} = 1 twisted sector}\\

\renewcommand{\extable}{1ex}
\begin{center}
\begin{tabular}{||c | c c c c c c     ||} 
\hline\hline
\multicolumn{7}{||c||}{
$\mathcal{I}^{(1)}_\alpha$ induced by 
$\mathbbm{Z}_{N_1} \times \txn{U}_\alpha(1) \times \mathbbm{Z}_{N_3}$
} \\ 
\hline
\rule{0pt}{3ex}  
\diagbox[width=3em,height=2.2em,font=\footnotesize\itshape]{$R_i$}{$N$}
&  $5$  &  $6$  & $7$  & $8$ & $9$ & $10$      \\ [0.0ex]
\hline\hline
\phantom{-}&  &    &    &     &  &    \\ [-2.2ex]
$R_{12}$   &      -3145 &      -2841 &  -9506 &  -3748 & -22563 & -16345 \\ [\extable]
$R_{13}$   &       -445 &      -2799 &  -9338 &  -3676 & -22113 & -16015 \\ [\extable]
$R_{14}$   &      -3175 &      -2883 &  -9674 &   -764 & -23013 & -16675 \\ [\extable]
$R_{15}$   &      -3205 &      -2925 &  -9842 &  -3892 &  -2133 & -17005 \\ [\extable]
$R_{16}$   &      -3235 &      -2967 & -10010 &  -3964 & -23913 &  17225 \\
\hline
\end{tabular}
\end{center}
\captionof{table}{ \label{I_alpha_Z_N1_anomaly_twisted_fields}
Phase induced in the partition function under a $\mathbbm{Z}_{N_1} \times \txn{U}_\alpha(1) \times \mathbbm{Z}_{N_3}$ transformation. 
}

\begingroup
\normalsize

\renewcommand{\extable}{2.5ex}
\begin{center}
\begin{tabular}{||c | c c c c c c     ||} 
\hline\hline
\multicolumn{7}{||c||}{
$\mathcal{I}^{(1)}_\beta$ induced by 
$\mathbbm{Z}_{N_1} \times \txn{U}_\alpha(1) \times \mathbbm{Z}_{N_3}$
} \\
\hline
\rule{0pt}{3ex}  
\diagbox[width=3em,height=2.2em,font=\footnotesize\itshape]{$R_i$}{$N$}
&  $5$  &  $6$  & $7$  & $8$ & $9$ & $10$      \\ [0.0ex]
\hline\hline
\phantom{-}&  &    &    &     &  &    \\ [-2.2ex]
$R_{12}$   &       -735 &       -768 &  -567 & -2400 &  -7623 &  -5760 \\ [\extable]
$R_{13}$   &       -735 &      -2304 & -8505 &  -800 & -22869 & -17280 \\ [\extable]
$R_{14}$   &        735 &        768 &  2835 &  2400 &   1089 &   5760 \\ [\extable]
$R_{15}$   &        735 &       2304 &  8505 &  2400 &  22869 &  17280 \\ [\extable]
$R_{16}$   &       3675 &        768 & 14175 & 12000 &  38115 & -28800 \\
\hline
\end{tabular}
\end{center}
\captionof{table}{ \label{I_beta_Z_N1_anomaly_twisted_fields}
Phase induced in the partition function under a $\mathbbm{Z}_{N_1} \times \txn{U}_\alpha(1) \times \mathbbm{Z}_{N_3}$ transformation. 
}

\endgroup


\newpage

$\bullet$ \textbf{Flavor center} $\mathbbm{Z}_{N_2}$\textbf{: \textit{s} = 2 twisted sector}\\

\renewcommand{\extable}{2.2ex}
\begin{center}
\begin{tabular}{||c | c c c c c c  ||} 
\hline\hline
\multicolumn{7}{||c||}{
$\mathcal{I}^{(2)}_\alpha$ induced by 
$\mathbbm{Z}_{N_2} \times \txn{U}_\beta(1) \times \mathbbm{Z}_{N_3}$
} \\
\hline
\rule{0pt}{3ex}  
\diagbox[width=3em,height=2.5em,font=\footnotesize\itshape]{$R_i$}{$N$} 
&  $5$  &  $6$  & $7$  & $8$ & $9$ & $10$   \\ [0.0ex]
\hline\hline
\phantom{-}&  &    &    &     &  &    \\ [-2.2ex]
$R_{4}$    &       -315 &       -384 & -1575 &  -288 &  -4851 & -3840 \\ [\extable]
$R_{5}$    &       -315 &        -96 & -1575 & -1440 &  -4851 & -1920 \\ [\extable]
$R_{6}$    &       -315 &       -384 & -1575 & -1440 &   -441 & -3840 \\ [\extable]
$R_{8}$    &        -45 &       -384 & -1575 & -1440 &  -4851 & -3840 \\ [\extable]
$R_{9}$    &       -315 &       -192 & -1575 &     - &   4851 &  1920 \\
\hline
\end{tabular}
\end{center}
\captionof{table}{ \label{I_alpha_Z_N2_anomaly_twisted_fields}
Phase induced in the partition function under a $\mathbbm{Z}_{N_2} \times \txn{U}_\beta(1) \times \mathbbm{Z}_{N_3}$ transformation. 
}

\renewcommand{\extable}{2.5ex}
\begin{center}
\begin{tabular}{||c | c c c c c c ||} 
\hline\hline
\multicolumn{7}{||c||}{
$\mathcal{I}^{(2)}_\beta$ induced by 
$\mathbbm{Z}_{N_2} \times \txn{U}_\beta(1) \times \mathbbm{Z}_{N_3}$
} \\
\hline
\rule{0pt}{3ex} 
\diagbox[width=3em,height=2.5em,font=\footnotesize\itshape]{$R_i$}{$N$} 
&  $5$  &  $6$  & $7$  & $8$ & $9$ & $10$      \\ [0.0ex]
\hline\hline
\phantom{-}&  &    &    &     &  &    \\ [-2.2ex]
$R_{4}$    &       -160 &       -279 & -1393 &  -724 &  -5346 &  -4535 \\ [\extable]
$R_{5}$    &        -95 &       -228 &  -441 &  -752 &  -1179 &  -1740 \\ [\extable]
$R_{6}$    &       -410 &       -633 & -3017 & -1532 & -11160 &  -9385 \\ [\extable]
$R_{8}$    &        200 &        339 &   329 &   836 &   6066 &   5075 \\ [\extable]
$R_{9}$    &        115 &        273 &  2583 &     - &  -9189 &  -3795 \\
\hline
\end{tabular}
\end{center}
\captionof{table}{ \label{I_beta_Z_N2_anomaly_twisted_fields}
Phase induced in the partition function under a $\mathbbm{Z}_{N_2} \times \txn{U}_\beta(1) \times \mathbbm{Z}_{N_3}$ transformation. 
}


\newpage

$\bullet$ \textbf{Flavor center} $\mathbbm{Z}_{N_3}$\textbf{: \textit{s} = 3 twisted sector}\\

\begingroup 
\normalsize 
{\scalefont{1} 

\renewcommand{\extable}{0.35ex}
\begin{center}
\begin{tabular}{||c |  c c c c c ||} 
\hline\hline
\multicolumn{6}{||c||}{
$\mathcal{I}^{(3)}_\alpha$ induced by 
$\mathbbm{Z}_{N_3} \times \txn{U}_\alpha(1) \times \txn{U}_\beta(1)$
} \\
\hline
\rule{0pt}{3ex}  
\diagbox[width=3.4em,height=2.4em,font=\scriptsize\itshape]{$R_i$}{$N$}& 
$5$  &  $6$  & $7$  & $8$ & $9$    \\ [0.0ex]
\hline\hline
\phantom{-}&  &    &    &     &     \\ [-2ex]
$R_{3}$    &  -12800  &  -6417  &  -46354  &  -38752  &  -122148   \\ [\extable]
$R_{4}$    &  -18700  &  -20349  &  -77707  &  -13528  &  -219834   \\ [\extable]
$R_{5}$    &  -25720  &  -7398  &  -117124  &  -104528  &  -346032   \\ [\extable]
$R_{6}$    &  -33860  &  -40563  &  -164605  &  -149416  &  -45522   \\ [\extable]
$R_{7}$    &  -4360  &  -747  &  -7840  &  -4976  &  -12312   \\
\hline
\end{tabular}
\end{center}
\captionof{table}{ \label{I_alpha_Z_N3_anomaly_twisted_fields}
Phase induced in the partition function under a $\mathbbm{Z}_{N_3} \times \txn{U}_\alpha(1) \times \txn{U}_\beta(1)$ transformation. 
}

}\endgroup

\begingroup 
\normalsize 
{\scalefont{1} 

\renewcommand{\extable}{0.35ex}
\begin{center}
\begin{tabular}{||c |  c c c c c ||} 
\hline\hline
\multicolumn{6}{||c||}{
$\mathcal{I}^{(3)}_\beta$ induced by 
$\mathbbm{Z}_{N_3} \times \txn{U}_\alpha(1) \times \txn{U}_\beta(1)$
} \\
\hline
\rule{0pt}{3ex}  
\diagbox[width=3.4em,height=2.4em,font=\scriptsize\itshape]{$R_i$}{$N$}& 
$5$  &  $6$  & $7$  & $8$ & $9$    \\ [0.0ex]
\hline\hline
\phantom{-}&  &    &    &     &     \\ [-2ex]
$R_{3}$    &      -5450 &      -3195 &  -25676 &  -23200 &   -77598 \\ [\extable]
$R_{4}$    &     -15950 &     -20286 &  -86156 &  -40520 &  -279450 \\ [\extable]
$R_{5}$    &     -10980 &     -22140 &  -38976 &  -62640 &   -94284 \\ [\extable]
$R_{6}$    &     -57860 &     -80964 & -365288 & -179120 & -1273644 \\ [\extable]
$R_{7}$    &       1840 &        738 &    4312 &    2960 &     7776 \\
\hline
\end{tabular}
\end{center}
\captionof{table}{ \label{I_beta_Z_N3_anomaly_twisted_fields}
Phase induced in the partition function under a $\mathbbm{Z}_{N_3} \times \txn{U}_\alpha(1) \times \txn{U}_\beta(1)$ transformation.
}

} \endgroup


\newpage

$\bullet$ \textbf{Color center} $\mathbbm{Z}_{N} \textbf{: \textit{s} = 4 twisted sector}$\\

\begingroup 
\normalsize 
{\scalefont{1} 

\renewcommand{\extable}{0.35ex}
\begin{center}
\begin{tabular}{||c |  c c c c c c ||} 
\hline\hline
\multicolumn{7}{||c||}{
$\mathcal{I}^{(4)}_\alpha$ induced by 
$\mathbbm{Z}_{N} \times \txn{U}_\alpha(1) \times \txn{U}_\beta(1) \times \mathbbm{Z}_{N_3}$
} \\
\hline
\rule{0pt}{3ex}  
\diagbox[width=3.4em,height=2.4em,font=\scriptsize\itshape]{$R_i$}{$N$}& 
$5$  &  $6$  & $7$  & $8$ & $9$ & $10$           \\ [0.0ex]
\hline\hline
\phantom{-}&  &    &    &     &   &   \\ [-2ex]
$R_{3}$    &       -158 &        -53 &  -274 &  -172 &  -422 &  -127 \\ [\extable]
$R_{4}$    &       -346 &       -233 &  -604 &   -76 &  -934 &  -563 \\ [\extable]
$R_{5}$    &       -604 &       -102 & -1060 &  -668 & -1644 &  -496 \\ [\extable]
$R_{6}$    &       -932 &       -631 & -1642 & -1036 &  -232 & -1541 \\ [\extable]
$R_{7}$    &          8 &          1 &     8 &     4 &     8 &     2 \\
\hline
\end{tabular}
\end{center}
\captionof{table}{ \label{I_alpha_ZN_anomaly_twisted_fields}
Phase induced in the partition function under a $\mathbbm{Z}_{N} \times \txn{U}_\alpha(1) \times \txn{U}_\beta(1) \times \mathbbm{Z}_{N_3}$ transformation.
}

}\endgroup

\begingroup 
\normalsize 
{\scalefont{1} 

\renewcommand{\extable}{0.5ex}
\begin{center}
\begin{tabular}{||c |  c c c c c c ||} 
\hline\hline
\multicolumn{7}{||c||}{
$\mathcal{I}^{(4)}_\beta$ induced by 
$\mathbbm{Z}_{N} \times \txn{U}_\alpha(1) \times \txn{U}_\beta(1) \times \mathbbm{Z}_{N_3}$
} \\
\hline
\rule{0pt}{3ex}  
\diagbox[width=3.4em,height=2.4em,font=\scriptsize\itshape]{$R_i$}{$N$}& 
$5$  &  $6$  & $7$  & $8$ & $9$ & $10$           \\ [0.0ex]
\hline\hline
\phantom{-}&  &    &    &     &   &   \\ [-2ex]
$R_{3}$    &        -62 &        -25 &  -146 &  -100 &  -262 &   -83 \\ [\extable]
$R_{4}$    &       -284 &       -226 &  -656 &  -224 & -1172 &  -742 \\ [\extable]
$R_{5}$    &       -252 &       -300 &  -348 &  -396 &  -444 &  -492 \\ [\extable]
$R_{6}$    &      -1568 &      -1244 & -3608 & -1232 & -6448 & -4084 \\ [\extable]
$R_{7}$    &         -8 &         -2 &    -8 &    -4 &    -8 &    -1 \\
\hline
\end{tabular}
\end{center}
\captionof{table}{ \label{I_beta_ZN_anomaly_twisted_fields}
Phase induced in the partition function under a $\mathbbm{Z}_{N} \times \txn{U}_\alpha(1) \times \txn{U}_\beta(1) \times \mathbbm{Z}_{N_3}$ transformation. 
}

} \endgroup

\newpage

$\bullet$ \textbf{Axial} $ \txn{U}_\alpha(1) \times \txn{U}_\beta(1)  $
\textbf{center}\textbf{: \textit{s} = 5 twisted sector}  \\

In some cases $\mathrm{g}_u = 1$, this renders the $\mathbbm{Z}_{\me_u}$ center trivial, for which case there is no twisted sector $s = 5$. These cases are shown for some values of $N$ in Table \ref{g_u_values}. \\

\renewcommand{\extable}{0.3ex}
\begin{center}
\begin{tabular}{||c | c c c c c c c c c c c c c c   ||} 
\hline\hline
\multicolumn{15}{||c||}{
${\mathrm{g}_u}$
} \\
\hline
\rule{0pt}{2ex}  
\diagbox[width=3em,height=2em,font=\scriptsize\itshape]{$R_i$}{$N$}& 
$5$ &$6$ &$7$ &$8$ &$9$ &$10$& $11$ &$12$ &$13$ &$14$& $15$ &$16$ &$17$ &$18$  \\ [0.5ex]
\hline\hline
\phantom{-}&  &  &  &  &  &  &  &  &  &  &  &  &  &       \\ [-2.2ex]
$R_{3}$   &         10 &          3 & 14 &  8 & 18 &  5 & 22 & 12 & 26 &  7 & 30 & 16 & 34 &  9  \\ [\extable]
$R_{4}$   &         11 &          7 & 17 &  1 & 23 & 13 & 29 &  8 &  7 & 19 & 41 & 11 & 47 &  5  \\ [\extable]
$R_{5}$   &          4 &          2 &  4 &  4 &  4 &  4 &  4 &  4 &  4 &  1 &  4 &  4 &  4 &  4  \\ [\extable]
$R_{6}$   &         13 &          9 & 23 &  7 &  3 & 19 & 43 &  6 & 53 & 29 & 63 & 17 & 73 & 39  \\ [\extable]
$R_{7}$   &          8 &          1 &  8 &  4 &  8 &  1 &  8 &  4 &  8 &  1 &  8 &  4 &  8 &  1  \\ [\extable]
$R_{8}$   &          1 &          3 &  1 &  1 &  3 &  1 &  1 &  - &  1 &  1 &  3 &  1 &  1 &  3  \\ [\extable]
$R_{9}$   &          2 &          1 &  2 &  - &  2 &  1 &  2 &  4 &  2 &  1 & 14 &  8 &  2 &  1  \\ [\extable]
$R_{10}$  &          5 &          1 &  1 &  1 &  1 &  5 &  1 &  2 & 19 & 11 & 25 &  1 & 31 & 17  \\ [\extable]
$R_{11}$  &          4 &          - &  4 &  4 & 12 &  2 &  4 & 12 & 28 &  2 & 36 &  4 & 44 &  3  \\ [\extable]
$R_{12}$  &         19 &         11 &  5 &  7 & 31 & 17 & 37 &  2 & 43 & 23 & 49 & 13 & 11 & 29  \\ [\extable]
$R_{13}$  &          1 &         13 & 31 &  1 & 41 & 23 & 17 &  2 & 61 & 11 & 71 & 19 &  9 & 43  \\ [\extable]
$R_{14}$  &         17 &          9 & 19 &  1 &  3 & 11 & 23 &  6 &  5 & 13 & 27 &  1 & 29 &  3  \\ [\extable]
$R_{15}$  &          5 &          7 & 13 &  1 &  1 &  5 &  1 &  2 &  7 &  1 &  5 &  1 &  1 &  1  \\ [\extable]
$R_{16}$  &         13 &          1 &  7 &  1 &  1 &  1 &  1 &  2 &  1 &  7 &  1 &  1 & 23 & 13  \\ [\extable]
$R_{17}$  &          4 &          4 &  4 &  4 &  4 &  2 &  4 &  4 &  4 &  4 &  4 &  4 &  4 &  1  \\ [\extable]
$R_{18}$  & $\diamond$ & $\diamond$ & 13 & 11 &  7 &  3 & 59 & 16 & 23 & 37 & 79 &  1 & 89 & 47  \\ [\extable]
$R_{19}$  &         26 &          7 &  2 & 16 & 34 &  1 & 38 &  4 & 14 & 11 & 46 &  8 & 10 & 13  \\ [\extable]
$R_{20}$  &         25 &         13 &  1 &  7 & 29 &  5 & 31 &  8 &  1 & 17 & 35 &  1 & 37 & 19  \\ [\extable]
$R_{21}$  & $\diamond$ &         11 &  7 &  1 & 19 &  1 & 17 &  2 &  1 &  7 &  1 &  1 & 11 &  1  \\ [\extable]
$R_{22}$  & $\diamond$ &          1 &  2 &  8 &  2 &  1 &  2 &  4 &  2 &  1 &  2 &  - &  2 &  1  \\ [\extable]
\hline
\diagbox[width=3em,height=2.2em,font=\footnotesize\itshape]{$R_i$}{$N$}
&$40$ &$41$ &$42$ &$43$ &$44$ &$45$& $46$ &$47$ &$48$ &$49$ &$50$ &$51$ &$52$ &$53$   \\ [0.5ex]
\hline\hline
\phantom{-}&  &  &  &  &  &  &  &  &  &  &  &  &  &      \\ [-2.2ex]
$R_{23}$  & 53 & 217 & 111 & 227 & 29 & 237 & 11 & 247 &  63 & 257 & 131 & 267 &  34 & 277  \\ [\extable]
$R_{24}$  &  5 &  11 &  73 & 149 & 19 & 155 & 79 &  23 &  41 & 167 &  85 & 173 &  22 & 179  \\ [\extable]
$R_{25}$  & 17 &  23 &   7 &  71 &  1 &  73 & 37 &   5 &  19 &  77 &   1 &  79 &   2 &   9  \\ [\extable]
$R_{26}$  &  1 &   1 &   3 &   7 &  1 &   9 &  1 &  11 &   3 &  13 &   7 &   3 &   2 &   1  \\ [\extable]
$R_{27}$  & 92 & 188 &  24 & 196 & 20 & 204 & 13 & 212 & 108 &  44 &  14 & 228 & 116 & 236  \\ [\extable]
$R_{28}$  & 37 & 151 &   7 & 157 &  8 & 163 & 83 & 169 &  43 &  35 &  89 & 181 &  46 &   1  \\ [\extable]
$R_{29}$  &  8 &  38 &  29 & 118 &  4 & 122 & 31 &   2 &  64 &  26 &  11 & 134 &  68 &  46  \\ [\extable]
$R_{30}$  &  1 &  77 &  39 &  79 &  2 &  81 & 41 &  83 &  21 &   1 &  43 &  87 &  11 &  89  \\ [\extable]
\hline
\end{tabular}
\end{center}
\captionof{table}{ \label{g_u_values}
Values for $\mathrm{g}_u$ at different $N$.
}

\begingroup 
\normalsize 
{\scalefont{0.87}

\renewcommand{\extable}{0.7ex}
\begin{center}
\begin{tabular}{||c |  c c c c c c ||} 
\hline\hline
\multicolumn{7}{||c||}{
$\mathcal{I}^{(5)}_\alpha$ induced by 
$\txn{U}_\alpha(1) \times \txn{U}_\beta(1)$
} \\
\hline
\rule{0pt}{3ex}  
\diagbox[width=3.4em,height=2.4em,font=\scriptsize\itshape]{$R_i$}{$N$}& 
$5$  &  $6$  & $7$  & $8$ & $9$ & $10$          \\ [0.0ex]
\hline\hline
\phantom{-}&  &    &    &     &   &   \\ [-2ex]
$R_{3}$    &       1350 &         63 &   9800 &  2592 &   39690 &   1100 \\ [\extable]
$R_{4}$    &       5940 &        147 &  47600 &     * &  202860 &  11440 \\ [\extable]
$R_{5}$    &      14580 &        378 &  14000 & 34992 &  555660 &   1760 \\ [\extable]
$R_{6}$    &      28080 &        756 & 257600 & 72576 &  105840 &  66880 \\ [\extable]
$R_{7}$    &       1080 &          * &   5600 &  1296 &   17640 &      * \\ [\extable]
$R_{8}$    &          * &         63 &      * &     * &   26460 &      * \\ [\extable]
$R_{9}$    &       7290 &          * &   1400 &     - &  -39690 &      * \\ [\extable]
$R_{10}$   &      10800 &          * &      * &     * &       * & -17600 \\ [\extable]
$R_{11}$   &      13500 &          - & -70000 & -1296 & -661500 & -22000 \\ [\extable]
$R_{12}$   &       5130 &        462 &  35000 &  4536 &  136710 &   1870 \\ [\extable]
$R_{13}$   &          * &       4914 &  43400 &     * & 1627290 &   2530 \\ [\extable]
$R_{14}$   &       4590 &        378 &  26600 &     * &   92610 &   1210 \\ [\extable]
$R_{15}$   &      36450 &       2646 &  18200 &     * &       * &    550 \\ [\extable]
$R_{16}$   &      87750 &          * & 245000 &     * &       * &      * \\ [\extable]
$R_{17}$   &        180 &        252 &   8400 &   432 &   26460 &   2640 \\ [\extable]
$R_{18}$   & $\diamond$ & $\diamond$ & 109200 &  9504 & 1296540 &   7920 \\ [\extable]
$R_{19}$   &       1170 &        441 &  21000 &  1728 &  224910 &      * \\ [\extable]
$R_{20}$   &       4500 &        819 &      * &  6048 &  767340 &  13200 \\ [\extable]
$R_{21}$   & $\diamond$ &       2772 & 235200 &     * & 2010960 &      * \\ [\extable]
$R_{22}$   & $\diamond$ &          * & 105000 &   864 & 2315250 &      * \\ [\extable]
\hline
\diagbox[width=3em,height=2.2em,font=\footnotesize\itshape]{$R_i$}{$N$}
          &  $40$ &  $41$ & $42$   & $43$   & $44$  & $45$       \\ [1.2ex]
\hline
\phantom{-}&  &    &    &     &  &     \\ [-2.2ex]
$R_{23}$   &   62756240 & 1136716308 &  10023300 &  1443925208 &   50644440 &  1814201820 \\ [\extable]
$R_{24}$   &    5920400 &  749080332 &   6591900 &   947774696 &   33180840 &  1186503300 \\ [\extable]
$R_{25}$   &   20129360 & 3252999204 &  28444500 &   451624184 &          * &  5029243020 \\ [\extable]
$R_{26}$   &          * &          * &  -6772500 &   -44526328 &          * & -1722343500 \\ [\extable]
$R_{27}$   &  136169200 & 1231006140 &    866880 &  1558421480 &   43659000 &  1951989300 \\ [\extable]
$R_{28}$   &  438109600 & 3954934620 &    505680 &  4993309640 &  139708800 &  6238710900 \\ [\extable]
$R_{29}$   &   11840800 & 6718150530 &   4713660 &   938233340 &  235758600 & 10506295350 \\ [\extable]
$R_{30}$   &          * & 8067018960 &  11269440 & 10050228320 &  279417600 & 12400873200 \\ [\extable]
\hline
\end{tabular}
\end{center}
\captionof{table}{ \label{I_alpha_Ua_Ub_anomaly_twisted_fields}
Phase induced in the partition function under a $\txn{U}_\alpha(1) \times \txn{U}_\beta(1)$ transformation. 
}

}\endgroup

\begingroup 
\normalsize 
{\scalefont{0.87} 

\renewcommand{\extable}{0.7ex}
\begin{center}
\begin{tabular}{||c |  c c c c c c ||} 
\hline\hline
\multicolumn{7}{||c||}{
$\mathcal{I}^{(5)}_\beta$ induced by 
$\txn{U}_\alpha(1) \times \txn{U}_\beta(1)$
} \\
\hline
\rule{0pt}{3ex} 
\diagbox[width=3.4em,height=2.4em,font=\scriptsize\itshape]{$R_i$}{$N$}& 
$5$  &  $6$  & $7$  & $8$ & $9$ & $10$        \\ [0.0ex]
\hline\hline
\phantom{-}&  &    &    &     &   &   \\ [-2ex]
$R_{3}$    &       4900 &        180 &    23814 &   5600 &    78408 &    2025 \\ [\extable]
$R_{4}$    &      10780 &        210 &    57834 &      * &   200376 &   10530 \\ [\extable]
$R_{5}$    &       5880 &        720 &     2268 &   8400 &    52272 &     540 \\ [\extable]
$R_{6}$    &      25480 &        540 &   156492 &  19600 &   574992 &   30780 \\ [\extable]
$R_{7}$    &      -3920 &          * &   -13608 &  -2800 &   -34848 &       * \\ [\extable]
$R_{8}$    &          * &        -90 &        * &      * &   -26136 &       * \\ [\extable]
$R_{9}$    &      -2940 &          * &    -1134 &      - &    26136 &       * \\ [\extable]
$R_{10}$   &      -9800 &          * &        * &      * &        * &    8100 \\ [\extable]
$R_{11}$   &      -9800 &          - &    34020 &    560 &   261360 &    4050 \\ [\extable]
$R_{12}$   &      37240 &       2640 &    34020 &  39200 &   540144 &    6885 \\ [\extable]
$R_{13}$   &          * &       9360 &    70308 &      * &  2143152 &    3105 \\ [\extable]
$R_{14}$   &     -33320 &      -2160 &  -129276 &      * &   -52272 &   -4455 \\ [\extable]
$R_{15}$   &     -29400 &      -5040 &   -29484 &      * &        * &    -675 \\ [\extable]
$R_{16}$   &    -127400 &          * &  -238140 &      * &        * &       * \\ [\extable]
$R_{17}$   &      13720 &       4320 &   551124 &  14000 &  1724976 &   21870 \\ [\extable]
$R_{18}$   & $\diamond$ & $\diamond$ &  1194102 &  15400 &   548856 &  196830 \\ [\extable]
$R_{19}$   &     -12740 &      -3780 &   -91854 & -11200 & -1332936 &       * \\ [\extable]
$R_{20}$   &     -24500 &      -3510 &        * &  -9800 & -2273832 & -109350 \\ [\extable]
$R_{21}$   & $\diamond$ &      -5940 & -1285956 &      * & -2979504 &       * \\ [\extable]
$R_{22}$   & $\diamond$ &          * & -1377810 &  -1120 &  -392040 &       * \\ [\extable]
\hline
\diagbox[width=3em,height=2.2em,font=\footnotesize\itshape]{$R_i$}{$N$}
          &  $40$ &  $41$ & $42$   & $43$   & $44$  & $45$       \\ [1.2ex]
\hline
\phantom{-}&  &    &    &     &  &     \\ [-2.2ex]
$R_{23}$   &  583390080 &   5264176960 &  46256364 &   6641384400 &   928805504 &   8292762720 \\ [\extable]
$R_{24}$   & -385257600 &   -266847680 & -30420852 &  -4359322800 &  -608527744 &  -5423536800 \\ [\extable]
$R_{25}$   &  -62375040 &  -1673862720 &  -8751204 &   -692420400 &           * &  -7662932640 \\ [\extable]
$R_{26}$   &          * &            * &   6250860 &     40960080 &           * &   1574575200 \\ [\extable]
$R_{27}$   & -791154000 &  -7126046000 &  -2500344 &  -8960017500 & -1251085000 & -11153241000 \\ [\extable]
$R_{28}$   & -636363000 & -11447159000 & -16043874 & -14354313750 & -1000868000 & -17823316500 \\ [\extable]
$R_{29}$   & -160524000 &  -4321113000 &  -9063747 &  -1798098750 &  -750651000 & -20010226500 \\ [\extable]
$R_{30}$   &          * & -11674586000 & -16252236 & -14445742500 &  -500434000 & -17713971000 \\ [\extable]
\hline
\end{tabular}
\end{center}
\captionof{table}{ \label{I_beta_Ua_Ub_anomaly_twisted_fields}
Phase induced in the partition function under a $\txn{U}_\alpha(1) \times \txn{U}_\beta(1)$ transformation. 
}

}\endgroup

%% file: Appendix_A.tex
\newpage

\section{The 3-loop beta function}
\label{appendix_beta_func}

The Dynkin index for the representation $\mathcal{R}_i$ of SU($N$) is defined as
\begin{align}
\txn{Tr}\big(   \mathcal{T}^a_i \mathcal{T}^b_i \big)
=
T_i \esp \delta^{ab}
\spa\spa,
\label{SU(N)_normalization_1}
\end{align}
where $\mathcal{T}^a_i$ are the generators of the group and $G$ denotes the adjoint representation. In this work we follow the same normalization condition for the SU($N$) generators as in \cite{Eichten:Kang:Koh:1982}, in which the Dynkin index for the fundamental representation is $T_F$ = 1/2. The quadratic Casimir operator is defined as 
\begin{align}
\mathcal{T}^a_i \mathcal{T}^a_i 
= C_2(\mathcal{R}_i) \esp \mathbb{1}_\mathcal{R}
\spa\spa,
\end{align}

which can be written as follows
\begin{align}
C_2(\mathcal{R}_i) 
=
\frac{\txn{dim}(G)}{\txn{dim}(\mathcal{R}_i)} \esp T_i
\spa\spa,
\end{align}

in particular, $\txn{dim}(G) = N^2 -1$ and $C_2(G) = N$.\\

The $\beta$ function with coupling $\alpha = g^2/4\pi$ for $N_i$ Weyl fermions in the representation $\mathcal{R}_i$ is 
\begin{align}
\beta(\alpha) 
= \mu^2 \frac{d\alpha}{d\mu^2}
=
\alpha \sum_{n=0}^\infty \beta_n \bigg( \frac{\alpha}{4\pi} \bigg)^n
\spa\spa,
\label{appendix_complete_beta_function}
\end{align}

up to three loops, the contributions are\hlfootnote{For Dirac fermions there would be an extra factor of 2 for each $N_i$.}  
\tcolor{\cite{Caswell:1974,Tarasov:Vladimirov:Zharkov:1980,Machacek:Vaughn:1983,Larin:Vermaseren:1993,Zoller:2016}}

\begin{align*}
\beta_0 &= 
-\frac{11}{3} C_2(G) + \sum_i \frac{2}{3} |N_i| T_i
\spa\spa,
\\
\beta_1 &= 
- \frac{34}{3} C_2^2(G) 
+ \sum_i |N_i| T_i \bigg( 
\frac{10}{3} C_2(G)  + 2 C_2(R_i) 
\bigg)
\spa\spa,
\numberthis
\label{beta_f_coefficients}
\\
\beta_2 &= 
-\frac{2857}{54} C_2^3(G)
+ \sum_i  |N_i| T_i 
\bigg( 
-  C_2^2(R_i) + \frac{205}{18}  C_2(G) C_2(R_i)
+ \frac{1415}{54} C_2^2(G)
\bigg)
\\
&-
\frac{1}{4} \sum_{i,j} |N_i||N_j| T_i   T_j
\bigg( \frac{44}{9} C_2(R_i) + \frac{158}{27} C_2(G) \bigg) 
\spa\spa.
\end{align*}

Note that when computing values of the 't Hooft coupling $\lambda = g^2 N$ at possible fixed points, in order to establish whether a coupling is small, we must specify the normalization condition for the generators being used. Recall that in the fundamental QCD vertex, for each factor of the coupling $g$ there is also a factor of the generators.

%% file: Appendix_bar_R_models.tex
\setlength{\jot}{13pt}

\section{Asymptotically free theories with finite $N$} \label{appendix_bar_R_models}

The authors of \cite{Eichten:Kang:Koh:1982} also list theories of the same kind as those studied in this work, but for which asymptotic freedom is guaranteed for a finite range of $N$ only. These models, labeled by $\overline{R}_i$, are shown in Table \ref{table_overline_R_i_models}. The values for $N$ in Table \ref{table_overline_R_i_models} differ marginally from those presented in \cite{Eichten:Kang:Koh:1982}. Here we list the values for which $\beta_0 < 0$ and $\beta_1 > 0$ when the inequality for $N$ is saturated such that asymptotic freedom is attained at one loop (see Appendix \ref{appendix_beta_func}). In some cases $\me_\gamma$ is not fully realized. Concretely, for $\overline{R}_5$ and $\overline{R}_{13}$ in practice $\me_\gamma = 1$ only. Similarly, for $\overline{R}_3$, $\overline{R}_9$ and $\overline{R}_{11}$, $\me_\gamma = 1,4$. Also, $\me_\gamma = 1$ for $\overline{R}_{22}$.\\

\renewcommand{\extable}{0.15ex}
\begin{center}
\begin{tabular}{||c c c c c c||} 
\hline\hline
\rule{0pt}{3ex}  
$\overline{R}_i$  & 
$N_1$  & 
$N_2$  &
$N_3$  &
$N$  &
$\me_\gamma$  \\ [0.6ex]
\hline\hline
\phantom{-} &  &  &   & &  \\ [-2ex]
$\overline{R}_1$ &  1  &  5  &  $6N-16$  & $ N \leq 23$ & B \\ [\extable]
$\overline{R}_2$ &  1  &  6  &  $7N-20$  & $ N \leq 9$ & A\\ [\extable]
$\overline{R}_3$ &  1  &  7  &  $8N-24$  & $ N \leq 7$ & B \\ [\extable]
$\overline{R}_4$ &  1  &  8  &  $9N-28$  & $ N = 5 $ & A \\ [\extable]
$\overline{R}_5$ &  1  &  9  &  $10N-32$  & $ N = 5$ & B \\ [\extable]
$\overline{R}_6$ &  1  &  -6  &  $-5N+28$  & $ N \leq 37$ & A \\ [\extable]
$\overline{R}_7$ &  1  &  -7  &  $-6N+32$  & $ N \leq 14$ & B \\ [\extable]
$\overline{R}_8$ &  1  &  -8  &  $-7N+36$  & $ N \leq 9$ & A \\ [\extable]
$\overline{R}_9$ &  1  &  -9  &  $-8N+40$  & $ N \leq 7$ & B \\ [\extable]
$\overline{R}_{10}$ &  1  &  -10  &  $-9N+44$  & $ N \leq 6$ & A \\ [\extable]
$\overline{R}_{11}$ &  1  &  -11  &  $-10N+48$  & $ N \leq 6$ & B \\ [\extable]
$\overline{R}_{12}$ &  1  &  -12  &  $-11N+52$  & $ N = 5$ & A \\ [\extable]
$\overline{R}_{13}$ &  1  &  -13  &  $-12N+56$  & $ N = 5$ & B \\ [\extable]
$\overline{R}_{14}$ &  1  &  -14  &  $-13N+60$  & $ N = 5$ & A \\ [\extable]
$\overline{R}_{15}$ &  2  &  5  &  $7N-12$  & $ N = 5$ & A \\ [\extable]
$\overline{R}_{16}$ &  2  &  -7  &  $-5N+36$  & $ N \leq 15$ & A \\ [\extable]
$\overline{R}_{17}$ &  2  &  -9  &  $-7N+44$  & $ N \leq 8$ & A \\ [\extable]
$\overline{R}_{18}$ &  2  &  -11  &  $-9N+52$  & $ N \leq 6$ & A \\ [\extable]
$\overline{R}_{19}$ &  3  &  -7  &  $-4N+40$  & $ 7 \leq N \leq 15$ & B \\ [\extable]
$\overline{R}_{20}$ &  3  &  -8  &  $-5N+44$  & $ 7 \leq N \leq 10$ & A \\ [\extable]
$\overline{R}_{21}$ &  4  &  -7  &  $-3N+44$  & $ 13 \leq N \leq 16$ & A \\ [\extable]
$\overline{R}_{22}$ &  5  &  -6  &  $-N+44$  & $ 43 \leq N \leq 45$ & A \\ [\extable]
\hline
\end{tabular}
\end{center}
\captionof{table}{ \label{table_overline_R_i_models}
Matter content that yields an anomaly-free and asymptotically free theory. Here  $\overline{R}_i$ labels each field theory. Negative $N_i$ indicates the appearance of $|N_i|$ copies of the associated complex-conjugate representation. The maximum multiplicity is $\ell = 1$ with $N \geq 5$. The $\textnormal{g}_\gamma$ factor determines the center symmetry $\mathbbm{Z}_{N_3 / \textsl{g}_\gamma} $ (see Table \tcolor{\ref{table_3flavors_Z_gamma}}).\\
}

There are no special cases in the $\overline{R}_i$ set such as $R_3$ or $R_7$. The constraint $N_1 = N_2 = 1$ with $N_3$ even yields a symmetry breaking pattern for $R_3$ in which the only flavor group SU$(N_3)$ breaks diagonally. For $R_7$, the fact that $N_1 + N_2 = 0$ and $N_3 = 8$ allows one to build a tower of operators distinguished by a single U(1) quantum number; these operators match the zero-form anomalies in a scenario where tumbling takes place.\\

Given that the matter content is fixed to satisfy the non-trivial gauge anomaly cancellation condition, and in particular that $N_3 \sim N$, it is not possible to take the large $N$ limit while preserving asymptotic freedom. This is contrary to the vector-like case, where for a theory with $N_f$ Weyls under both SU$(N)$ irreps $\mathbf{R}$ and $\mathbf{R}$*, we can take $N \to \infty$ with $N_f/N$ fixed. This way the theory remains asymptotically free and gauge anomaly cancellation is trivially satisfied for any $N_f$. For the allowed $N$ values such that the $\overline{R}_i$ theories remain asymptotically free, we can compute the 't Hooft coupling at the two-loop fixed point. In Table \ref{table_Banks_Zaks_overline_R} we report the smallest coupling in the allowed $N$ range for each model. For $N \sim \mathcal{O}(1)$, some of these couplings are around 5 times smaller than the smallest coupling reported in Table \ref{table_Banks_Zaks} for the $R_i$ family (computed in the $N \to \infty$ limit).\\

Some of these models contain 't Hooft anomalies of the Witten type (see Section \ref{section_witten_anomalies}). We list these anomalies in Table \ref{table_appendix_witten_anomalies_overline_R}. We can also compute the total Dirac indices in each twisted sector for the $\overline{R}_i$ family. While the anomalies computed here are also valid for these models, we have not studied any of them in detail.\\

\renewcommand{\extable}{0.7ex}
\begin{center}
\begin{tabular}{||c | c   ||} 
\hline\hline
\rule{0pt}{3ex}  
$N_i$ & $(\overline{R}_i , N )$  \\ [0.5ex]
\hline\hline
\phantom{-} &   \\ [-2ex]
$N_1$  & $(\overline{R}_{15} , 5 )$ \esp,\esp $(\overline{R}_{16} , (5,6,9,10,13,14) )$ \esp,\esp $(\overline{R}_{17} , (5,6) )$ \esp,\esp $(\overline{R}_{18} , (5,6) )$ \\ [\extable]
$N_2$  & - \\ [\extable]
$N_3$  & $(\overline{R}_{7} , 5 )$ \esp,\esp $(\overline{R}_{11} , 5 )$ \\ [\extable]
\hline
\end{tabular}
\end{center}
\captionof{table}{ \label{table_appendix_witten_anomalies_overline_R}
Models $\overline{R}_i$ for which there is a Witten anomaly at particular $N$.\\
}

\phantom{-}

\renewcommand{\extable}{0.8ex}
\begin{minipage}[c]{0.5\textwidth}
\begin{center}
\begin{tabular}{||c c c  c  c||} 
\hline\hline
\rule{0pt}{3ex}  
$\overline{R}_i$ & $N$ & $N_3$ & $\lambda_0$ & $\eps$ \\ [0.3ex]
\hline\hline
\phantom{-} &  & & &  \\ [-2ex]
$\overline{R}_{1}$   &  23 & 122 & 0.000963  &  0.00400     \\ [\extable]
$\overline{R}_{2}$   &  9 & 43 &  0.007844  &  0.03377    \\ [\extable]
$\overline{R}_{3}$   &  7 & 32 &  0.003284  &  0.01396     \\ [\extable]
$\overline{R}_{4}$   &  5 & 17 &  0.041215  &  0.20030    \\ [\extable]
$\overline{R}_{5}$   &  5 & 18 &  0.015170  &  0.06809  \\ [\extable]
$\overline{R}_{6}$   &  37 & -157 & 0.000575  &  0.00233     \\ [\extable]
$\overline{R}_{7}$   &  14 & -52&  0.003106 &  0.01275    \\ [\extable]
$\overline{R}_{8}$   &  9 & -27 &  0.012986  &  0.05565  \\ [\extable]
$\overline{R}_{9}$   &  7 & -16 &  0.025411  &  0.11407  \\ [\extable]
$\overline{R}_{10}$  &  6 & -10 &  0.036267  &  0.16876  \\ [\extable]
$\overline{R}_{11}$  &  6 & -12 &  0.007609  &  0.03222 \\ [\extable]
\hline
\end{tabular}
\end{center}
\end{minipage}
\begin{minipage}[c]{0.5\textwidth}
\begin{center}
\begin{tabular}{||c c c  c  c||} 
\hline\hline
\rule{0pt}{3ex}  
$\overline{R}_i$ & $N$ & $N_3$ & $\lambda_0$ & $\eps$ \\ [0.3ex]
\hline\hline
\phantom{-} &  & & &  \\ [-2ex]
$\overline{R}_{12}$  &  5 & -3 &  0.054348  &  0.26661  \\ [\extable]
$\overline{R}_{13}$  &  5 & -4 &  0.025837  &  0.11719 \\ [\extable]
$\overline{R}_{14}$  &  5 & -5 &  0.004516  &  0.01901  \\ [\extable]
$\overline{R}_{15}$  &  5 & 23 &  0.014810 &  0.06538  \\ [\extable]
$\overline{R}_{16}$  &  15 & -39 & 0.001373  &  0.00545 \\ [\extable]
$\overline{R}_{17}$  &  8 & -12 &  0.005291  &  0.02139  \\ [\extable]
$\overline{R}_{18}$  &  5 & 7 &  0.004378  &  0.01806  \\ [\extable]
$\overline{R}_{19}$  &  15 & -20 &  0.004032  &  0.01583  \\ [\extable]
$\overline{R}_{20}$  &  7 & 9  & 0.002920  &  0.01155     \\ [\extable]
$\overline{R}_{21}$  &  13 & 5 & 0.001476  &  0.00562   \\ [\extable]
$\overline{R}_{22}$  &  45 & -1 & 0.000414  &  0.00155     \\ [\extable]
\hline
\end{tabular}
\end{center}
\end{minipage}
\captionof{table}{
\label{table_Banks_Zaks_overline_R}
't Hooft coupling $\lambda_0$ at the two-loop fixed point and the stability parameter $\eps$ for the $\overline{R}_i$ family.
}

%% file: Appendix_B.tex
\newcommand{\TwoTwoIrrep}{
\raisebox{2.5pt}{
\scaleobj{0.28}{
\begin{ytableau}
~ &  \\ 
~ &
\end{ytableau}}}
}

\newcommand{\TwoOneOneIrrep}{
\raisebox{6.5pt}{
\scaleobj{0.28}{
\begin{ytableau}
~ ~ &  \\ 
~  \\
~ 
\end{ytableau}}}
}

\newcommand{\OneOneOneOneIrrep}{
\raisebox{9.75pt}{
\scaleobj{0.28}{
\begin{ytableau}
~  \\ 
~  \\
~  \\
~  
\end{ytableau}}}
}

\newpage

\section{The Most Attractive Channel hypothesis}
\label{appendix_clases_MAC}

One possible scenario for the low energy dynamics is one where the gauge group is broken. For the class of chiral gauge theories studied in this work,
this is briefly discussed in Section \tcolor{\ref{large_N_negative_N2_and_tumbling}}. For instance, see \tcolor{\cite{Goity:Peccei:Zeppenfeld:1985}} for an example where a set of massless composite baryons that follow from tumbling and complementarity \tcolor{\cite{Fradkin:Shenker:1979,Elitzur:1975,Banks:Rabinovici:1979,Osterwalder:Seiler:1978}} is presented as a solution for the 't Hooft anomaly matching conditions for the theory $R_7$.\\

The breaking of the gauge group can take place in a single step where the gauge group gets completely broken, or in a series of so-called tumbling steps \tcolor{\cite{Raby:Dimopoulos:Susskind:1980:tumbling}}, depending on the particular colored bosonic operator that condenses and breaks the gauge group. Given a set of UV Weyl fermion operators, it has been conjectured that the colored fermion bilinear that can be built out of these operators and that is responsible for the breaking of the gauge group is that in the Most Attractive Channel (MAC) \tcolor{\cite{Raby:Dimopoulos:Susskind:1980:tumbling,Dimopoulos:Raby:Susskind:1980:light_composite}}. The MAC hypothesis states that the colored composite that breaks the gauge group is the one where the tree-level potential $V$ between a pair of Weyl fermions is minimized

\begin{align}
V \sim g^2(m) \esp \big( C_c - C_1 - C_2 \big)
\spa\spa.
\end{align}

Here $C_c$ denotes the Casimir operator for the colored composite and $C_i$ the Casimir operator for the Weyl fermions that form the composite. This channel is attractive when $C_c < C_1 + C_2$, so the MAC is given by the minimum $V$.\\

For vector-like theories where Weyl fermions are available in both a representation $\mathbf{R}$ and its complex conjugate $\overline{\mathbf{R}}$,  we can build composites such as $\mathbf{R} \otimes \overline{\mathbf{R}}$. Usually, on general group-theoretical grounds, $V$ is minimized by singlet operators $\mathbf{R} \otimes \overline{\mathbf{R}}$. The breaking of the gauge group is triggered by colored operators, and thus if a color singlet minimizes $V$, then the breaking of the gauge group is not likely to take place, at least as the MAC hypothesis states.  For chiral gauge theories, we can't build composites like $\mathbf{R} \otimes \overline{\mathbf{R}}$, so colored composites have a chance \say{to compete} to have the minimum $V$ against other color singlets. In this sense, the breaking of the gauge group is more likely to occur in chiral gauge theories, as opposed to vector-like theories where the singlet in $\mathbf{R} \otimes \overline{\mathbf{R}}$ usually minimizes $V$.\\

The color and flavor indices carried by the fermions for the models studied here are

\setlength{\jot}{20pt}
\begin{equation}
\begin{gathered}
\psi^{n_1 , ij} \esp \in
\raisebox{-3.5pt}{
\scaleobj{0.75}{
\begin{ytableau}
~ & 
\end{ytableau}}}
\spa\spa \simtworep
\spa\spa,\spa\spa
\lambda^{n_2 , ij} \esp \in
\raisebox{3.5pt}{
\scaleobj{0.75}{
\begin{ytableau}
~ \\
~
\end{ytableau}}}
\spa\spa \asimtworep
\spa\spa,\spa\spa
\lambda^{n_2}_{ij} \esp \in
\raisebox{3.5pt}{
\scaleobj{0.75}{
\begin{ytableau}
\phantom{.}\overline{\phantom{\Big|aa.}} 
~ \\
~
\end{ytableau}}}
\spa\spa \ccasimtworep
\\
\eta^{n_3}_i \esp \in
\raisebox{-3.2pt}{
\scaleobj{0.75}{
\begin{ytableau}
\overline{\phantom{\Big|aa.}} 
\end{ytableau}}}
\spa\spa \ccfundrep
\spa\spa,\spa\spa
\eta^{n_3,i} \esp \in
\raisebox{-3.2pt}{
\scaleobj{0.75}{
\begin{ytableau}
\phantom{\Big|aa.}
\end{ytableau}}}
\spa\spa \fundrep
\spa\spa,
\\
i,j = 1, \esp..., \esp N
\spa\spa,\spa\spa
n_i = 1, \esp..., \esp N_i
\spa\spa.
\numberthis \label{regular_fermion_operators_MAC_appendix}
\end{gathered}
\end{equation}
\setlength{\jot}{13pt}

\newpage

For 11 out of the 30 theories in Table \ref{table_N1_N2_N3_values}, the $\lambda$ operators transform under the two-index anti-symmetric representation $\txn{\textbf{A}}_{2}$, no complex-conjugation required. In contrast, in 19 of these models (for which $N_3$ is still non-zero in general) these fermions transform in the complex-conjugated representation $\overline{\txn{\textbf{A}}}_{2}$ (for these cases $N_2$ is negative in Table \ref{table_N1_N2_N3_values}). Note also that for $R_9$, $R_8$, $R_{10}$, $R_{11}$, $R_{15}$, $R_{16}$, $R_{21}$, $R_{22}$ and $R_{26}$, $N_3 < 0$ for some values of $N$, so in these cases the $\eta^i$ Weyl fermions transform in the fundamental representation. There are three possible scenarios that define the color representations for the fermions, and depending on these the MAC ordering can change. In what follows we show the minimum value of $V$ for each possible colored fermion bilinear together with the representation that defines it.

\subsection{$N_2 > 0$ \& $N_3>0$}

The representations for the fermions are 
\begin{align}
N_1 \txn{\textbf{S}}_{2} \oplus N_2 \txn{\textbf{A}}_{2} \oplus N_3 \overline{\txn{\textbf{F}}}_1
\spa\spa.
\end{align}
The condensate that minimizes $V$ is $\langle \psi \eta \rangle$ in the fundamental. For large values of $N$, $\langle \lambda \eta \rangle$ in the fundamental develops a value for $V$ similar to that of $\langle \psi \eta \rangle$.\\

\renewcommand{\extable}{0.75ex}
\begin{center}
\begin{tabular}{||c | c c c ||} 
\hline\hline
\rule{0pt}{2ex}  
\phantom{-}$\mathcal{O}(x)$\phantom{-}  & 
\phantom{-}irrep\phantom{-}  &
\phantom{--}min$(V)$\phantom{--} &
\phantom{--}$N\longrightarrow \infty$\phantom{--} \\ [0.0ex]
\hline\hline
\phantom{-}&  &  &       \\ [-2.5ex]
$\psi\psi$       &  \TwoTwoIrrep       & $-2(N+2)/N$ &  -2 \\ [\extable]
$\psi\lambda$    &  \TwoOneOneIrrep    & $-2(N+2)/N$ &  -2  \\ [\extable]
$\psi\eta$       &  $\txn{\textbf{F}}$ & $-(N-1)(N+2)/N$  & $-\infty$ \\ [\extable]
$\lambda\lambda$ & $\txn{\textbf{A}}_4$ & $-4(N+1)/N$  & -4 \\ [\extable]
$\lambda\eta$    & $\txn{\textbf{F}}$  & $-(N+1)(N-2)/N$  & $-\infty$ \\ [\extable]
$\eta\eta$       &$\overline{\txn{\textbf{A}}}_2$ &   $-(N+1)/N$ & -1  \\ [\extable]
\hline
\end{tabular}
\end{center}
\captionof{table}{ \label{appendix_table_MAC_1_S_A_F*_formulas}
Properties for the condensate that minimizes $V$.
}

\renewcommand{\extable}{0.75ex}
\begin{center}
\begin{tabular}{||c | c c c c c c  ||} 
\hline\hline
\rule{0pt}{2ex}  
\diagbox[width=4em,height=2em,font=\scriptsize\itshape]{$\mathcal{O}(x)$}{$N$}& 
\phantom{-}$5$\phantom{-}  & 
\phantom{-}$6$\phantom{-}  &
\phantom{--}$7$\phantom{--}  &
\phantom{--}$8$\phantom{--}  &
\phantom{--}$9$\phantom{--}  &
\phantom{--}$10$\phantom{--}  \\ [0.0ex]
\hline\hline
\phantom{-}&  &  &   &     & &  \\ [-2.5ex]
$\psi \psi$   &   $\big( \bm{ \overline{50} }, -\frac{14}{5}\big) $   &   $\big( \bm{ \overline{105}' }, -\frac{8}{3}\big) $   &   $\big( \bm{ \overline{196} }, -\frac{18}{7}\big) $   &   $\big( \bm{ 336 }, -\frac{5}{2}\big) $   &   $\big( \bm{ 540 }, -\frac{22}{9}\big) $   &   $\big( \bm{ 825 }, -\frac{12}{5}\big) $  \\ [\extable]
$\psi \lambda$   &   $\big( \bm{ \overline{45} }, -\frac{14}{5}\big) $   &   $\big( \bm{ \overline{105} }, -\frac{8}{3}\big) $   &   $\big( \bm{ \overline{210} }, -\frac{18}{7}\big) $   &   $\big( \bm{ 378 }, -\frac{5}{2}\big) $   &   $\big( \bm{ 630 }, -\frac{22}{9}\big) $   &   $\big( \bm{ 990 }, -\frac{12}{5}\big) $  \\ [\extable]
$\psi \eta$   &   $\big( \bm{ 5 }, -\frac{28}{5}\big) $   &   $\big( \bm{ 6 }, -\frac{20}{3}\big) $   &   $\big( \bm{ 7 }, -\frac{54}{7}\big) $   &   $\big( \bm{ 8 }, -\frac{35}{4}\big) $   &   $\big( \bm{ 9 }, -\frac{88}{9}\big) $   &   $\big( \bm{ 10 }, -\frac{54}{5}\big) $  \\ [\extable]
$\lambda \lambda$   &   $\big( \bm{ \bar{5} }, -\frac{24}{5}\big) $   &   $\big( \bm{ \overline{15} }, -\frac{14}{3}\big) $   &   $\big( \bm{ \overline{35} }, -\frac{32}{7}\big) $   &   $\big( \bm{ 70 }, -\frac{9}{2}\big) $   &   $\big( \bm{ 126 }, -\frac{40}{9}\big) $   &   $\big( \bm{ 210 }, -\frac{22}{5}\big) $  \\ [\extable]
$\lambda \eta$   &   $\big( \bm{ 5 }, -\frac{18}{5}\big) $   &   $\big( \bm{ 6 }, -\frac{14}{3}\big) $   &   $\big( \bm{ 7 }, -\frac{40}{7}\big) $   &   $\big( \bm{ 8 }, -\frac{27}{4}\big) $   &   $\big( \bm{ 9 }, -\frac{70}{9}\big) $   &   $\big( \bm{ 10 }, -\frac{44}{5}\big) $  \\ [\extable]
$\eta\eta$   &   $\big( \bm{ \overline{10} }, -\frac{6}{5}\big) $   &   $\big( \bm{ \overline{15} }, -\frac{7}{6}\big) $   &   $\big( \bm{ \overline{21} }, -\frac{8}{7}\big) $   &   $\big( \bm{ \overline{28} }, -\frac{9}{8}\big) $   &   $\big( \bm{ \overline{36} }, -\frac{10}{9}\big) $   &   $\big( \bm{ \overline{45} }, -\frac{11}{10}\big) $  \\ [\extable]
\txn{min} & -28/5 & -20/3 & -54/7 & -35/4 & -88/9 & -54/5 \\ [\extable]
\hline
\end{tabular}
\end{center}
\captionof{table}{ \label{appendix_table_MAC_1_S_A_F*}
Minimum $V$ value for each possible bilinear composite operator for different $N$. The first entry in each pair inside the parentheses indicates the irrep that gives that minimum $V$. The last row denotes the minimum $V$ among all possible operators for each particular $N$.
}

\begin{figure}[H]
\includegraphics[scale=0.17]{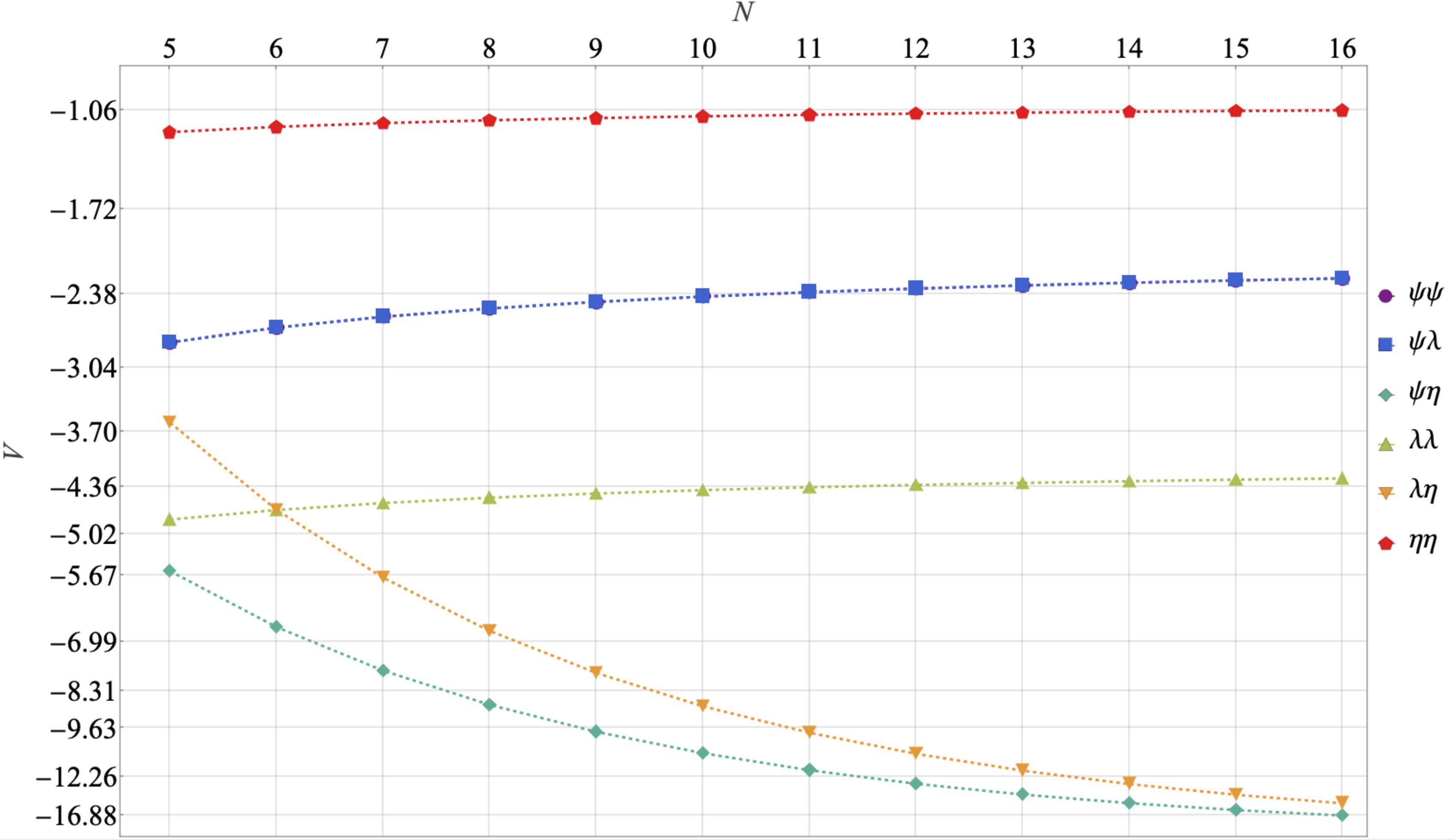}
\centering
\caption{Minimum $V$ for each bilinear composite operator for different $N$. Note that $\langle \psi \psi \rangle$ and $\langle \psi \lambda \rangle$ overlap.}
\label{figure_MAC_plot_1_2_fermion_op_S_A_Fcc}
\end{figure}

\phantom{-}\\

\subsection{$N_2 < 0$ \& $N_3>0$}

\newcommand*{\myov}[1]{\overbracket[2pt][-0pt]{#1}}

\newcommand{\OneOneOneOneCCIrrep}{
\raisebox{9.75pt}{
\scaleobj{0.28}{
\begin{ytableau}
\phantom{.}\myov{\phantom{\bigg|aa..}}
~  \\ 
~  \\
~  \\
~  
\end{ytableau}}}
}

The representations for the fermions are 

\begin{align}
N_1 \txn{\textbf{S}}_{2} \oplus N_2 \overline{\txn{\textbf{A}}}_{2} \oplus N_3 \overline{\txn{\textbf{F}}}_1
\spa\spa.
\end{align}

The condensate that minimizes $V$ is $\langle \psi \eta \rangle$ in the fundamental of SU($N$). For large values of $N$, $\langle \psi \lambda \rangle$ in the adjoint develops a value for $V$ similar to that of $\langle \psi \eta \rangle$. \\

\renewcommand{\extable}{0.95ex}
\begin{center}
\begin{tabular}{||c | c c c ||} 
\hline\hline
\rule{0pt}{2ex}  
\phantom{-}$\mathcal{O}(x)$\phantom{-}  & 
\phantom{-}irrep\phantom{-}  &
\phantom{--}min$(V)$\phantom{--} &
\phantom{--}$N\longrightarrow \infty$\phantom{--} \\ [0.0ex]
\hline\hline
\phantom{-}&  &  &       \\ [-2.5ex]
$\psi\psi$       &  \TwoTwoIrrep   & $-2(N+2)/N$ &  -2 \\ [\extable]
$\psi\lambda$    &  $\txn{\textbf{Adj}}$ & $-(N-2)(N+2)/N$ &  $-\infty$  \\ [\extable]
$\psi\eta$       &  $\txn{\textbf{F}}$ & $-(N-1)(N+2)/N$  & $-\infty$ \\ [\extable]
$\lambda\lambda$ &  $\overline{\txn{\textbf{A}}}_4$ & $-4(N+1)/N$  & -4 \\ [\extable]
$\lambda\eta$    & $\overline{\txn{\textbf{A}}}_3$ &  $-2(N+1)/N$  & -2 \\ [\extable]
$\eta\eta$       &$\overline{\txn{\textbf{A}}}_2$ &   $-(N+1)/N$ & -1  \\ [\extable]
\hline
\end{tabular}
\end{center}
\captionof{table}{ \label{appendix_table_MAC_2_S_A*_F*_formulas}
Properties for the condensate that minimizes $V$.
}

\renewcommand{\extable}{0.7ex}
\begin{center}
\begin{tabular}{||c | c c c c c c  ||} 
\hline\hline
\rule{0pt}{2ex}  
\diagbox[width=4em,height=2em,font=\scriptsize\itshape]{$\mathcal{O}(x)$}{$N$}& 
\phantom{-}$5$\phantom{-}  & 
\phantom{-}$6$\phantom{-}  &
\phantom{--}$7$\phantom{--}  &
\phantom{--}$8$\phantom{--}  &
\phantom{--}$9$\phantom{--}  &
\phantom{--}$10$\phantom{--}  \\ [0.0ex]
\hline\hline
\phantom{-}&  &  &   &     & &  \\ [-2.5ex]
$\psi \psi$   &   $\big( \bm{ \overline{50} }, -\frac{14}{5}\big) $   &   $\big( \bm{ \overline{105}' }, -\frac{8}{3}\big) $   &   $\big( \bm{ \overline{196} }, -\frac{18}{7}\big) $   &   $\big( \bm{ 336 }, -\frac{5}{2}\big) $   &   $\big( \bm{ 540 }, -\frac{22}{9}\big) $   &   $\big( \bm{ 825 }, -\frac{12}{5}\big) $  \\ [\extable]
$\psi \lambda$   &   $\big( \bm{ 24 }, -\frac{21}{5}\big) $   &   $\big( \bm{ 35 }, -\frac{16}{3}\big) $   &   $\big( \bm{ 48 }, -\frac{45}{7}\big) $   &   $\big( \bm{ 63 }, -\frac{15}{2}\big) $   &   $\big( \bm{ 80 }, -\frac{77}{9}\big) $   &   $\big( \bm{ 99 }, -\frac{48}{5}\big) $  \\ [\extable]
$\psi \eta$   &   $\big( \bm{ 5 }, -\frac{28}{5}\big) $   &   $\big( \bm{ 6 }, -\frac{20}{3}\big) $   &   $\big( \bm{ 7 }, -\frac{54}{7}\big) $   &   $\big( \bm{ 8 }, -\frac{35}{4}\big) $   &   $\big( \bm{ 9 }, -\frac{88}{9}\big) $   &   $\big( \bm{ 10 }, -\frac{54}{5}\big) $  \\ [\extable]
$\lambda\lambda$   &   $\big( \bm{ 5 }, -\frac{24}{5}\big) $   &   $\big( \bm{ 15 }, -\frac{14}{3}\big) $   &   $\big( \bm{ 35 }, -\frac{32}{7}\big) $   &   $\big( \bm{ 70 }, -\frac{9}{2}\big) $   &   $\big( \bm{ \overline{126} }, -\frac{40}{9}\big) $   &   $\big( \bm{ \overline{210} }, -\frac{22}{5}\big) $  \\ [\extable]
$\lambda\eta$   &   $\big( \bm{ 10 }, -\frac{12}{5}\big) $   &   $\big( \bm{ 20 }, -\frac{7}{3}\big) $   &   $\big( \bm{ \overline{35} }, -\frac{16}{7}\big) $   &   $\big( \bm{ \overline{56} }, -\frac{9}{4}\big) $   &   $\big( \bm{ \overline{84} }, -\frac{20}{9}\big) $   &   $\big( \bm{ \overline{120} }, -\frac{11}{5}\big) $  \\ [\extable]
$\eta\eta$   &   $\big( \bm{ \overline{10} }, -\frac{6}{5}\big) $   &   $\big( \bm{ \overline{15} }, -\frac{7}{6}\big) $   &   $\big( \bm{ \overline{21} }, -\frac{8}{7}\big) $   &   $\big( \bm{ \overline{28} }, -\frac{9}{8}\big) $   &   $\big( \bm{ \overline{36} }, -\frac{10}{9}\big) $   &   $\big( \bm{ \overline{45} }, -\frac{11}{10}\big) $  \\ [\extable]
\txn{min} & -28/5 & -20/3 & -54/7 & -35/4 & -88/9 & -54/5 \\ [\extable]
\hline
\end{tabular}
\end{center}
\captionof{table}{ \label{appendix_table_MAC_2_S_A_F*}
Minimum $V$ value for each possible bilinear composite operator for different $N$. The first entry in each pair inside the parentheses indicates the irrep that gives that minimum $V$. The last row denotes the minimum $V$ among all possible operators for each particular $N$.
}

\begin{figure}[H]
\includegraphics[scale=0.17]{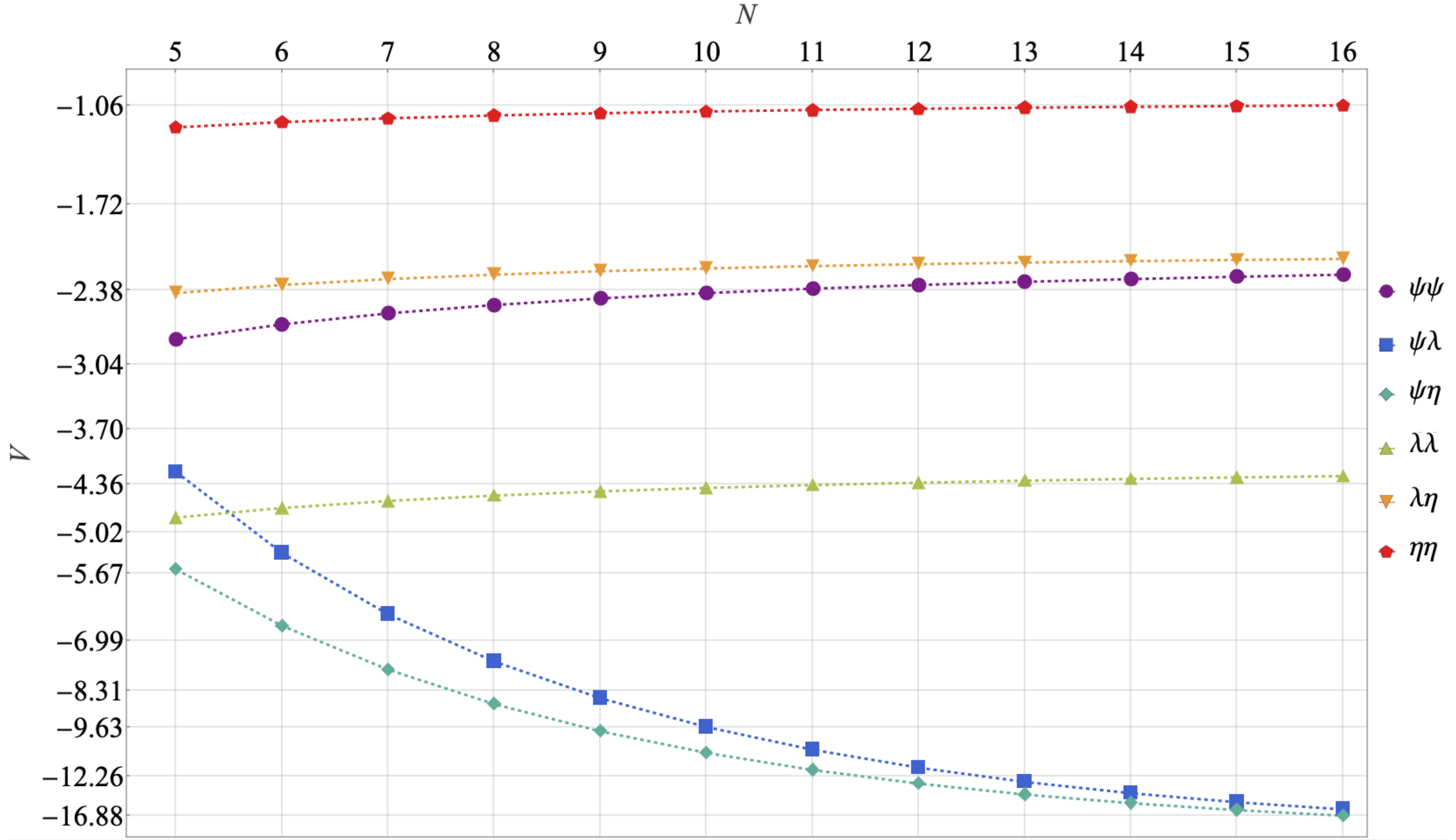}
\centering
\caption{Minimum $V$ value for each possible bilinear composite operator for different $N$.}
\label{figure_MAC_plot_2_2_fermion_op_S_Acc_Fcc}
\end{figure}

\subsection{$N_2 < 0$ \& $N_3 < 0$}

\newcommand{\TwoOneIrrep}{
\raisebox{2.5pt}{
\scaleobj{0.28}{
\begin{ytableau}
~ &  \\ 
~ 
\end{ytableau}}}
}

The representations for the fermions are 

\begin{align}
N_1 \txn{\textbf{S}}_{2} \oplus N_2 \overline{\txn{\textbf{A}}}_{2} \oplus N_3 \txn{\textbf{F}}_1
\spa\spa.
\end{align}

For SU(5) the condensate that minimizes $V$ is $\langle \lambda\lambda \rangle$ in the 4-index anti-symmetric representation $\overline{\txn{\textbf{A}}}_4$. However, for $N\geq6$ $\langle \psi\lambda \rangle$ minimizes $V$ in the adjoint. For large values of $N$, $\langle \lambda \eta \rangle$ in the anti-fundamental develops a value for $V$ similar to that of $\langle \psi \lambda \rangle$.\\

\renewcommand{\extable}{0.5ex}
\begin{center}
\begin{tabular}{||c | c c c ||} 
\hline\hline
\rule{0pt}{2ex}  
\phantom{-}$\mathcal{O}(x)$\phantom{-}  & 
\phantom{-}irrep\phantom{-}  &
\phantom{--}min$(V)$\phantom{--} &
\phantom{--}$N\longrightarrow \infty$\phantom{--} \\ [0.0ex]
\hline\hline
\phantom{-}&  &  &       \\ [-2.5ex]
$\psi\psi$       &  \TwoTwoIrrep   & $-2(N+2)/N$ &  -2 \\ [\extable]
$\psi\lambda$    &  $\txn{\textbf{Adj}}$ & $-(N-2)(N+2)/N$ &  $-\infty$  \\ [\extable]
$\psi\eta$       & \TwoOneIrrep  & $-(N+2)/N$  & -1 \\ [\extable]
$\lambda\lambda$ & $\overline{\txn{\textbf{A}}}_4$  & $-4(N+1)/N$  & -4 \\ [\extable]
$\lambda\eta$    & $\overline{\txn{\textbf{F}}}$ &  $-(N+1)(N-2)/N$  & $-\infty$ \\ [\extable]
$\eta\eta$       & $\txn{\textbf{A}}_2$ &   $-(N+1)/N$ & -1  \\ [\extable]
\hline
\end{tabular}
\end{center}
\captionof{table}{ \label{appendix_table_MAC_3_S_A*_F_formulas}
Properties for the condensate that minimizes $V$.
}

\renewcommand{\extable}{0.5ex}
\begin{center}
\begin{tabular}{||c | c c c c c c  ||} 
\hline\hline
\rule{0pt}{2ex}  
\diagbox[width=4em,height=2em,font=\scriptsize\itshape]{$\mathcal{O}(x)$}{$N$}& 
\phantom{-}$5$\phantom{-}  & 
\phantom{-}$6$\phantom{-}  &
\phantom{--}$7$\phantom{--}  &
\phantom{--}$8$\phantom{--}  &
\phantom{--}$9$\phantom{--}  &
\phantom{--}$10$\phantom{--}  \\ [0.0ex]
\hline\hline
\phantom{-}&  &  &   &     & &  \\ [-2.5ex]
$\psi \psi$   &   $\big( \bm{ \overline{50} }, -\frac{14}{5}\big) $   &   $\big( \bm{ \overline{105}' }, -\frac{8}{3}\big) $   &   $\big( \bm{ \overline{196} }, -\frac{18}{7}\big) $   &   $\big( \bm{ 336 }, -\frac{5}{2}\big) $   &   $\big( \bm{ 540 }, -\frac{22}{9}\big) $   &   $\big( \bm{ 825 }, -\frac{12}{5}\big) $  \\ [\extable]
$\psi \lambda$   &   $\big( \bm{ 24 }, -\frac{21}{5}\big) $   &   $\big( \bm{ 35 }, -\frac{16}{3}\big) $   &   $\big( \bm{ 48 }, -\frac{45}{7}\big) $   &   $\big( \bm{ 63 }, -\frac{15}{2}\big) $   &   $\big( \bm{ 80 }, -\frac{77}{9}\big) $   &   $\big( \bm{ 99 }, -\frac{48}{5}\big) $  \\ [\extable]
$\psi \eta$   &   $\big( \bm{ \overline{40} }, -\frac{7}{5}\big) $   &   $\big( \bm{ 70 }, -\frac{4}{3}\big) $   &   $\big( \bm{ 112 }, -\frac{9}{7}\big) $   &   $\big( \bm{ 168 }, -\frac{5}{4}\big) $   &   $\big( \bm{ 240 }, -\frac{11}{9}\big) $   &   $\big( \bm{ 330 }, -\frac{6}{5}\big) $  \\ [\extable]
$\lambda\lambda$   &   $\big( \bm{ 5 }, -\frac{24}{5}\big) $   &   $\big( \bm{ 15 }, -\frac{14}{3}\big) $   &   $\big( \bm{ 35 }, -\frac{32}{7}\big) $   &   $\big( \bm{ 70 }, -\frac{9}{2}\big) $   &   $\big( \bm{ \overline{126} }, -\frac{40}{9}\big) $   &   $\big( \bm{ \overline{210} }, -\frac{22}{5}\big) $  \\ [\extable]
$\lambda\eta$   &   $\big( \bm{ \bar{5} }, -\frac{18}{5}\big) $   &   $\big( \bm{ \bar{6} }, -\frac{14}{3}\big) $   &   $\big( \bm{ \bar{7} }, -\frac{40}{7}\big) $   &   $\big( \bm{ \bar{8} }, -\frac{27}{4}\big) $   &   $\big( \bm{ \bar{9} }, -\frac{70}{9}\big) $   &   $\big( \bm{ \overline{10} }, -\frac{44}{5}\big) $  \\ [\extable]
$\eta \eta$   &   $\big( \bm{ 10 }, -\frac{6}{5}\big) $   &   $\big( \bm{ 15 }, -\frac{7}{6}\big) $   &   $\big( \bm{ 21 }, -\frac{8}{7}\big) $   &   $\big( \bm{ 28 }, -\frac{9}{8}\big) $   &   $\big( \bm{ 36 }, -\frac{10}{9}\big) $   &   $\big( \bm{ 45 }, -\frac{11}{10}\big) $  \\ [\extable]
\txn{min} & -24/5 & -16/3 & -45/7 & -15/2 & -77/9 & -48/5 \\ [\extable]
\hline
\end{tabular}
\end{center}
\captionof{table}{ \label{appendix_table_MAC_2_S_A_F*}
Minimum $V$ value for each possible bilinear composite operator for different $N$. The first entry in each pair inside the parentheses indicates the irrep that gives that minimum $V$. The last row denotes the minimum $V$ among all possible operators for each particular $N$.
}

\begin{figure}[H]
\includegraphics[scale=0.166]{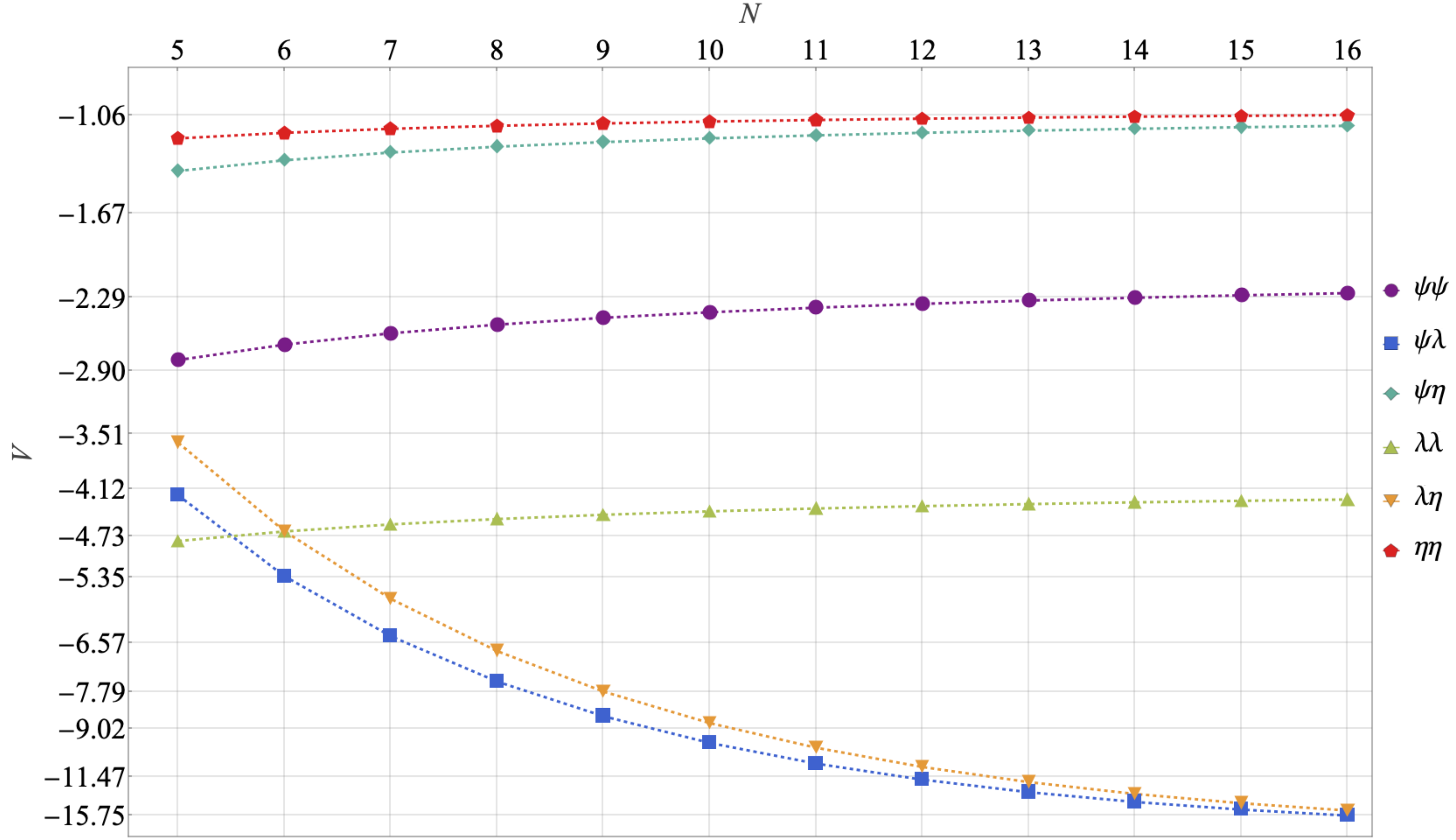}
\centering
\caption{Minimum $V$ value for each possible bilinear composite operator for different $N$.}
\label{figure_MAC_plot_3_2_fermion_op_S_Acc_F}
\end{figure}

\subsection{$N_2 < 0$ \& $N_3 = 0$}
\label{MAC_appendix_two_species_models}

The representations for the fermions are 
\begin{align}
N_1 \txn{\textbf{S}}_{2} \oplus N_2 \overline{\txn{\textbf{A}}}_{2}
\spa\spa.
\end{align}
For SU(5) the condensate that minimizes $V$ is $\langle \lambda\lambda \rangle$ in the 4-index anti-symmetric representation $\overline{\txn{\textbf{A}}}_4$. However, for $N\geq6$ $\langle \psi\lambda \rangle$ minimizes $V$ in the adjoint. In this case there is no candidate order parameter in the fundamental. For these theories $N$ is fixed as shown in Table \tcolor{\ref{two_species_table_appendix}}.\\

\renewcommand{\extable}{0.3ex}
\begin{center}
\begin{tabular}{||c | c c c ||} 
\hline\hline
\rule{0pt}{2ex}  
\phantom{-}$\mathcal{O}(x)$\phantom{-}  & 
\phantom{-}irrep\phantom{-}  &
\phantom{--}min$(V)$\phantom{--} &
\phantom{--}$N\longrightarrow \infty$\phantom{--} \\ [0.0ex]
\hline\hline
\phantom{-}&  &  &       \\ [-2.5ex]
$\psi\psi$       &  \TwoTwoIrrep   & $-2(N+2)/N$ &  -2 \\ [\extable]
$\psi\lambda$    &  $\txn{\textbf{Adj}}$ & $-(N-2)(N+2)/N$ &  $-\infty$  \\ [\extable]
$\lambda\lambda$ & $\overline{\txn{\textbf{A}}}_4$  & $-4(N+1)/N$  & -4 \\ [\extable]
\hline\hline
\end{tabular}
\end{center}
\captionof{table}{ \label{appendix_table_MAC_4_S_A*formulas}
Properties for the condensate that minimizes $V$.}
\renewcommand{\extable}{0.27ex}
\begin{center}
\begin{tabular}{||c | c c c c c c  ||} 
\hline\hline
\rule{0pt}{2ex}  
\diagbox[width=4em,height=2em,font=\scriptsize\itshape]{$\mathcal{O}(x)$}{$N$}& 
\phantom{-}$5$\phantom{-}  & 
\phantom{-}$6$\phantom{-}  &
\phantom{--}$7$\phantom{--}  &
\phantom{--}$8$\phantom{--}  &
\phantom{--}$9$\phantom{--}  &
\phantom{--}$10$\phantom{--}  \\ [0.0ex]
\hline\hline
\phantom{-}&  &  &   &     & &  \\ [-2.5ex]
$\psi \psi$   &   $\big( \bm{ \overline{50} }, -\frac{14}{5}\big) $   &   $\big( \bm{ \overline{105}' }, -\frac{8}{3}\big) $   &   $\big( \bm{ \overline{196} }, -\frac{18}{7}\big) $   &   $\big( \bm{ 336 }, -\frac{5}{2}\big) $   &   $\big( \bm{ 540 }, -\frac{22}{9}\big) $   &   $\big( \bm{ 825 }, -\frac{12}{5}\big) $  \\ [\extable]
$\psi \lambda$   &   $\big( \bm{ 24 }, -\frac{21}{5}\big) $   &   $\big( \bm{ 35 }, -\frac{16}{3}\big) $   &   $\big( \bm{ 48 }, -\frac{45}{7}\big) $   &   $\big( \bm{ 63 }, -\frac{15}{2}\big) $   &   $\big( \bm{ 80 }, -\frac{77}{9}\big) $   &   $\big( \bm{ 99 }, -\frac{48}{5}\big) $  \\ [\extable]
$\lambda\lambda$   &   $\big( \bm{ 5 }, -\frac{24}{5}\big) $   &   $\big( \bm{ 15 }, -\frac{14}{3}\big) $   &   $\big( \bm{ 35 }, -\frac{32}{7}\big) $   &   $\big( \bm{ 70 }, -\frac{9}{2}\big) $   &   $\big( \bm{ \overline{126} }, -\frac{40}{9}\big) $   &   $\big( \bm{ \overline{210} }, -\frac{22}{5}\big) $  \\ [\extable]
\txn{min} & -24/5 & -16/3 & -45/7 & -15/2 & -77/9 & -48/5 \\ [\extable]
\hline
\end{tabular}
\end{center}
\captionof{table}{ \label{appendix_table_MAC_2_S_A_F*}
Minimum $V$ value for each possible bilinear composite operator for different $N$. The first entry in each pair inside the parentheses indicates the irrep that gives that minimum $V$. The last row denotes the minimum $V$ among all possible operators for each particular $N$.}
\begin{figure}[H]
\includegraphics[scale=0.149]{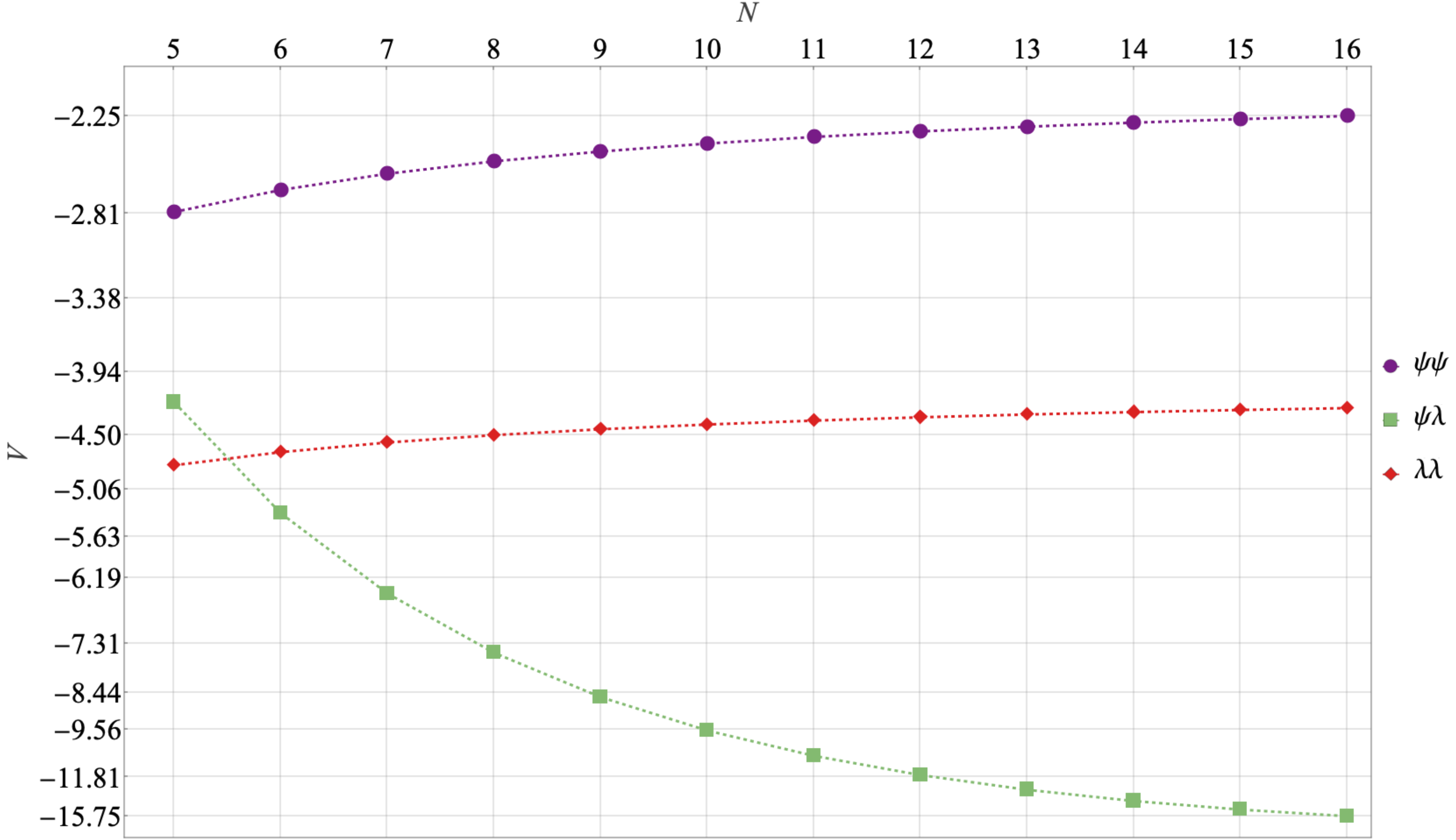}
\centering
\caption{Minimum $V$ value for each possible bilinear composite operator for different $N$.}
\label{figure_MAC_plot_4_2_fermion_op_S_Acc}
\end{figure}

%% file: Appendix_C.tex
\newpage

\section{Classes of chiral gauge theories}
\label{appendix_clases_CGT}

In this appendix we summarize a few different classes of chiral gauge theories in four and two dimensions that have been studied in the literature.

\subsection{Abelian theories}

It is possible to construct four-dimensional U(1) chiral gauge theories with $N_f$ flavors of fermions by finding a charge assignment that yields an anomaly-free theory. Since in $4d$ the anti-particle of a right-handed fermion is left-handed we can consider a set of only left-handed
fermions with charges $Q_i$. The theory is anomaly-free only if 

\begin{align}
\sum_{i = 1}^{N_f} Q_i^3 = 0
\spa\spa,
\end{align}

see \tcolor{\cite{Tong:GaugeTheory:2018}} for a brief review on this class of theories. One of the simplest solutions comes from considering $N_f = 4$ Weyl fermions with three of the charges being positive and one negative

\begin{align}
3^3 +  4^3 +  5^3 +  (-6)^3 = 0 
\spa\spa.
\label{abelian_3456_theory_appendix}
\end{align}

Similarly, we can consider two of the charges being positive and two negative

\begin{gather*}
1^3 +  12^3 +  (-9)^3 +  (-10)^3 = 0 
\spa\spa \txn{or equivalently} \spa\spa
1^3 +  12^3 = 9^3 + 10^3 = 1729
\numberthis
\spa\spa,
\end{gather*}

which is the first number that can be expressed as the sum of two cubes in two different ways\hlfootnote{If we also consider negative integers this number would be $91 = 6^3 + (-5)^3 = 4^3 + 3^3$, which corresponds to the anomaly free theory in (\tcolor{\ref{abelian_3456_theory_appendix}}).}. Note that this theory suffers from a mixed gauge-gravitational anomaly given by $\sum_i Q_i = -6$, and thus this theory is inconsistent when placed on curved spacetime.\\

In $2d$ there is a well known anomaly-free chiral gauge theory usually called the 3450 model. This theory consists of two left-moving plus two right-moving Weyl fermions with charges 3, 4, 5 and 0 respectively. Since $3^2 + 4^2 - (5^2 + 0^2 ) = 0$, the theory is anomaly free. See \cite{Wang:Weng:2018,Berkowitz:Cherman:Jacobson:2023,Seifnashri:2026,Lu:Seifnashri:Shao:2026,Thorngren:Preskill:Fidkowski:2026,Zakharov:Ueda:Verstraete:Beenakker:2026,Baig:Chen:Cherman:Neuzil:2026,Dang:Karur:Sen:2026,Lamm:Roggero:Singh:Spagnoli:2026} for recent progress on simulating chiral abelian theories on the lattice. For non-abelian theories see \cite{Onoda:2026}.

\subsection{4-dimensional theories}

As pointed out in \cite{Bolognesi:Konishi:Luzio:2023:GeneralIrrep}, we can consider the following direct sum of irreducible representations of SU($N$)

\begin{align}
N_{\psi}
\raisebox{-3.5pt}{
\scaleobj{0.75}{
\begin{ytableau}
~ & 
\end{ytableau}}}
\oplus
N_{\widetilde{\psi}}
\raisebox{-3.5pt}{
\scaleobj{0.75}{
\begin{ytableau}
\phantom{aaa.}\overline{\phantom{\Big|aaaa-}} 
~ & 
\end{ytableau}}}
\oplus N_{\widetilde{\chi}}
\raisebox{3.5pt}{
\scaleobj{0.75}{
\begin{ytableau}
~ \\
~
\end{ytableau}}}
\oplus N_{\chi}
\raisebox{3.5pt}{
\scaleobj{0.75}{
\begin{ytableau}
\phantom{.}\overline{\phantom{\Big|a...}} 
~ \\
~
\end{ytableau}}}
\oplus N_{\widetilde{\eta}}
\raisebox{-3.2pt}{
\scaleobj{0.75}{
\begin{ytableau}
~
\end{ytableau}}}
\oplus N_\eta
\raisebox{-3.2pt}{
\scaleobj{0.75}{
\begin{ytableau}
\overline{\phantom{\Big|aa.}} 
\end{ytableau}}}
\oplus N_\lambda
\raisebox{22.5pt}{
\scaleobj{0.75}{
\begin{ytableau}
~ & \\
~ \\
\none[\raisebox{-1.75pt}{\vdots}] \\
~ \\
~ 
\end{ytableau}}}
\spa\spa,
\label{appendix_general_models}
\end{align}

in order for gauge anomalies to vanish it must be true that

\begin{align}
\big( N_{\psi} - N_{\widetilde{\psi}} \big) \big(N+4\big)
+
\big( N_{\widetilde{\chi}} - N_{\chi} \big) \big(N-4\big)
+
\big( N_{\widetilde{\eta}} - N_{\eta} \big) 
= 0
\spa\spa.
\end{align}

For the theory to be asymptotically free the condition is 

\begin{align}
\beta_0 =
11N 
- \big( N_{\psi} + N_{\widetilde{\psi}} \big)\big(N+2\big)
- \big( N_{\widetilde{\chi}} + N_{\chi} \big) \big(N-2\big)
- \big( N_{\widetilde{\eta}} + N_{\eta} \big)
- 2 N N_\lambda > 0
\esp.
\end{align}

Some combinations of the flavor numbers can yield vector-like theories. For instance, for $N_{\widetilde{\eta}} = N_{\eta}$ and all the others set to zero, we recover ordinary QCD. Other choices of the matter content render anomaly-free and asymptotically free chiral gauge theories.

\subsubsection{Three fermion species}

First introduced in \tcolor{\cite{Eichten:Kang:Koh:1982}}, this is a list of 30 SU$(N)$ models that satisfy gauge anomaly cancellation and asymptotic freedom for arbitrarily large $N$. These are the models studied in this work. The representations are as follows 

\begin{align}
R_i = 
N_1
\raisebox{-3.5pt}{
\scaleobj{0.75}{
\begin{ytableau}
~ & 
\end{ytableau}}}
\oplus N_2
\raisebox{3.5pt}{
\scaleobj{0.75}{
\begin{ytableau}
~ \\
~
\end{ytableau}}}
\oplus N_3
\raisebox{-3.2pt}{
\scaleobj{0.75}{
\begin{ytableau}
\overline{\phantom{\Big|aa.}} 
\end{ytableau}}}
\spa\spa.
\label{appendix_Eichten_etal_models}
\end{align}

See Tables \tcolor{\ref{table_N1_N2_N3_values}}, \tcolor{\ref{table_rep_multiplictly_1}} and \tcolor{\ref{table_rep_multiplictly_2}} for a list of the precise matter content $N_1$, $N_2$ and $N_3$, as well as the representation multiplicity $\ell$ needed to satisfy asymptotic freedom as a function of $N$. See \cite{Eichten:Peccei:Preskill:Zeppenfeld:1986} for a study of the $R_3$ theory by means of the large $N$ limit, and \cite{Goity:Peccei:Zeppenfeld:1985,Eichten:Peccei:Preskill:Zeppenfeld:1986,Armoni:Shifman:2012,Bolognesi:Konishi:Shifman:2017,Sheu:Shifman:2022,Bolognesi:Konishi:Luzio:2022} for a series of analyses of the $R_7$ model.\\

In Appendix \ref{appendix_bar_R_models} we present an overview of models of the same kind as those in (\ref{appendix_Eichten_etal_models}), but for which asymptotic freedom is guaranteed for a finite range of $N$ only. These theories were also introduced in \cite{Eichten:Kang:Koh:1982}.\\

Recent work \tcolor{\cite{Gripaios:Nguyen:2024,Gripaios:Nguyen:2025,Gripaios:Nguyen:2025_2}} has pointed out that the list reported in \tcolor{\cite{Eichten:Kang:Koh:1982}} is missing some irreps; see also \tcolor{\cite{Cacciapaglia:Deandrea:Sannino:2025}}. In \cite{Eichten:Feinberg:1982}, the authors list other anomaly-free and asymptotically free SU($N$) chiral gauge theories with $N > 17$.

\subsubsection{Two fermion species}
\label{appendix_4d_two_species_models}

For these models the gauge group is SU$(N)$ with $N$ fixed  such that asymptotic freedom is satisfied. Formally, since $N$ is fixed, these theories do not admit a large $N$ limit. See \tcolor{\cite{Anber:Chan:2023,Anber:Hong:Son:2022,Bolognesi:Konishi:Luzio:Orso:2025}} for a recent study of these models. The matter content is given by 14 possible combinations of chiral fermions transforming under the two-index symmetric and two-index complex-conjugate anti-symmetric representation as follows

\begin{align}
N_1
\raisebox{-3.5pt}{
\scaleobj{0.75}{
\begin{ytableau}
~ & 
\end{ytableau}}}
\oplus N_2
\raisebox{3.5pt}{
\scaleobj{0.75}{
\begin{ytableau}
\phantom{.}\overline{\phantom{\Big|a...}} 
~ \\
~
\end{ytableau}}}
\spa\spa,
\label{appendix_two_species_models}
\end{align}

where 

\begin{align}
N_1 = \frac{N-4}{k}
\spa\spa,\spa\spa
N_2 = \frac{N+4}{k}
\spa\spa,
\end{align}

with $k$ a common divisor of $N-4$ and $N+4$. The allowed values for $N$ and $k$ are listed in Table \tcolor{\ref{two_species_table_appendix}}. These models are dubbed \say{2-index} chiral gauge theories.

\renewcommand{\extable}{0.5ex}
\begin{table}[H]
\begin{center}
\begin{tabular}{||c | c c c c c c c c c c c c c c||} 
\hline\hline
\rule{0pt}{2.5ex}  
$R_i / \overline{R}_i$ & $\overline{R}_9$ & - & $R_{11}$ & - & $R_{9}$ & $\overline{R}_{19}$  & -  & $R_{8}$ &$R_{22}$ & -   &$R_{15}$ & $R_{21}$  & $R_{26}$  & $\overline{R}_{22}$ \\ [\extable]
$N$                    & 5 & 6 & 6        & 8 & 8       & 10 & 12 & 12      & 16      & 20  & 20      & 28        & 36        & 44 \\ [\extable]
$k$                    & 1 & 1 & 2        & 2 & 4       & 2  & 4  & 8       & 4       & 4   & 8       & 8         & 8         & 8 \\ [\extable]
Type                   & F & F & F        & B & B       & F  & B  & B       & B       & B   & B       & B         & B         & B \\ [\extable]
\hline
\end{tabular}
\end{center}
\captionof{table}{ \label{two_species_table_appendix}
Matter content of the two-fermion species models. The $R_i / \overline{R}_i$ row relates each 2-index theory with a particular model in Table \tcolor{\ref{table_N1_N2_N3_values}} or \tcolor{\ref{table_overline_R_i_models}} with $N_3 = 0$. The type label denotes whether the theory is classified as fermionic (F) or bosonic (B) as discussed in this section.
}
\end{table}

These theories can be classified as \say{fermionic} or \say{bosonic}. The matter content of the fermionic theories allows us to form gauge invariant fermionic local operators. On the other hand, the gauge invariant operators allowed by the matter content of the bosonic theories do not admit a spinor index \tcolor{\cite{Anber:Chan:2023}}.\\ 

For 9 out of the 30 models in Table \tcolor{\ref{table_N1_N2_N3_values}}, there is a fixed $N$ such that $N_3 = 0$, see  equation (\tcolor{\ref{N3_as_function_of_N1_N2_N}}). These are $R_2, R_5, R_8, R_9, R_{11}, R_{15}, R_{21}, R_{22}$ and $R_{26}$. However, $N = 4,2$ for $R_2$ and $R_5$ respectively, and thus these theories cease to be asymptotically free. Only 7 of these models are a particular case of the general three fermion species theories in (\tcolor{\ref{appendix_Eichten_etal_models}}). Similarly, there are 3 cases for which $N_3 = 0$ in Table \ref{table_overline_R_i_models}; these are $\overline{R}_{9}$, $\overline{R}_{19}$ and $\overline{R}_{22}$. For all of these cases $N_2$ is negative, indicating that the $\lambda$ Weyl fermions transform under the complex-conjugate two-index anti-symmetric representation of SU$(N)$.\\

In terms of symmetries, there are two important differences with the three-fermion species models. Some of the two-fermion species models enjoy a genuine zero-form $\mathbbm{Z}_{2,4}$ discrete symmetry (disconnected component of the U(1)$_\alpha$), and for $N$ even, for all theories a one-form $\mathbbm{Z}_{2}^{(1)}$ symmetry that acts on Wilson lines is realized. In contrast, the three-species models do not exhibit a separate zero-form discrete symmetry and a one-form symmetry is present only when the full global symmetry group is gauged via background fields.  The presence of the $\mathbbm{Z}_{2,4}$ discrete symmetry makes an important difference when conjecturing the IR phase by matching the 't Hooft anomalies.\\ 

Given the explicit relation between the two and three fermion models illustrated in Table \tcolor{\ref{two_species_table_appendix}}, it is natural to ask if there are more models of the three-species kind that were missed in \tcolor{\cite{Eichten:Kang:Koh:1982}} that resemble a two-species model when the number of anti-fundamentals vanishes for a fixed $N$. We do not attempt to obtain such relations here.

\subsubsection{Symmetric Bars-Yankielowicz models}

These models with gauge group SU$(N)$ were originally studied in \tcolor{\cite{Bars:Yankielowicz:1981}}, where the authors reported a set of free fermions in the IR that match the zero-form UV 't Hooft anomalies while preserving the global symmetry group. More recently these theories have also been studied in \tcolor{\cite{Karasik:Onder:Tong:2022}} by means of the large $N$ limit and non-supersymmetric
dualities. See also \tcolor{\cite{Appelquist:Duan:Sannino:2000,Bolognesi:Konishi:Luzio:2020,Csaki:Murayama:Telem:2021, Csaki:Murayama:Telem:2022, Smith:Karasik:Lohitsiri:Tong:2022,Karasik:2022,Bolognesi:Konishi:Luzio:2021,Li:Alvaro:Vatani:2026}}. The conditions for \say{complete} asymptotic freedom of these theories, when they in addition feature a scalar field transforming either in the fundamental or in the adjoint of the gauge group, are studied in \cite{Cacciapaglia:Sannino:Wagner:2026}. Below we write the matter content and the conditions on the flavor numbers such that the theory remains anomaly-free and asymptotically free respectively
\begin{align}
\raisebox{-3.5pt}{
\scaleobj{0.75}{
\begin{ytableau}
~ & 
\end{ytableau}}}
\oplus p
\raisebox{-3.2pt}{
\scaleobj{0.75}{
\begin{ytableau}
~ \\
\end{ytableau}}}
\oplus q
\raisebox{-3.2pt}{
\scaleobj{0.75}{
\begin{ytableau}
\overline{\phantom{\Big|aa.}} 
\end{ytableau}}}
\spa\spa,\spa\spa
q = p + N+4
\spa\spa,\spa\spa
p \leq \frac{9}{2} N - 3 
\spa\spa.
\label{appendix_4d_symm_BY_models}
\end{align}

Note that one of the simplest of these models, which corresponds to $p = 0$ and $ q = N+4$, is the same as the simplest model of the three-species type in (\tcolor{\ref{appendix_Eichten_etal_models}}), i.e. $R_1$ in Table \tcolor{\ref{table_N1_N2_N3_values}}.

\subsubsection{Anti-symmetric Bars-Yankielowicz models}

Besides the two-index symmetric representation, the complex-conjugated anti-symmetric can be used instead. Below we write the matter content and conditions on the flavor numbers such that the theory remains anomaly-free and asymptotically free respectively

\begin{align}
\raisebox{3.5pt}{
\scaleobj{0.75}{
\begin{ytableau}
\phantom{.}\overline{\phantom{\Big|aa.}} 
~ \\
~
\end{ytableau}}}
\oplus p
\raisebox{-3.2pt}{
\scaleobj{0.75}{
\begin{ytableau}
~ \\
\end{ytableau}}}
\oplus q
\raisebox{-3.2pt}{
\scaleobj{0.75}{
\begin{ytableau}
\overline{\phantom{\Big|aa.}} 
\end{ytableau}}}
\spa\spa,\spa\spa
p = q + N-4
\spa\spa,\spa\spa
q \leq \frac{9}{2} N + 3 
\spa\spa.
\label{appendix_4d_anti_symm_BY_models}
\end{align}

The gauge group is SU$(N)$. We can obtain the anti-symmetric models by setting $N \to - N$ in the symmetric ones, and thus the 't Hooft anomalies are the same. Formally, the anti-symmetric theory is the continuation of the symmetric one to negative values of $N$ \tcolor{\cite{Karasik:Onder:Tong:2022}}.\\

Similarly to the symmetric case, one of the simplest of these models, which corresponds to $q = 0$ and $ p = N-4$, is \textit{almost} the same as the second simplest model of the three-species type in (\tcolor{\ref{appendix_Eichten_etal_models}}), i.e. $R_2$ in Table \tcolor{\ref{table_N1_N2_N3_values}}. Instead of having one fermion in the anti-symmetric and $N-4$ in the anti-fundamental the complex conjugation is flipped, so there is one fermion in the complex-conjugate anti-symmetric and $N-4$ in the fundamental.

\subsubsection{Georgi-Glashow models}

Similarly to the anti-symmetric Bars-Yankielowicz models, we can employ the two-index anti-symmetric irrep (no complex-conjugation) and interchange fundamentals with anti-fundamentals. The conditions for anomaly cancellation and asymptotic freedom are 

\begin{align}
\raisebox{3.5pt}{
\scaleobj{0.75}{
\begin{ytableau}
~ \\
~
\end{ytableau}}}
\oplus p
\raisebox{-3.2pt}{
\scaleobj{0.75}{
\begin{ytableau}
~ \\
\end{ytableau}}}
\oplus q
\raisebox{-3.2pt}{
\scaleobj{0.75}{
\begin{ytableau}
\overline{\phantom{\Big|aa.}} 
\end{ytableau}}}
\spa\spa,\spa\spa
q = p + N-4
\spa\spa,\spa\spa
p \leq \frac{9}{2} N + 3 
\spa\spa.
\label{appendix_4d_anti_Georgi_Glashow_models}
\end{align}

A detailed study of the simplest case $p = 0$ and $N = 5$ can be found in \cite{Bai:Stolarski:2021}. This model has the same group structure as the simplest SU(5) grand unified theory of the Standard Model. In \cite{Csaki:Murayama:Telem:2021,Csaki:Murayama:Telem:2022} this class of theories is studied by means of anomaly-mediated supersymmetry breaking. The conclusions drawn from this analysis differ from the dynamics suggested by the tumbling hypothesis \cite{Raby:Dimopoulos:Susskind:1980:tumbling}. See also \cite{Li:Alvaro:Vatani:2025} for a study of dynamical symmetry breaking in these theories, where the effective action and functional renormalization group formalism was employed. The conditions for \say{complete} asymptotic freedom of these theories, when they in addition feature a scalar field transforming either in the fundamental or in the adjoint of the gauge group, are studied in \cite{Cacciapaglia:Sannino:Wagner:2026}.

\subsubsection{Other gauge groups}

Besides SU($N$), the only other simple Lie groups that admit complex representations are $E_6$ and SO(4$n$+2). For $n\geq 2$, theories where the gauge group is either of these are automatically anomaly free. The only constraint on the matter content is that of asymptotic freedom. Chiral gauge theories based on these gauge groups have been studied by means of  anomaly-mediated supersymmetry breaking. In \cite{Kondo:Murayama:Sylber:2022} the authors consider SO(10) as the gauge group with $N_f$ Weyl fermions in the spinor \textbf{16} representation. Similarly, theories where $E_6$ is the gauge group with $N_f$ Weyl fermions in the fundamental \textbf{27} representation were studied in \cite{Goh:Murayama:Singh:Suter:Wong:2025}. For both theories, the results differ from the dynamics suggested by the tumbling hypothesis \cite{Raby:Dimopoulos:Susskind:1980:tumbling}.

\subsection{2-dimensional theories}

\subsubsection{Two fermion species}

See \tcolor{\cite{Onder:2023}} for a recent study of these theories. For this class of models where the gauge group is SU($N$), there can be a single right-moving Weyl fermion in the symmetric representation in addition to $q_L$ left-moving fermions in the anti-fundamental. The explicit matter content and condition that renders the theory anomaly-free are

\begin{align}
\raisebox{-3.5pt}{
\scaleobj{0.75}{
\begin{ytableau}
~ & 
\end{ytableau}}}
\oplus q_L
\raisebox{-3.2pt}{
\scaleobj{0.75}{
\begin{ytableau}
\overline{\phantom{\Big|aa.}} 
\end{ytableau}}}
\spa\spa,\spa\spa
q_L = N+2
\spa\spa.
\label{appendix_2d_symm_models}
\end{align}

Recall that in two dimensions the fermion contributions to the gauge anomaly are proportional to the Dynkin indices $T$ and not the anomaly coefficients $A$ like in the four dimensional case.\\

Similarly, there can be a single right-moving Weyl fermion in the anti-symmetric together with $q_L$ left-moving fermions in the anti-fundamental. The explicit matter content and condition on the flavor numbers such that the theory remains anomaly-free are 

\begin{align}
\raisebox{3.5pt}{
\scaleobj{0.75}{
\begin{ytableau}
\phantom{.}\overline{\phantom{\Big|aa.}} 
~ \\
~
\end{ytableau}}}
\oplus q_L
\raisebox{-3.2pt}{
\scaleobj{0.75}{
\begin{ytableau}
~ \\
\end{ytableau}}}
\spa\spa,\spa\spa
q_L = N - 2
\spa\spa.
\label{appendix_2d_anti_symm_models}
\end{align}

\subsubsection{Three fermion species}

These models have been studied in \tcolor{\cite{Onder:2023}}. Below we specify the matter content and condition on the flavor numbers such that the theory remains anomaly-free 

\begin{align}
\raisebox{3.5pt}{
\scaleobj{0.75}{
\begin{ytableau} 
~ \\
~
\end{ytableau}}}
\oplus p_R
\raisebox{-3.2pt}{
\scaleobj{0.75}{
\begin{ytableau}
\overline{\phantom{\Big|aa.}} 
\end{ytableau}}}
\oplus q_L
\raisebox{-3.2pt}{
\scaleobj{0.75}{
\begin{ytableau}
\overline{\phantom{\Big|aa.}} 
\end{ytableau}}}
\spa\spa,\spa\spa
q_L = p_R + N-2
\spa\spa,
\label{appendix_2d_three_species_models}
\end{align}

there is a single right-moving Weyl fermion in the anti-symmetric, $p_R$ right-moving fermions and $q_L$ left-moving fermions both in the anti-fundamental. A similar story applies for the symmetric version of the theory.

\subsection{The Standard Model}

The Standard Model of particle physics is a chiral gauge theory and thus gauge anomaly cancellation is non-trivial, but, as is well known, the charge assignment for the chiral fields renders an anomaly-free theory. Could the charge assignments have been different? Given the matter content of the Standard Model, the answer is no. The authors of \cite{Lohitsiri:Tong:2020} answer this question by considering an arbitrary hypercharge assignment for the chiral fields, showing that the gauge anomaly cancellation equations can be recast in the form of Fermat's last theorem, which famously has only trivial solutions. These trivial solutions reproduce the hypercharge assignment we see in nature.\\

Perhaps less widely appreciated, the Standard Model is part of a two-parameter family of chiral gauge theories with gauge group \tcolor{\cite{Tong:GaugeTheory:2018,Lohitsiri:Tong:2019}}

\begin{align}
G = \txn{SU}(N) \times \txn{U}(1) \times \txn{Sp}(r)
\label{extended_SM_gauge_group}
\spa\spa,
\end{align}

where $N$ is odd. There are $r$ copies of each of the right-handed fermions in the Standard Model, now including the right-handed neutrino. The gauge group representations under which the fermions transform are shown in Table \tcolor{\ref{reps_Standard_Model_Sp(r)_theory}}, where the subscript $L$ stands for left-handed and $R$ for right-handed.\\

See also \cite{Cacciapaglia:Deandrea:Kollias:Sannino:2026}, where an atlas of simple gauge theories is used to chart which grand-unifiable models can accommodate the Standard Model at low energies.

\begingroup 
\normalsize 
{\scalefont{1} 
\renewcommand{\extable}{0.7ex}
\begin{table}[H]
\begin{center}
\begin{tabular}{||c | c c c ||}
\hline\hline
\rule{0pt}{3ex}  
Chiral fields   &  U(1) & Sp($r$) & SU($N$)  \\ [0.2ex]
\hline\hline  
 &   &   &  \\ [-2ex]
$\ell_L$ &  $-N$ &  \textbf{2r} & \textbf{1} \\ [\extable]
$q_L$    &  $+1$ &  \textbf{2r} & \textbf{N} \\ [\extable]
\hline\hline
 &   &   &  \\ [-2ex]
$(e_Y)_R$ &  $-2YN$ &  \textbf{1} & \textbf{1} \\ [\extable]
$(u_Y)_R$ &  $1+(2Y-1)N$ &  \textbf{1} & \textbf{N} \\ [0.5ex]
\hline\hline
 &   &   &  \\ [-2ex]
 $(\nu_Y)_R$ &  $2(Y-1)N$ &  \textbf{1} & \textbf{1} \\ [\extable]
 $(d_Y)_R$ &  $1-(2Y-1)N$ &  \textbf{1} & \textbf{N} \\ [\extable]
 \hline
\end{tabular}
\end{center}
\captionof{table}{ \label{reps_Standard_Model_Sp(r)_theory}
Transformation properties for chiral fields under the gauge group (\ref{extended_SM_gauge_group}).
}
\end{table}
}\endgroup